\documentclass{emulateapj}
\usepackage{amsmath}
\usepackage{rotating}

\def\lsim{\mathrel{\rlap{\lower4pt\hbox{\hskip1pt$\sim$}}
    \raise1pt\hbox{$<$}}}                
\def\gsim{\mathrel{\rlap{\lower4pt\hbox{\hskip1pt$\sim$}}
    \raise1pt\hbox{$>$}}}                

\def\plotonespecial#1{\centering \leavevmode
    \includegraphics[angle=90,width=1.35\columnwidth]{#1}}

\def\plottwospecial#1#2{\centering \leavevmode
    \includegraphics[angle=90,width=1.01\columnwidth]{#1} \hfil
    \includegraphics[angle=90,width=1.01\columnwidth]{#2}}

\def\plottwosuperspecial#1#2{\centering \leavevmode
    \includegraphics[angle=0,height=0.88\columnwidth]{#1} \hfil
    \includegraphics[angle=90,width=1.18\columnwidth]{#2}}

\def\plottwokindaspecial#1#2{\centering \leavevmode
    \includegraphics[angle=90,width=1.0\columnwidth]{#1} \hfil
    \includegraphics[angle=90,width=1.02\columnwidth]{#2}}

\begin{document}

\title{The Assembly of the Red Sequence at $z\sim1$: The Color and Spectral Properties of Galaxies in the C\lowercase{l}1604 Supercluster}
\author{Lemaux, B.~C., Gal, R.~R.\altaffilmark{1}, Lubin, L.~M., Kocevski, D.~D.\altaffilmark{2}, Fassnacht, C.~D., McGrath, E.~J.\altaffilmark{2}, Squires, G.~K.\altaffilmark{3}, Surace, J.~A.\altaffilmark{3}, \& Lacy, M.\altaffilmark{4}}
\affil{Department of Physics, University of California, Davis, 1 Shields Avenue, Davis, CA 95616, USA}
\email{lemaux@physics.ucdavis.edu}
\altaffiltext{1}{University of Hawai'i, Institute for Astronomy, 2680 Woodlawn Drive, Honolulu, HI 96822, USA}
\altaffiltext{2}{University of California Observatories/Lick Observatory, University of California, Santa Cruz, CA 95064, USA}
\altaffiltext{3}{Spitzer Science Center, California Institute of Technology, M/S 220-6, 1200 E. California Blvd., Pasadena, CA 91125, USA}
\altaffiltext{4}{NRAO, 520 Edgemont Road, Charlottesville, VA 22903, USA}
\begin{abstract}
We investigate the properties of the 525 spectroscopically confirmed members of the Cl1604 supercluster at $z\sim0.9$ as 
part of the Observations of Redshift Evolution in Large Scale Environments (ORELSE) survey. In particular, we focus on the photometric, 
stellar mass, morphological, and spectral properties of the 305 member galaxies of the eight clusters and groups that comprise the Cl1604 
supercluster. Using an extensive Keck LRIS/DEIMOS spectroscopic database in conjunction with ten-band ground-based, \emph{Spitzer}, 
and \emph{Hubble Space Telescope} imaging, we investigate the buildup of the red sequence in groups and clusters at high redshift. Nearly
all of the brightest and most massive red-sequence galaxies present in the supercluster environment are found to lie within the bounds of the 
cluster and group systems, with a surprisingly large number of such galaxies present in low-mass group systems. Despite the prevalence
of these red-sequence galaxies, we find that the average cluster galaxy has a spectrum indicative of a star-forming galaxy, with 
a star formation rate between those of $z\sim1$ field galaxies and moderate redshift cluster galaxies. The average group 
galaxy is even more active, exhibiting spectral properties indicative of a starburst. The presence of massive, red galaxies and the
high fraction of starbursting galaxies present in the group environment suggest that significant processing is occurring in 
group environments at $z\sim1$ and earlier. There is a deficit of low-luminosity red-sequence galaxies in all Cl1604 clusters and 
groups, suggesting that such galaxies transition to the red sequence at later times. Extremely massive 
($\sim10^{12}$ $\mathcal{M}_{\odot}$) red-sequence galaxies routinely observed in rich clusters at $z\sim0$ are also absent from the Cl1604 clusters
and groups. We suggest that such galaxies form at later times through merging processes. There are significant populations of 
transition galaxies at intermediate stellar masses [$\log(\mathcal{M}_{\ast})=10.25$-$10.75$] present in the group and cluster environments, suggesting that this range is important 
for the buildup of the red-sequence mass function at $z\sim1$. Through a comparison of the transitional populations present in 
the Cl1604 cluster and group systems, we find evidence that massive blue cloud galaxies are quenched earliest 
in the most dynamically relaxed systems and at progressively later times in dynamically unrelaxed systems. 
 
\end{abstract}


\section{Introduction}

In the local universe, the Sloan Digital Sky Survey (SDSS) has greatly enhanced our understanding of galaxy properties. 
Studies of SDSS data have revealed insights into the nature of star formation and quenching (e.g., Goto et al.\ 2003;
Brinchmann et al.\ 2004; Kauffmann et al.\ 2004), properties of clusters and their member galaxies (e.g, G{\'o}mez et al.\ 2003;
Hansen et al.\ 2009; von der Linden et al.\ 2010), relationships between fundamental observable quantities (e.g., Bernardi et al.\ 2003;
Tremonti et al.\ 2004; Blanton et al.\ 2005, La Barbera et al.\ 2010), and the properties of active galactic nuclei (AGNs) and their host galaxies
(e.g., Kauffmann et al.\ 2003; Kewley et al.\ 2006; Yan et al.\ 2006). Such studies, however, are, by themselves, of limited use in the
context of galaxy evolution as they provide only a snapshot of the result of galaxy processing over a Hubble time. It is only by
comparing such galaxies to those at higher redshifts that galaxy evolution can be fully investigated.

At high redshifts, there exist several surveys (e.g., zCOSMOS, DEEP2, VVDS) that contain both large samples of UV/optical spectra necessary
to characterize star-formation activity, stellar ages, and metallicities and the high-resolution multiwavelength data necessary to characterize
morphologies, AGN contributions, and stellar masses. These surveys have been instrumental in probing the nature of galaxy evolution both in the
field and intermediate-density regimes (e.g., Ilbert et al.\ 2005; Cooper et al.\ 2006, 2007, 2008; Faber et al.\ 2007; P{\'e}rez-Montero et al.\ 2009; Cucciati
et al.\ 2010). By comparing the evolution of fundamental relationships, such as, e.g., mass-metallicity, morphology/color$-$density, and
star formation rate (SFR)$-$density, between $z\sim1$ and the present day, a picture of galaxy evolution in such environments has begun
to emerge. The number density of
star-forming, blue late-type galaxies in group environments decreases significantly from $z\sim1$ to $z\sim0$, with a corresponding
rise in the number density of red, quiescent early-type galaxies (ETGs; Poggianti et al.\ 2008; Balogh et al.\ 2009; Tasca et al.\ 2009;
Wilman et al.\ 2009; McGee et al.\ 2011). Similarly, the red/early-type fraction correlates weakly with
local density at $z\sim1$, with red galaxies only slightly preferring overdense environments. At lower
redshifts this correlation becomes stronger; the fraction of red galaxies in low-mass group environments increases
significantly from $z\sim1$ to lower redshifts, while remaining essentially unchanged in field environments (e.g., Cooper et al.\ 2007).
These results suggest that the highest density environments present in such surveys (i.e., low-mass groups) play the most prominent
role in this picture.

However, the limited range of environments present in such surveys limits the conclusions that can be drawn from these data. Due to the 
scarcity of massive galaxy associations, these surveys contain limited information on intermediate-density
(i.e., rich group) and high-density (i.e., cluster) environments. This is problematic for galaxy evolution studies, as it has long been
known (Butcher \& Oemler 1984) that such environments are instrumental in the transformation of galaxies. In the last half
decade, surveys of higher redshift clusters extending to several times
the virial radius at $z\sim$0.5 (e.g., Treu et al.\ 2003; Dressler et al.\ 2004; Poggianti et al.\ 2006; Ma et al.\ 2008, 2010; Oemler et al.\ 2009) 
and the innermost cores of clusters at $z\sim1$ (e.g., Postman et al.\ 2005) seem to support this claim, as galaxies in rich groups
and clusters show strong differential evolution relative to the field over the last $\sim5$ Gyr.

The lack of comprehensive data sets of cluster galaxies
at the same redshift range as these surveys means that the processes responsible for driving this evolution in galaxy clusters and
high mass groups are still not well understood. This is partly due to the sheer number of processes that galaxies are subjected to
in high-density environments that are either not present or less effective in the field (e.g., ram-pressure stripping, harassment, strangulation, tidally induced
merging, and tidal stripping). The overlapping spheres of influence of each effect and the requirement of high signal-to-noise ratio
(S/N) spectral data necessary to precisely quantify star formation histories (SFHs), separate from the effects of stellar mass and
metallicity, make disentangling these processes both extremely complicated and observationally expensive. Though several studies
have attempted such analyses, there is significant disagreement as to the primary mechanism responsible for driving galaxy
evolution in cluster environments (see, e.g., the literature review in Oemler et al.\ 2009). This disagreement is likely due, at least
in part, to the spread in global properties of cluster galaxy populations, as well as the varying galaxy selection functions for each
study and the clustocentric radius to which they
extend. Furthermore, as discussed extensively in Moran et al.\ (2007), the dominant mechanism responsible for galaxy transformation in the cluster
environment is likely to vary from cluster to cluster. Processes like ram-pressure stripping will be more effective in virialized, massive
clusters, while merging and low-velocity tidal interactions should be more prevalent in lower mass systems. As such, in order to
gain a comprehensive picture of galaxy evolution in these environments from $z\sim1$ to the present day, it is necessary to study the galaxy
populations of high-redshift clusters that encompass a wide range of both dynamical states and masses.

One manifestation of cluster-specific evolution is the cluster red sequence. At lower redshifts, massive virialized galaxy
clusters are marked by a tight sequence of red galaxies observed in color$-$magnitude space (e.g., Bower et al.\ 1992; van Dokkum et al.\
1998; Terlevich et al.\ 2001; L\'{o}pez-Cruz et al.\ 2004; Haines et al.\ 2006). At higher redshift, clusters observed to be dynamically young and X-ray
underluminous show increasing scatter in their red sequences as well as a significant deficit of low-luminosity red-sequence galaxies
(RSGs; e.g., De Lucia et al.\ 2004, Homeier et al.\ 2006a; Mei et al.\ 2009), a clear sign that the red sequence is still being assembled.
The question of which primary mechanism is responsible for
building up the red sequence, a question intimately related to the transformation of blue late-type galaxies into quiescent ETGs, 
is, however, still not settled (e.g., Faber et al.\ 2007). Standard galaxy scenarios predict the bulk of star formation to occur in
brightest cluster galaxies (BCGs) primarily at $z\sim3$, with fainter luminosity red-sequence members forming the bulk of their stars at 
progressively later epochs. Thus, by studying galaxy clusters at $z\sim1$, only 4 Gyr after the nominal formation epoch of BCGs, it is 
possible to observe clusters in their early stages of assembly.

This paper is the first in a series studying the spectral, color, and morphological properties of galaxy clusters and high- to
intermediate-mass groups at $z\sim1$. In this paper we present the \emph{Hubble Space Telescope (HST)} Advanced Camera for Surveys (ACS) magnitude, color, and morphological
properties as well as the composite Keck I/II Low-Resolution Imaging Spectrometer (LRIS; Oke et al.\ 1995) and DEep Imaging Multi-Object
Spectrograph (DEIMOS; Faber et al.\ 2003)
spectral properties of the 525 spectroscopically confirmed members of the Cl1604 supercluster at $z\sim0.9$. In addition, we
present stellar masses of the galaxy populations that comprise the eight groups and clusters of the Cl1604 supercluster. These stellar masses are
used to investigate the buildup of the red sequence in these structures independent of star-formation effects, as even small amounts
of star formation can significantly alter galaxy magnitudes (Bruzual 2007). In future papers we will extend this work
to investigate the SFR$-$density and morphology/color$-$density relationships of Cl1604 galaxies as well as galaxy populations of
other $z\sim1$ large scale structures as part of the Observations of Redshift Evolution in Large Scale Environments (ORELSE) survey 
(Lubin et al.\ 2009, hereafter L09). The virtues of the current observational data sets in the ORELSE fields include 
multiwavelength imaging and spectroscopy across large areas, extending to several virial radii in most fields, and uniform field-to-field
selection functions used to target galaxies for spectroscopy. In addition, the
ORELSE structures span a large range in mass, X-ray/optical properties, richness, and dynamical states allowing investigations of galaxy
evolution over a variety of different regimes at high-redshift.

The paper is organized as follows. \S2 presents the observation and reduction of our optical and near-infrared (NIR) imaging
and spectral data, \S3 presents the data analysis, \S4 presents our results, \S5 discusses the implication
of our results, and \S6 presents our conclusions. Throughout this paper we adopt a concordance $\Lambda$CDM cosmology, with
$H_{0}$ = 70 km s$^{-1}$ Mpc$^{-1}$, $\Omega_M$=0.27, and $\Omega_{\Lambda}$=0.73. All equivalent width (EW) measurements
are presented in the rest frame. We adopt the convention that negative EWs are used for 
features observed in emission and positive EWs for those in absorption. All magnitudes are given in the
AB system (Oke \& Gunn 1983; Fukugita et al.\ 1996).

\section{Observations and Reduction}

The Cl1604 supercluster was observed as part of the ORELSE survey (L09). The environments present within the Cl1604 supercluster span from rich 
($\sim800$ km s$^{-1}$), virialized clusters dominated by red, ETGs and a hot intracluster medium, to moderate mass ($\sim$300-500 
km s$^{-1}$) groups and sparse chains of galaxies dominated by starbursts and luminous AGN (Gal et al.\ 2008; Kocevski et al.\ 2009b; 2011a). 
This structure and some of the associated data have been described in great detail in other papers (Gal \& Lubin 2004;
Gal et al.\ 2008; Kocevski et al.\ 2009a, 2009b, 2011a, 2011b; Lemaux et al.\ 2009, 2010). In the following section we review all data obtained
on the Cl1604 supercluster to date, including new data that has not been previously presented.  

\subsection{Optical and Near-infrared Imaging}
\label{imaging}

Initial wide-field $r\arcmin i\arcmin z\arcmin$ optical imaging of the Cl1604 supercluster was taken with the Large Format
Camera (LFC; Simcoe et al.\ 2000) mounted on the Palomar Hale 5-m telescope. These data were reduced using Image Reduction and
Analysis Facility (IRAF; Tody 1993) with a set of publicly available routines. Photometry was performed using Source Extractor
(SExtractor; Bertin \& Arnouts 1996) and is described in further detail in \S\ref{photometry} and Appendix A. Further details of the
observation and reduction are described in Gal et al.\ (2005, 2008). The LFC images reach
5$\sigma$ point source limiting magnitudes of 25.2, 24.8, \& 23.3 mags in the $r\arcmin$, $i\arcmin$, and $z\arcmin$ bands, respectively.

Wide-field NIR imaging of the Cl1604 field was obtained with two different sets of observations. Imaging in the $K_{s}$
band was obtained with the Wide-field InfraRed Camera (WIRC; Wilson et al.\ 2003) mounted at prime focus 
on the Palomar Hale 5-m telescope on 2006 August 8th and
9th UTC. Conditions were photometric and seeing ranged from 0.9$\arcsec$-1.3$\arcsec$ in the $K_{s}$ band. The WIRC data were processed
using a combination of scripts written in IDL and IRAF. All frames were corrected for dark current and flat fielded using dome flats.
The sky background in each frame was fit using a third order polynomials in both coordinates and subtracted. Known bad pixels and satellite
trails were masked. Astrometry was obtained by fitting to stars from the USNO A2 catalog using the task {\em msccmatch}. The dark-corrected,
flat-fielded, sky-subtracted, bad-pixel-masked images at each pointing were then median combined using the IRAF task {\em imcombine}.
A second astrometric correction was applied to the final image in the same manner as for the individual exposures. These data were primarily
used to perform our spectral energy density (SED) fitting for the purpose of obtaining stellar masses (see \S\ref{SEDfitting})
and reach a 5$\sigma$ point source limiting magnitude of $K_{s}$=21.3.

Imaging in the $K$ band was also obtained with
the Wide-Field Camera (WFCAM; Casali et al.\ 2007) mounted on the United Kingdom Infrared Telescope (UKIRT) on 2007 April 29-30 UTC in photometric
conditions and 0.6$\arcsec$-0.7$\arcsec$ seeing. These data were processed using the standard UKIRT processing pipeline
courtesy of the Cambridge Astronomy Survey Unit\footnote{http://casu.ast.cam.ac.uk/surveys-projects/wfcam/technical}. These data
were used to obtain $K$-band stellar masses when SED fitting was not available or poorly constrained (see \S\ref{SEDfitting}).
The UKIRT imaging is deeper than the WIRC imaging, reaching a 5$\sigma$ point source limiting magnitude of 22.4, equivalent
to a 0.2$L^{\ast}$ cluster elliptical at $z\sim0.9$.

A portion of the Cl1604 field is spanned by a 17 pointing \emph{HST} ACS (Ford et al.\ 1998) mosaic. 
Fifteen of these pointings are single orbit observations in both the \emph{F}606\emph{W} and 
\emph{F}814\emph{W} filters, reaching 5$\sigma$ point source limiting magnitudes of 27.2 \& 26.8 mags, respectively.
Two of the pointings, centered on clusters Cl1604+4304 and Cl1604+4321, are deeper, reaching
5$\sigma$ point source limiting magnitudes of 28.1 \& 27.6 mags in \emph{F}606\emph{W} \& \emph{F}814\emph{W}, respectively.
Further details on the observation and reduction of these data can be found in Kocevski et al.\ (2009b).

Deep \emph{Spitzer} InfraRed Array Camera (IRAC; Fazio et al.\ 2004)) 3.6/4.5/5.8/8.0 $\mu$m imaging has also been obtained for the entire Cl1604 field
as part of the \emph{Spitzer} program GO-30455 (PI L.\ M.\ Lubin). Data were reduced using the standard \emph{Spitzer} Science Center reduction
pipeline and further processed using a modified version of the SWIRE survey pipeline (Surace et al.\ 2005).
The total exposure time of the mosaic is
1080 s per pixel, which results in 5$\sigma$ point source limiting magnitudes of 24.0, 23.7, 22.2, and 21.9 mags in IRAC
channels 1-4, respectively. Additional observations with \emph{Spitzer} were obtained with the Multiband Imaging
Photometer for SIRTF (MIPS; Rieke et al.\ 2004) at 24$\mu$m and cover a large fraction of the supercluster field. The effective exposure time
of the observations in the area covering the supercluster members is 1200 s per pixel, which results in a
5$\sigma$ point source limiting magnitude of $m_{24\mu\rm{m}}$= 19.4, or roughly $L_{TIR}=3\times10^{10}L_{\odot}$
at $z\sim0.9$.  Further details on the observation and reduction of both the IRAC and
MIPS data can be found in Kocevski et al.\ (2011a; hereafter K11).

\subsection{Optical Spectroscopy}
\label{spectroscopy}

The original spectroscopic data in the Cl1604 field were obtained using LRIS
on the Keck 10-m telescopes. The initial LRIS campaign consisted of a magnitude limited survey (R $<$ 23) that targeted galaxies
in the vicinity of two of the constituent clusters of the Cl1604 supercluster system, Cl1604+4304 and Cl1604+4321 (see Oke, Postman, 
\& Lubin 1998 for further details). Following the original survey, a follow-up LRIS spectroscopic campaign
of six slitmasks was undertaken in the Cl1604 field in the vicinity of clusters Cl1604+4314 and Cl1604+4321
(see Gal \& Lubin 2004 for details). In total
85 high-quality redshifts were obtained with LRIS between 0.84 $\leq$ $z$ $\leq$ 0.96, the
adopted redshift range of the Cl1604 supercluster.

The bulk of the redshifts in the Cl1604 field come from observations
of 18 slitmasks with DEIMOS on the Keck II 10-m telescope between 2003 May and 2010 June. The
details of the observations and spectroscopic selection of 12 of these masks are described in
Gal et al.\ (2008; hereafter G08) and Lemaux et al. (2010; hereafter L10). The remaining six DEIMOS masks 
(referred to hereafter as ``completeness masks") were designed to obtain a magnitude limited sample to a depth 
of $F814W\sim23.5$ across a 16.7$\arcmin$ $\times$ 5$\arcmin$ subsection
of the field running roughly north to south and encompassing clusters Cl1604+4314 and Cl1604+4321
(hereafter clusters B and D, adopting the naming convention of G08). In total, we targeted 90\% of galaxies
with $F814W\le23.5$ in the subsection of the Cl1604 field covered by the completeness masks and
obtained high quality redshifts for 75\% of galaxies brighter than this limit. In the remainder
of the field our spectroscopic completeness limit was slightly shallower, roughly complete to a depth of
$F814W\sim22.5$ and reaching a limiting magnitude of $F814W\sim27.1$.

All DEIMOS slitmasks were observed with the 1200 l mm$^{-1}$ grating with an FWHM
resolution of $\sim$1.7\AA\ (68 km s$^{-1}$), with a typical wavelength coverage
of 6385-9015\AA. The slitmasks were observed with differing total integration times
depending on weather and seeing conditions and varied from 3600s to 14400s in seeing that
ranged from 0.45$\arcsec$-1.4$\arcsec$. The initial 12 masks were observed 
for an average total integration time per mask of $\sim$2.75 hr, while the completeness masks are
much shallower, averaging just under 1.5 hr of total integration per mask. The exposure frames
for each DEIMOS slitmask were combined using a modified version of the DEEP2 \emph{spec2d} package
(Davis et al.\ 2003). The details of this package as well as the reduction process are
described further in Lemaux et al.\ (2009; hereafter Lem09). In total, 1340 total high-quality (Q $\geq$ 3;
see G08 for detailed explanations of the quality codes) extragalactic
DEIMOS spectra were obtained in the Cl1604 field, with 440 objects having measured redshifts within the adopted redshift
range of the supercluster. Combined with the additional redshifts obtained in the two LRIS campaigns,
525 high quality spectra have been obtained for members of the Cl1604 supercluster.

\section{Analysis}

\subsection{Imaging Measurements}

\subsubsection{Photometry}
\label{photometry}

Nearly all of the results presented in this paper rely heavily on the magnitudes of the
Cl1604 supercluster members as measured in our ten-band imaging. Thus, 
both the accuracy and precision of our
absolute photometric measurements and the self-consistency of these measurements
from band to band are extremely important. The latter is of particular concern,
as poorly matched apertures from multiband imaging can result in significant issues with differential photometry
(Vanzella et al.\ 2001; Coe et al.\ 2006), which introduces bias into the SED fitting process. In Appendix A we 
describe the processes used to obtain reliable photometry from our LFC, WIRC, WFCAM, IRAC, and ACS imaging, as well
as discuss our choice of apertures for each band and systematics associated with these choices. We refer interested
readers to Appendix A.

\subsubsection{Spectral Energy Distribution Fitting and Stellar Masses}
\label{SEDfitting}

Synthetic stellar templates were fit to the optical/IR SED of each galaxy in the Cl1604 field with the Le PHARE (Arnouts et al.\ 1999; Ilbert et al.\ 2006)
code using the single stellar population (SSP) models of Bruzual \& Charlot (2003) with a Chabrier initial mass function (IMF; Chabrier 2003).
For galaxies with spectroscopic redshifts, the redshift was used
as a prior to constrain the range of best-fit templates. For each galaxy, $\chi^2$ minimization is performed by Le PHARE relative
to six parameters: the redshift (in the case of no spectroscopic redshift prior), stellar mass, stellar age, extinction,
metallicity, and $\tau$, the e-folding time of the star-formation event in the galaxy. Extinction values were bounded by
the range $E(B-V)$ = 0 to $E(B-V)$ = 0.5 in six bins of size $\delta E(B-V)=0.1$ using a Calzetti et al. (2000) reddening
law (though the stellar mass, the most important parameter derived from this fitting, is relatively insensitive to this 
choice; see, e.g., Swindle et al.\ 2011). Metallicity values were chosen to be 0.2$Z_{\odot}$, 0.4$Z_{\odot}$, and $Z_{\odot}$, 
consistent with the range of metallicity values used for other high-redshift surveys (e.g., Ilbert et al.\ 2010). Additionally, the 
BC03 SSP models contained nine different values of $\tau$, ranging from a near-instantaneous burst ($\tau=0.1$ Gyr) 
to a model consistent with that of a dwarf spiral galaxy ($\tau=30$ Gyr). The stellar mass and mean stellar age of each galaxy was not
discretized, but was rather constrained by the SFH of the best-fit template and scaled by the
observed luminosity. Errors on each parameter are estimated through Monte-Carlo simulations in which each broadband magnitude
is varied by its formal error to simulate random errors in the photometry and does not account for any systematic bias. 
For this paper, we require that a galaxy be detected in at least the $r\arcmin, i\arcmin, z\arcmin,$ \& $K_{s}$ bands and have 
a secure spectroscopic redshift to consider the SED stellar mass reliable. The resulting average stellar mass error for the 
$\sim375$ Cl1604 members that fulfill these criteria is 0.14 dex.

For those Cl1604 member galaxies that went undetected in any of the LFC bands ($r\arcmin$/$i\arcmin$/$z\arcmin$) or
the WIRC $K_{s}$ imaging, stellar masses were derived using our UKIRT $K$-band imaging. The observed UKIRT $K$ magnitude
for each detected galaxy was converted to a rest-frame $K$-band luminosity by applying an evolutionary \emph{k}-correction of
$-1.5$ (using a BC03 $\tau=0.6$ Gyr, $z_{f}$=3 SSP model, see L09). Interpolated values of
$K$-band mass-to-light ($\mathcal{M}$/$L$) ratios at $z=0.9$ (Drory et al.\ 2004) are multiplied by the resulting $K$-band luminosity to obtain
stellar mass estimates. Errors in these estimates are derived from the formal errors in our UKIRT photometry. The resulting
average stellar mass error for the Cl1604 members detected in the UKIRT imaging is 0.07 dex.

Stellar masses derived from UKIRT data were compared to those estimated by Le PHARE for the subset of Cl1604 members detected in
at least four bands. The scatter of the stellar masses derived from the two methods is reasonably small ($\sim0.23$
dex\footnote{This scatter is significantly increased relative to the quadrature sum of the formal errors of the two mass estimates.
The quoted errors on the two mass estimators are random errors only and do not include the systematic errors associated with
our choice of templates for the SED fitting, imperfect \emph{k}-corrections, and our ignorance of the rest-frame $K$-band $\mathcal{M}$/$L$
ratios. It appears that these systematic errors and random errors discussed earlier contribute roughly equally.}, and,
perhaps more importantly, there exists no bias between the two methods as a function of stellar mass.
For this paper, Le PHARE-derived stellar masses are given preference over UKIRT-derived stellar masses in cases where both
mass estimates were available and reliable. In total, the two methods resulted in reliable stellar mass measurements for 452
of the 525 members of the Cl1604 supercluster system of which 399 are detected in our ACS imaging.

\subsubsection{Group and Cluster Membership}
\label{membership}

Since many of the results presented in this paper rely on the comparison of the member galaxy populations of the eight
spectroscopically confirmed clusters and groups in the Cl1604 supercluster, we define here our criteria for cluster or
group membership. Our general philosophy is to err on the side of being overly inclusive, such that we include all galaxies
that could potentially be associated with the cluster and to include a large range of environments. For the majority of
the cluster and group systems we consider a galaxy a member of a particular system if it satisfies
$(1)$ $\delta v<\pm3\sigma_{v}$, where $\delta v$ is the velocity offset of a galaxy from the systemic velocity of the
group or cluster and $\sigma_{v}$ is the group or cluster velocity dispersion, and $(2)$ $r_{\rm{proj}}\leq2R_{\rm{vir}}$, where $r_{\rm{proj}}$
is the projected radial offset of a galaxy from the group or cluster center and $R_{\rm{vir}}$ is the virial radius.
The center of each system is defined as the centroid (as determined by SExtractor) of the smoothed red galaxy overdensity of each system
and is described in detail in G08. The errors on these centroids, estimated by comparing centroids derived in this manner to 
X-ray centroids for all X-ray bright ORELSE clusters with the requisite data, ranges from 25 to 150 kpc (3-20$\arcsec$). While the upper limit of
this error is somewhat large, we stress that the optically derived centroid is the more relevant quantity for determining the local density in 
systems that are X-ray underluminous and still in the process of formation (as most of the sc1604 systems are). Thus, we choose to ignore
this error for the remainder of the paper. The virial radius for each system is defined in terms of the radius 
at which the mean density is equal to 200 times the critical density of the universe at the redshift of that group or cluster ($R_{200}$), 
such that $R_{\rm{vir}}=R_{200}/1.14$ (Biviano et al.\ 2006; Poggianti et al.\ 2009). The value of $R_{200}$ is calculated by (Carlberg et al. 1997):

\begin{equation}
R_{200} = \frac{\sqrt{3}\sigma_{v}}{10 H(z)}
\label{eqn:R200}
\end{equation}

\noindent where $H(z)$ is the value of the Hubble parameter at the redshift of interest. The values of $\sigma_{v}$ are taken
from K11 and G08. For each cluster and group system we also calculate a virial mass, given by:

\begin{equation}
M_{\rm{vir}} = \frac{3\sqrt{3}\sigma_{v}^3}{11.4 G \, H(z)}
\label{eqn:Mvir}
\end{equation}

\noindent where $G$ is Newton's gravitational constant.

The one exception to these criteria is group C, in which we see a continuum of galaxies
at $\pm3\sigma_{v}$ spanning from the group core to well past $2R_{\rm{vir}}$. This observation is consistent with preliminary
velocity dispersion measurements of the members of group C using new data obtained since the publication of K11. In this measurement
we observe a significant increase in the velocity dispersion relative to the value reported in K11, suggesting that we have
underestimated the virial radius of group C by adopting the K11 value. As such, we relax the projected radius criterion
for this group, considering any galaxy a member if it lies within $\delta v<\pm3\sigma_{v}$ and $2.5R_{\rm{vir}}$ of the
group center. Using these and the above criteria results in 305 of the 525 member galaxies of the Cl1604 supercluster ($\sim57$\%) being
classified as either group or cluster members. The remaining 220 galaxies that are not associated with a particular cluster
or group system will be referred to hereafter as the ``superfield" population. The number of members, as well as the name,
location, mean redshift, velocity dispersion, virial radius, and virial mass for each Cl1604 cluster and group are given in
Table \ref{tab:globalproperties}.

As discussed in detail
in G08 and K11, the groups and clusters of the Cl1604 supercluster span a large range of masses and dynamical states.
Our two most massive clusters, A and B, are of nearly identical (optically derived) mass, but show significant
differences in their galaxy populations, X-ray properties, and the radial distributions of their RSGs.
In addition, as discussed briefly in K11, and as will be discussed later in this paper, the group systems also show similar
variance in the radial distribution and the color properties of their constituent galaxies. Since it takes a cluster or
group galaxy traveling 1000 km s$^{-1}$ less than 2 Gyr to fall into the cores of these structures from the 
maximum projected radii (our largest projected cutoff is 2.10 $h^{-1}$ Mpc; cluster B), being liberal in our membership
criteria includes galaxies that may eventually be virialized into the cluster or group
cores by $z\sim0$. This way we can study both the current assembly of blue-cloud and RSGs in these structures as
well as discuss the likely evolution of these systems over the next several Gyr. For some parts of our analysis we will be
interested only in the former point and will restrict our study to galaxies at smaller projected radii.

\begin{deluxetable*}{ccccccccc}
\tablecaption{Properties of the Galaxy Groups and Clusters in the Cl1604 Supercluster \label{tab:globalproperties}}
\tablehead{\colhead{} & \colhead{} & \colhead{} & \colhead{} & \colhead{} & \colhead{} & \colhead{$\sigma_v$} & \colhead{$R_{\rm{vir}}$} & \colhead{$M_{vir}$} \\
\colhead{Name} & \colhead{ID} & \colhead{$\alpha_{\rm{J}2000}$} & \colhead{$\delta_{\rm{J}2000}$} & \colhead{$\langle z \rangle$} & \colhead{\rm{N}$_{\rm{mem}}$\tablenotemark{a}} & \colhead{(km s$^{-1}$)} & \colhead{($h_{70}^{-1}$ Mpc)} & \colhead{$(\times10^{14} h_{70}^{-1} \mathcal{M}_{\odot})\tablenotemark{b}$}} 

\startdata
A & Cl1604+4304 & 241.0975 & 43.0812 & 0.898 & 41 & 703$\pm$110 & 0.92 & 3.28$\pm$1.53 \\[4pt]
B & Cl1604+4314 & 241.1051 & 43.2396 & 0.865 & 85 & 783$\pm$74  & 1.05 & 4.61$\pm$1.31 \\[4pt] 
C & Cl1604+4316 & 241.0316 & 43.2631 & 0.935 & 21 & 304$\pm$36  & 0.39 & 0.26$\pm$0.09 \\[4pt]
D & Cl1604+4321 & 241.1387 & 43.3534 & 0.923 & 96 & 582$\pm$167 & 0.75 & 1.83$\pm$1.57 \\[4pt]
F & Cl1605+4322 & 241.2131 & 43.3709 & 0.936 & 28 & 543$\pm$220 & 0.70 & 1.48$\pm$1.80 \\[4pt] 
G & Cl1604+4324 & 240.9251 & 43.4017 & 0.901 & 17 & 409$\pm$86  & 0.53 & 0.64$\pm$0.41 \\[4pt] 
H & Cl1604+4322 & 240.8965 & 43.3731 & 0.852 & 10 & 302$\pm$64  & 0.42 & 0.27$\pm$0.17 \\[4pt] 
I & Cl1603+4323 & 240.7969 & 43.3918 & 0.902 & 7 & 359$\pm$140 & 0.47 & 0.44$\pm$0.51  \\[4pt] 
\tablenotetext{a}{Within $R<2R_{\rm{vir}}$ for all systems except C. For group C we allow $R<1h_{70}^{-1}$ Mpc, see \S\ref{membership}.} 
\tablenotetext{b}{Errors in $M_{\rm{vir}}$ are calculated from errors in $\sigma_v$}
\enddata
\end{deluxetable*}

\subsubsection{Morphology}
\label{morphology}

For all Cl1604 supercluster members observed in our 17 pointing \emph{HST} ACS mosaic, morphological classification
was assigned through visual inspection of the data by one of the authors (LML). Briefly, ACS cutout images of each supercluster
member galaxy were generated and presented to the inspector without prior knowledge of their location in the
supercluster. For each inspected galaxy a
primary morphological type was assigned using standard Hubble classification, as well as information on
merger and interaction signatures, tidal features, etc. For this paper, we adopt the convention that all galaxies classified
as spirals as well as those classified as irregular or amorphous (Sandage \& Brucato 1979) are defined as late-type systems, while
galaxies classified as either elliptical or S0 are early-type (though we discriminate between these two classes later in the paper).
Merging systems, which were typically separated in the ACS imaging, were
assigned the morphological classification of the galaxy associated with the DEIMOS/LRIS spectrum. In cases where the merging system
was not separated in our imaging, or in cases where the primary galaxy was obscured by the merging process we did not assign an 
late-/early-type morphological classification. Such cases were rare, however, only comprising $\sim2.6$\% of the cluster and group 
members with reliable stellar masses. For completeness, we include such systems when analyzing the color/stellar 
mass/morphological properties of member galaxies in \S\ref{massmorph}, but we leave their morphological classification ambiguous. 
Visual inspections are preferred here over
statistical quantities (i.e., Gini, M20, compactness, etc.) due to the added information that can be included when visually classifying
galaxies and due the relatively small number of galaxies of the sample, which makes visual inspection
feasible. Regardless, we find good agreement with the morphologies derived through visual inspections and those
derived through more automated statistical methods (see discussion in K11). 

In order to estimate the precision associated with the visual classification process, a random subset of 150 supercluster 
galaxies was presented to two of the authors (LML, RRG) for classification. These galaxies were presented blindly, in that 
neither author had knowledge of the original morphological classification of the galaxy. This process was used to test both
the consistency of visual classification of a single observer and to test the objectivity of the 
process by including multiple classifiers. In both cases the results were comparable to the original classification, with 
the rms corresponding to roughly half a class, where a class refers to late-type, S0, and
elliptical\footnote{More specifically, we assigned a number to each galaxy with elliptical=0, S0=1, and late-type=2 and 
took the difference between each trial for each galaxy. The resulting rms was $\sigma=0.51$ when comparing multiple
classifications by a single observer and $\sigma=0.68$ when comparing results from multiple observers.} Thus, we expect 
roughly $5$\%$-10$\% of our sample to be morphologically misclassified. Since none of the results presented in this study are 
sensitive to changes of this level we choose to ignore this uncertainty for the remainder of the paper. 

\subsubsection{Red-sequence Fitting}
\label{RSfitting}

For many of our studies we must divide systems not only into categories
defined by their morphological classification but also to differentiate between red and blue galaxy
populations. This will be especially important in \S\ref{radial} and \S\ref{massmorph} when comparisons are made
between the red and blue galaxy populations of groups and clusters of very different masses in the Cl1604 supercluster. As
such, we use the color$-$magnitude properties of each system to formally define the \emph{HST} ACS
``red sequence" and ``blue cloud" galaxy populations in each of the constituent systems of the Cl1604
supercluster. The process of determining a formal red sequence for each system is similar to that used 
in Gladders et al.\ 1998 and Stott et al.\ 2009 and is described in detail in Appendix B. The slope, intercept, and
width of the red sequence for each of the three Cl11604 clusters, as well as the composite ``Groups" sample (see Appendix B), 
are given in Table \ref{tab:RSparameters}. In addition, these red-sequence fits
are plotted, along with the color and magnitude properties of the constituent galaxies of each system, in
\S\ref{colormag} and \S\ref{colormass}.

\begin{deluxetable}{cccc}
\tablecaption{Red-sequence Fitting Parameters of the Cl1604 Galaxy Groups and Clusters \label{tab:RSparameters}}
\tablehead{\colhead{Name} & \colhead{Intercept} & \colhead{Slope} & $1\sigma$ Width\tablenotemark{a}}
\startdata
Cluster A & 2.20$\pm$0.02 & $-0.020$\tablenotemark{b} & 0.046 \\[4pt]
Cluster B & 3.24$\pm$0.15 & $-0.065$\tablenotemark{b} & 0.048 \\[4pt]
Cluster D & 3.21$\pm$0.30 & $-0.062$\tablenotemark{b} & 0.045 \\[4pt]
Groups    & 2.95$\pm$0.19 & $-0.051$\tablenotemark{b} & 0.076 \\[4pt]
\tablenotetext{a}{For all clusters $\pm3\sigma$ from the best fit color$-$magnitude relation was adopted for the width
of the red sequence. For the group systems $\pm2\sigma$ was adopted for the width (see Appendix B).}
\tablenotetext{b}{The formal error in the red-sequence slope of all systems is smaller than 10$^{-3}$.}
\enddata
\end{deluxetable}

\subsection{Spectral Measurements}

In this section we present the method used to extract measurements from our spectra,
estimate their errors, and generate composite spectra of the galaxy populations of these systems.

\subsubsection{Composite Spectra}
\label{composite}

Composite spectra were generated for the member galaxies of each Cl1604 cluster and group system 
following the method of Lem09. Composite spectra were created separately for members observed with 
DEIMOS and those observed with LRIS so as to not degrade the higher resolution DEIMOS data. 
Our use of variance weighting (see Lem09), in principle, will give higher average weight to brighter galaxies in the sample
(due to galaxies effectively being weighted by their S/N ratio). The primary motivation for this weighting scheme
is to down-weight those pixels that have been affected by poor night sky subtraction or which fall in the 10 \AA\ chip
gap between the red and blue CCD arrays on DEIMOS. While the difference between continuum S/N of the brightest and faintest
galaxies in any individual system is, on average, a factor of 2-3, the differences between S/N ratios near spectral
features of interest (i.e., [\ion{O}{2}] and H$\delta$) is significantly less. We, therefore, choose to ignore this effect
for any equivalent width measurements made on composite spectra. 

However, for $D_n(4000)$ measurements (see next section) this effect may be more pronounced since
$D_n(4000)$ is a quantity that relies on direct measurement of the significant portions of the continuum. As a
result of our weighting scheme, composite spectra produced in such a manner will be slightly biased to higher $D_n(4000)$
values (i.e., older average stellar populations). In order to determine the magnitude of this effect we have compared the
$D_n(4000)$ measurements of composite spectra created using other weighting schemes (e.g., luminosity weighting, clipped
variance weighting) and found the resulting difference to be of order $\delta D_n(4000)=0.03$. While this difference is
certainly not trivial, the conclusions presented in \S\ref{discussion} are robust to changes of this level to $D_n(4000)$. We, therefore,
choose to ignore this bias for $D_n(4000)$ measurements as well.

\subsubsection{Equivalent Width and $D_n(4000)$ Measurements}
\label{EWnD4000}

Equivalent widths (EWs) of the [\ion{O}{2}] $\lambda$3727\AA\ and $H\delta$ $\lambda$4101\AA\ features were measured from
composite spectra of group and cluster galaxies following the bandpass
method of L10. While fitting methods generally lead to more precise results in the case of high S/N
spectra, the process of combining group or cluster galaxies into a single
composite spectrum tends to blur out small scale features in the constituent spectra, which diminishes the effectiveness
and usefulness of such methods. Bandpasses for both the [\ion{O}{2}] and $H\delta$ features were adopted from 
Fisher et al.\ (1998). For further details on the method used to calculate EWs see L10.
Since composite spectra from DEIMOS and LRIS were generated separately (see \S\ref{composite}), EW measurements were performed on each
set of spectra separately. For EW measurements of the spectrum of a given galaxy population, the final EW value
was calculated by number-weighting the individual EW
measurements from the DEIMOS and LRIS composite spectra. Errors on these quantities were similarly calculated. Table \ref{tab:EWnD40001} gives
the EW([\ion{O}{2}]) and EW(H$\delta$) measurements from the composite spectra of the eight Cl1604 groups and clusters. For all EW measurements
we ignore the effect of differential extinction, which generally has a small effect on EW measurements (see discussion in L10).

In addition to EWs, the strength of the continuum break at 4000\AA\ ($D_n(4000)$) is measured from composite spectra using the ratio of
the blue and red continua as defined by Balogh et al.\ (1999). Mean flux density values are calculated from the $\sigma$-clipped
spectrum of each region, with the $D_n(4000)$ index defined as $D_n(4000)=\langle F_{\lambda,r} \rangle/\langle F_{\lambda,b} \rangle$.
Errors on the $D_n(4000)$ index are calculated from the variance spectrum in each region, again using $\sigma$-clipping
to avoid regions of poor night sky subtraction or regions that fell within the 10 \AA\ CCD chip gap. As with EWs, measurements
of the $D_n(4000)$ value were performed separately on DEIMOS and LRIS composite spectra for each group and cluster system and 
combined by a number-weighted average. $D_n(4000)$ measurements from the composite spectra of the member
galaxies of the eight groups and clusters of the Cl1604 system are given in Table \ref{tab:EWnD40001}. 

The effects of reddening on
$D_n(4000)$ are not negligible. An approximately 1 Gyr old SSP 
with no dust [i.e., $E(B-V)=0$] has a $D_n(4000)$ that is 10\% smaller than that of a identical age SSP with 
significant dust [i.e., $E(B-V)=0.5$]. Differences in metallicity have a similar effect, changing $D_n(4000)$ by roughly 6\%
in $\sim1$ Gyr old SSPs when metallicity changes by a factor of two.
Though these effects are reasonably large, the quantitative work involving the $D_n(4000)$ index in this paper
relies not only on the $D_n(4000)$ index but also on the EW measurements described above. Through such analysis we are able 
to mitigate the effects of dust and metallicity differences when interpreting the evolutionary state of a particular system of galaxies. 
More importantly, in all cases throughout the paper our main conclusions do not change if the dust or metallicity properties of the 
galaxies are altered significantly.
 
\subsubsection{Spectroscopic Selection and Completeness}
\label{bootstrap}

\begin{figure*}
\epsscale{1.17}
\plotone{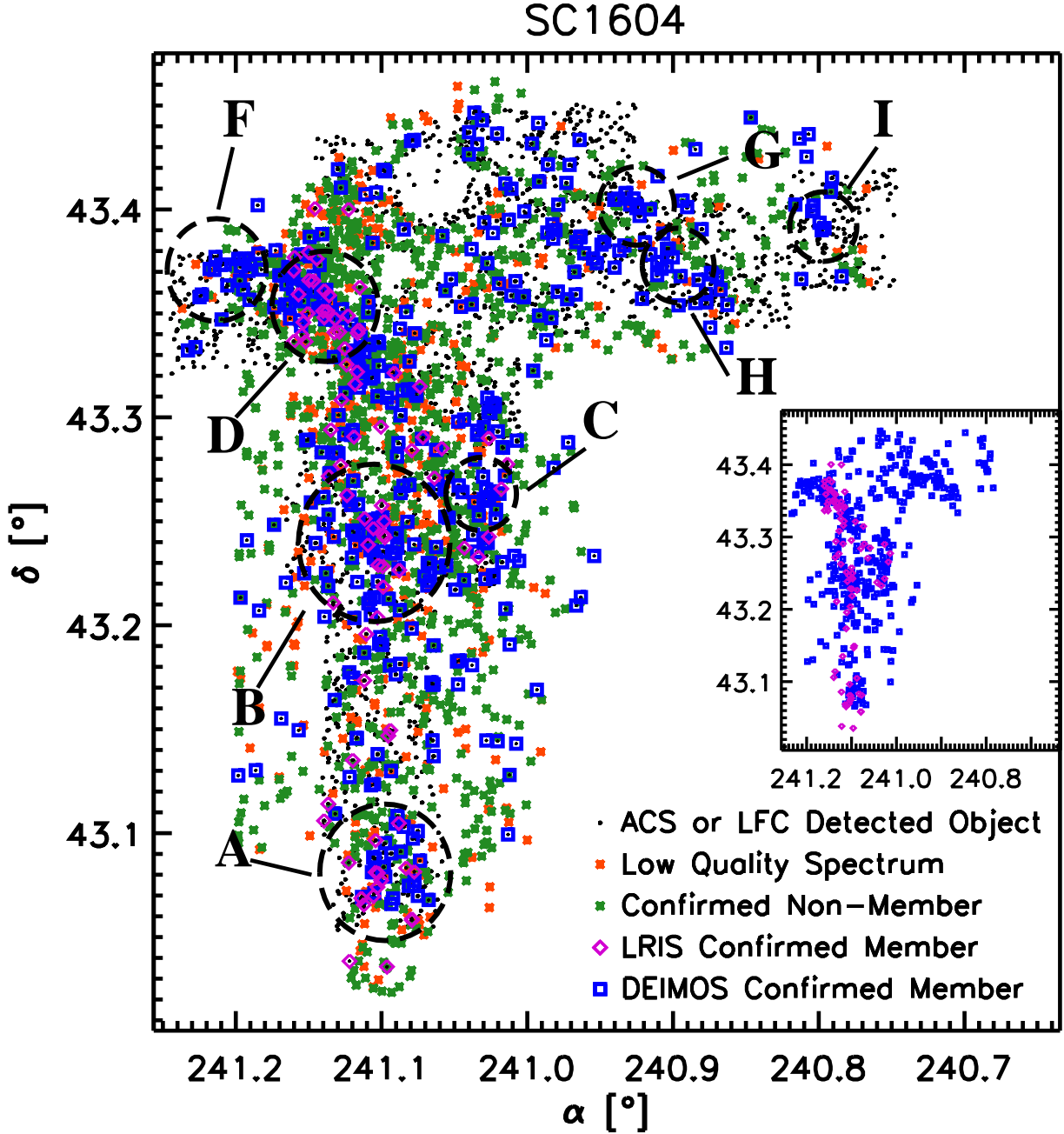}
\caption{The Cl1604 supercluster at $z\sim0.9$. All photometric objects within the \emph{HST} ACS field of view brighter than
$F814W<23.5$ or $i\arcmin<23.5$ are plotted as small black dots. In addition, we plot all photometric objects for which we
have obtained spectroscopic information. The 525 confirmed members of the Cl1604 are circumscribed by blue squares or
magenta diamonds. Small green $\times$s denote stars and galaxies outside of the redshift range of the supercluster.
Spectroscopic objects for which we were unable to obtain a high-quality redshift are shown as small orange $\times$s.
The name of each cluster and group is labeled. Dashed lines indicate the virial radius of each system. The inset on the right
side of the plot shows spatial distribution of the supercluster members only.}
\label{fig:RAdecincompleteness}
\end{figure*}

With over 500 spectroscopically confirmed members, the Cl1604 supercluster is one of the most well-studied large scale structures
at intermediate redshifts. Despite this fact, there exist a significant number of galaxies both within the superfield and within the
truncation radius of the constituent groups and clusters for which we do not have spectroscopic
information (see Figure \ref{fig:RAdecincompleteness}). This issue is further complicated by our method of selecting
targets for spectroscopy, which has evolved considerably over the course of the spectroscopic campaign.
These selections have resulted in certain areas of the supercluster that are roughly spectroscopically complete, either to $R<23$
(clusters A and D; see Oke, Postman, \& Lubin 1998) or to $F814W<23.5$ (clusters B and D and the superfield spanning the two structures; see
Figure \ref{fig:RAdecincompleteness} and \S\ref{colormag}), while other areas, like those that include the five Cl1604 groups,
have sparser spectroscopic coverage. In order to investigate the effects that spectroscopic incompleteness and selection have
on our results bootstrap analysis was performed on the composite spectra of all Cl1604 systems. This analysis, which is described in 
detail in Appendix C, uses a combination of the \emph{HST} ACS photometry and the DEIMOS/LRIS spectroscopic information in such a way
so as to simulate the \emph{maximum possible variance} of the composite EW and $D_n(4000)$ values due to spectroscopic sampling alone. 

While these ``incompleteness errors" can be quite large relative to the formal random errors (see Table \ref{tab:EWnD40001}), we 
stress that the errors generated by this process properly account for the effects of differing spectroscopic selection functions and 
spectroscopic coverage. In such a way, any statistically significant
differences that we observe between the composite Cl1604 galaxy populations and that of low-redshift samples with similar spectroscopic
coverage to that of Cl1604 or high-redshift samples with sparser spectroscopic coverage \emph{must be the result of true differences in the
properties of the galaxies}. Similarly, this is true when making comparisons between the composite galaxy properties of individual clusters
or groups within the supercluster. For details on the methodology used to estimate these incompleteness errors see Appendix C.

\section{Results}

\subsection{Color$-$Magnitude Properties}
\label{colormag}

\begin{figure*}
\epsscale{0.8}
\plotone{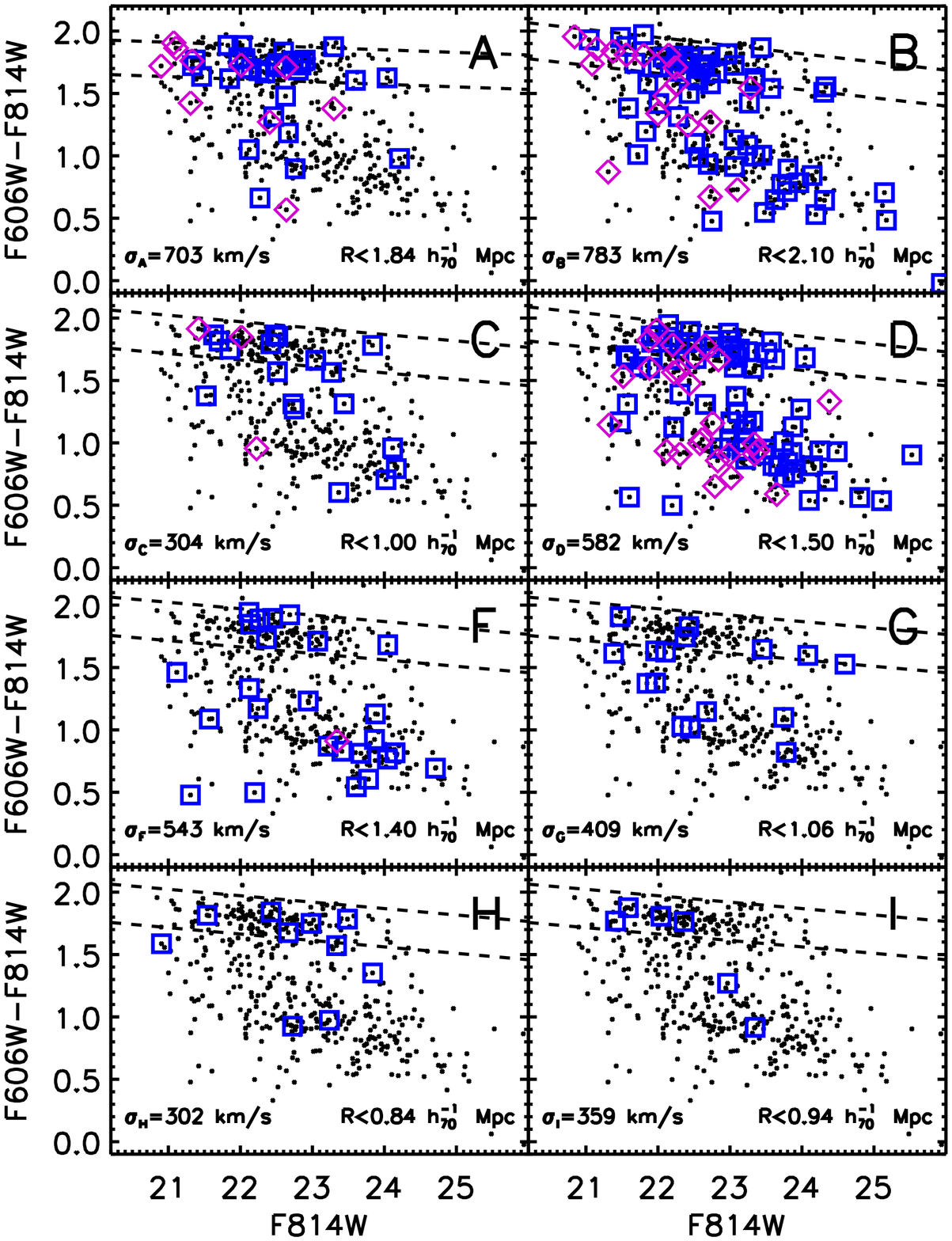}
\caption{\emph{HST} ACS color$-$magnitude diagram of the members of the Cl1604 supercluster. Plotted in small black points in each panel
are the 460 spectroscopically confirmed members of the supercluster detected in both ACS bands. Galaxies circumscribed by blue squares (DEIMOS
confirmed) or magenta diamonds (LRIS confirmed) denote the members of a particular group or cluster and dashed lines indicate the
bounds of red sequence for each system (see \S\ref{RSfitting} and Table \ref{tab:RSparameters}). The name of the group
or cluster is given in each panel, along with the velocity dispersion of each system and the projected radial cutoff used
for membership (see \S\ref{membership}). Despite being at nearly the same epoch, large variations in color$-$magnitude properties are observable
in the three clusters (A, B, \& D) and the five groups (C, F, G, H, \& I) of the Cl1604 supercluster.}
\label{fig:individualCMDs}
\end{figure*}

In Figure \ref{fig:individualCMDs} we plot the ACS color$-$magnitude diagrams (CMDs) for the three clusters (A, B, \& D) and
five groups (C, F, G, H, \& I) of the Cl1604 supercluster. In each panel we plot the 460 Cl1604 members which 
fall in the ACS field of view (small black points) to highlight the range of colors and magnitudes spanned by the member
galaxies of the supercluster system.
The magenta diamonds (LRIS confirmed) and blue squares (DEIMOS confirmed) in each panel indicate the members of that particular system.
Cluster and group membership is defined by the criteria given in \S\ref{membership}. In total, using a truncation radius of $2R_{\rm{vir}}$,
the Cl1604 groups and clusters contain 288 of the 467 (62\%) ACS detected spectroscopically confirmed members of the
supercluster.

\begin{figure*}
\plottwokindaspecial{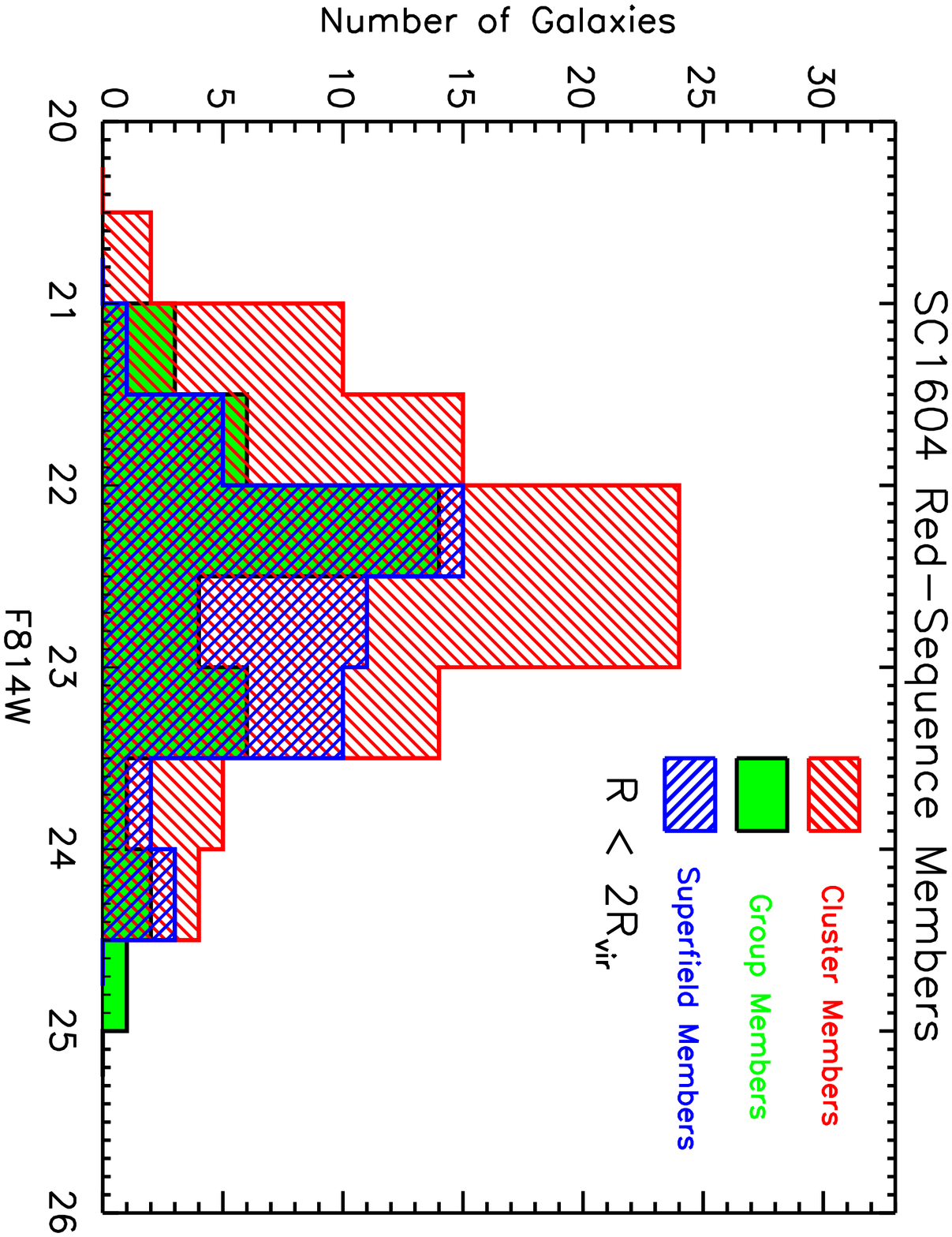}{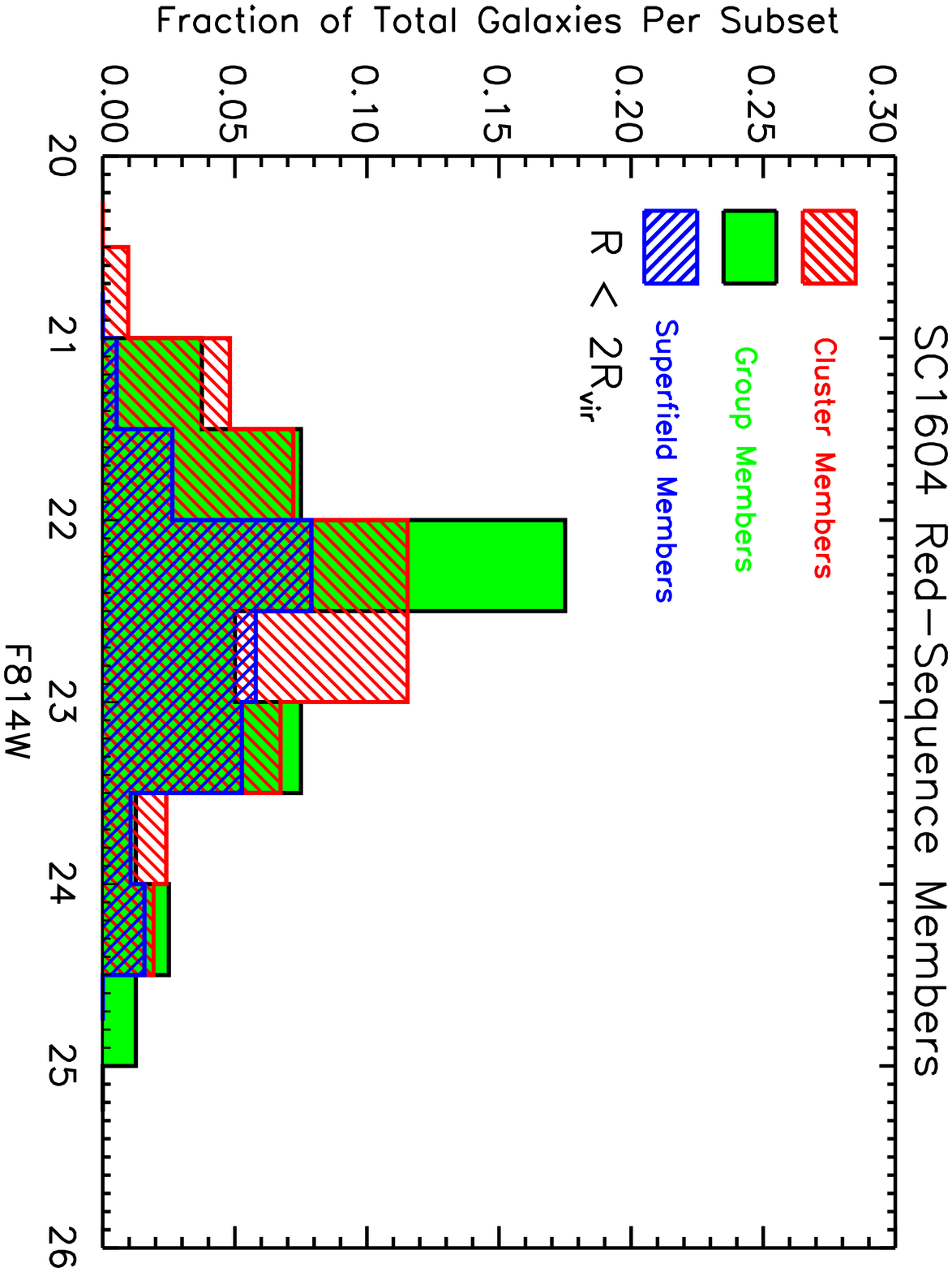}
\caption{\emph{Left:} Histogram of the number of RSGs spectroscopically confirmed in the cluster, group, and superfield 
samples as a function of $F814W$. The radial criterion for group and cluster membership is shown on the plot. The superfield population is 
comprised of all Cl1604 member galaxies that do not belong to a particular cluster or group. While the total number of galaxies in the 
clusters and superfield samples are roughly similar, the group sample contains roughly half the number of galaxies. Despite this, at
brighter ($F814W\lsim22.5$) magnitudes the number of group RSGs matches or exceeds those of the superfield RSGs. \emph{Right:} Fractional 
contribution of RSGs to the total cluster, group, and superfield population as a function of $F814W$. At nearly all magnitudes 
RSGs contribute more to the total cluster and group populations than in the superfield. This difference is especially noticeable 
for brighter ($F814W\lsim22.5$) RSGs.}
\label{fig:fractionalmag}
\end{figure*}

Looking at the CMDs, a few observations are immediately clear.
A large fraction ($\sim$76\%) of RSGs and virtually all of the bright RSGs
in the Cl1604 supercluster are contained within the groups and clusters. This can also be seen in Figure \ref{fig:fractionalmag} where we
plot both the total number of RSGs, as well as the fractional contribution of RSGs, as a function of $F814W$ magnitude for the cluster, group, 
and superfield samples. The fraction of RSGs 
in the combined Cl1604 clusters and groups sample is 47\%, while it is only 23\% for superfield
galaxies. Additionally, at nearly every magnitude, the fractional contribution of RSGs is significantly more in cluster and group environments 
than in the Cl1604 superfield. Despite the fact that these groups and clusters
are optically selected, generally X-ray underluminous (see Kocevski et al.\ 2009a), and still in the process of formation, the member galaxies
of the groups and clusters are already beginning to distinguish themselves from their field counterparts. Surprisingly, the red sequence
fraction of the galaxy population in the Cl1604 clusters is 47\%, \emph{identical} to the fraction for just the  
Cl1604 group galaxies. This suggests that significant processing has occurred, and at similar levels, in both group
and cluster environments at $z\sim0.9$. The considerable processing observed in the Cl1604 group environments will be a recurring point
in later sections.

What is further striking in Figure \ref{fig:individualCMDs} is the large variance in the color and magnitude
properties of the cluster and group galaxies from structure to structure. While the three clusters differ in their (optically
derived) virial mass ($M_{\rm{vir}}$) by only a little over a factor of two (and are consistent within the errors, see Table
\ref{tab:globalproperties}), both the fraction of RSGs and the number of bright blue galaxies change
drastically from cluster to cluster. In cluster A, a cluster that is relatively relaxed
and dominated by a bright ICM (see Kocevski et al.\ 2009a), the fraction of RSGs is quite high ($\sim70$\%)
and essentially no bright blue-cloud galaxies are observed. In the X-ray underluminous clusters B and D, the red-sequence fraction
is significantly lower, 49\% and 36\%, respectively, and a large number of bright blue galaxies are observed (though these two
populations have significantly different properties, see \S\ref{colormass}).

\begin{figure*}
\epsscale{0.8}
\plotone{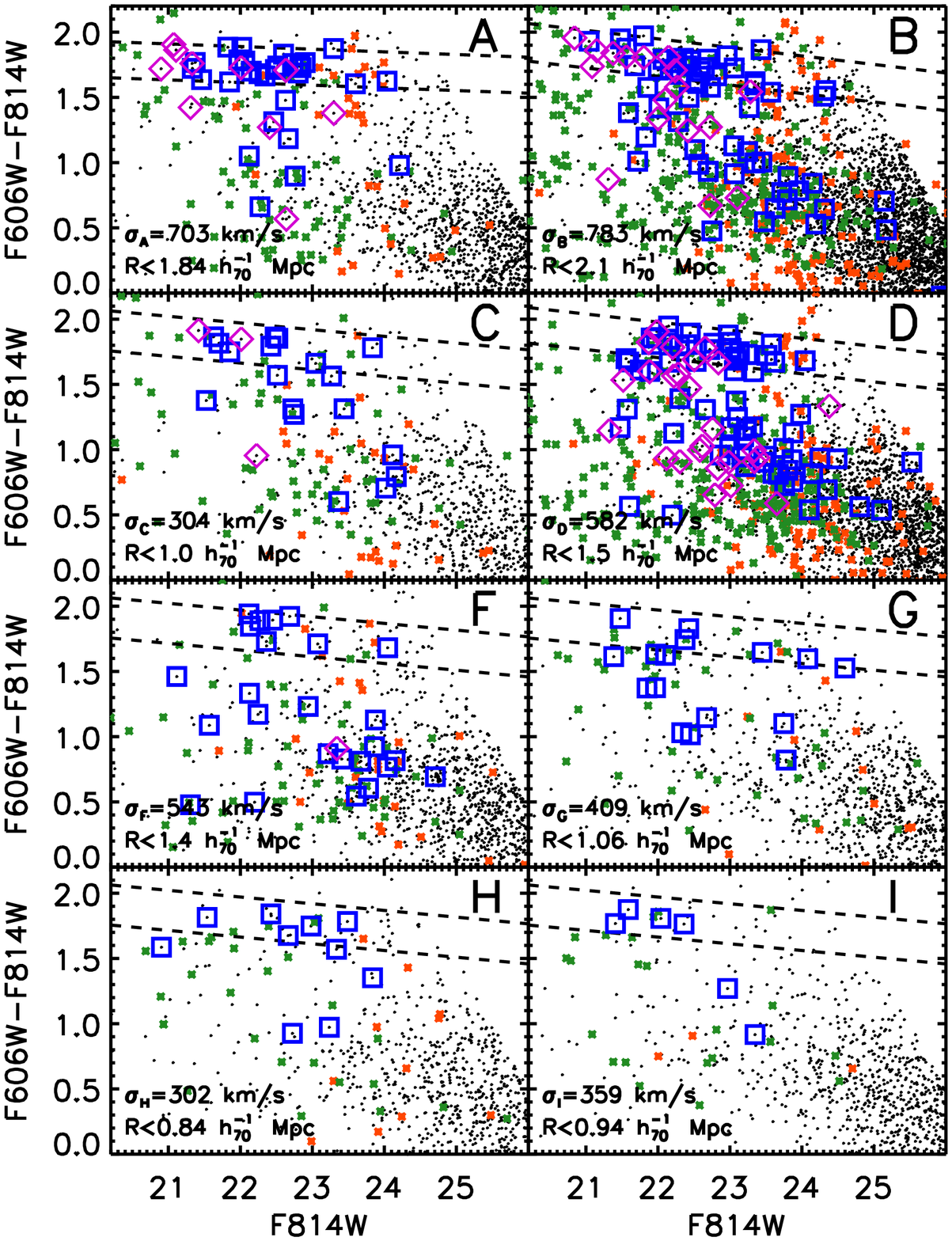}
\caption{\emph{HST} ACS color$-$magnitude diagram of all photometric objects lying within $R<2R_{\rm{vir}}$ of any group or
cluster center. The meaning of the circumscribed squares and magenta diamonds in each panel meanings are identical to
those in Figure \ref{fig:individualCMDs}. Small green $\times$s denote stars and galaxies outside of the redshift range of a particular
cluster or group and small orange $\times$s denote Spectroscopic objects for which we were unable to obtain a high-quality redshift.
Additionally, no magnitude cut is imposed on photometric objects (small black points). As in Figure \ref{fig:individualCMDs}, 
the name of each group or cluster as well as the radial cutoff for membership and associated velocity dispersion is shown in each
panel. Only those galaxies that are detected in both ACS bands and are brighter than $F606W<28$ and $F814W<28$ are shown. Nearly 
all photometric objects brighter than $F814W<23.5$ within $R<2R_{\rm{vir}}$ of the center of clusters B \& D were targeted for 
spectroscopy, while for cluster A and the group systems this is true only for galaxies on the red sequence.}
\label{fig:individualCMDswcompleteness}
\end{figure*}

In the group systems, the variance of the color$-$magnitude properties of the
member galaxies is even more pronounced. The most massive group in the Cl1604 system (group F) has
the lowest observed fraction of RSGs (31\%) of any structure in the supercluster and a large fraction of
bright, blue, 24$\mu m$-detected starburst galaxies (see K11). Conversely, the two lowest mass
group systems in Cl1604 (groups C and H) have observed
red-sequence fractions that are $\gsim50$\% and contain a large fraction of the brightest RSGs observed in the
group systems. The errors on the virial mass estimates of the group systems are, however, quite large (see Table
\ref{tab:globalproperties}). Further increasing our uncertainty is the large fraction of blue galaxies in groups F and G,
which may be artificially inflating the observed velocity dispersion relative to groups comprised
primarily of RSGs (as in, e.g., Zabludoff \& Franx 1993). Considering these large uncertainties, if we
instead assume all group systems belong to roughly the same mass halo, the variance in the colors and magnitudes of
the group members observed from system to system is still surprising. From Figure \ref{fig:individualCMDswcompleteness}
we conclude that this variance and the variance of the color$-$magnitude properties of the cluster members is not due to incomplete
spectral sampling, but rather represents \emph{real differences in the galaxy populations of the Cl1604 groups and clusters}.

\subsection{Global Spectral Properties}
\label{spectral}

\begin{figure}
\epsscale{1.2}
\plotone{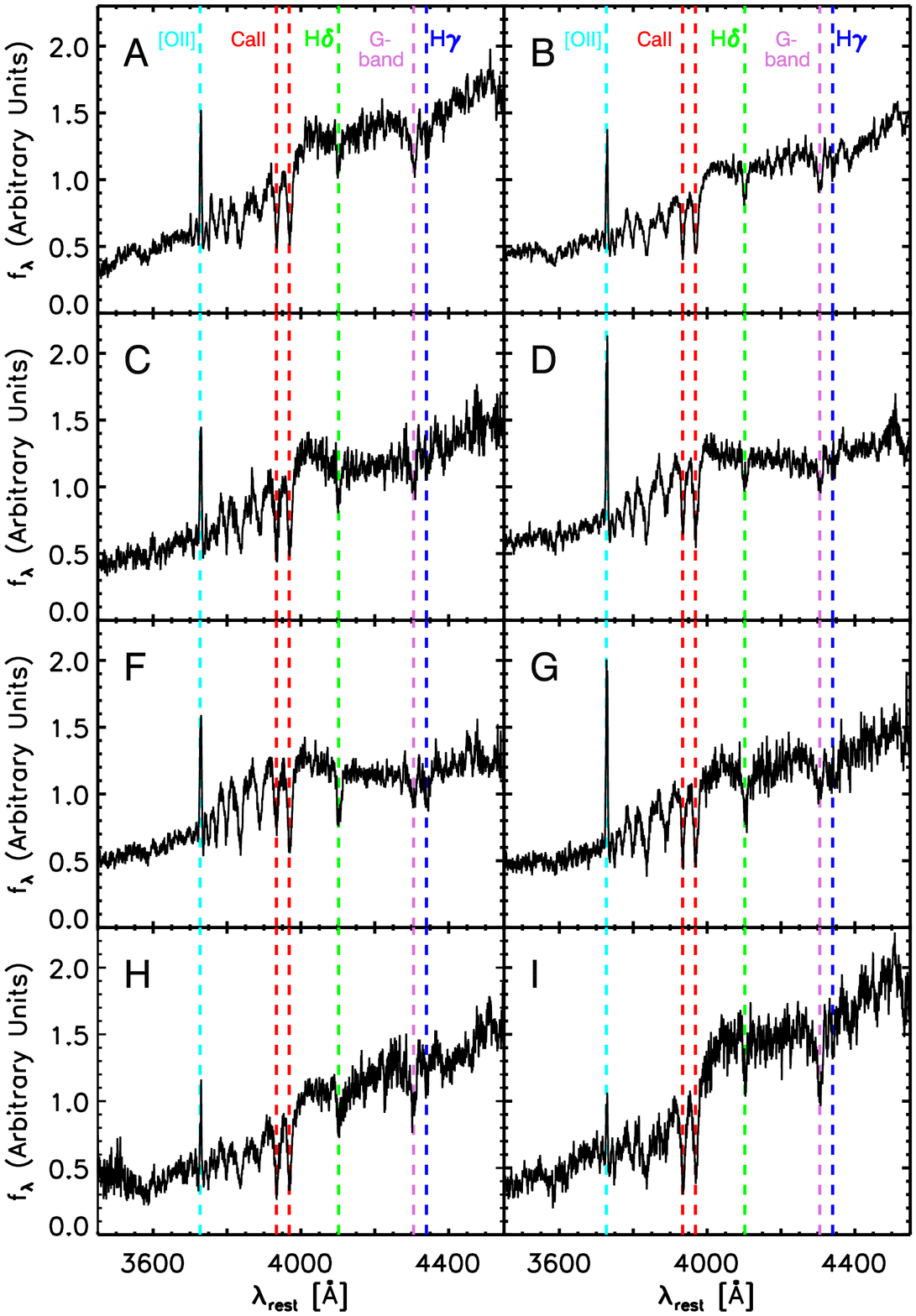}
\caption{DEIMOS composite spectra of member galaxies of each of the eight Cl1604 group and cluster systems. Composites
of group and cluster members using spectra obtained with LRIS are generated separately and are not included here.
Important spectral features are marked and the name of each cluster or group system is given in the top right corner of
each panel. Spectra are smoothed with a Gaussian kernel of $\sigma=2.2$ pixels (0.36\AA\ at the restframe of the supercluster).
Significant differences are apparent in the spectra of the average group members. The average member in groups F \& G
is young [small $D_n(4000)$], with a several strong features indicative of recently formed stars (H$\delta$ and H$\gamma$).
In contrast, the continuum of the average member of group I is dominated by older stellar populations.}
\label{fig:spectralmosaic}
\end{figure}

The differences in the galaxy populations between the Cl1604 groups and clusters are not limited to their broadband
properties. In Figure \ref{fig:spectralmosaic} we plot the composite UV/optical spectra of member galaxies of the clusters
and groups observed with DEIMOS. Important spectral emission and absorption features are overlaid in the plot (for
a review of these features see Burstein et al. 1984; Rose 1985; L10). Since the DEIMOS spectra make up
$\gsim70$\% of the spectrally confirmed members in the Cl1604
clusters and nearly all of the confirmed members of the groups (see Figure \ref{fig:individualCMDs}), we plot the
DEIMOS composite spectra here to highlight the general spectral properties of the cluster and group populations.
The LRIS composite spectra, which we will include later in the section when measuring spectral quantities, are generated separately
from the DEIMOS spectra (see \S\ref{composite}) and are not shown here.

Just as significant variance was observed in the color$-$magnitude properties of the group and cluster members, we 
observe that variance manifested here in the spectral properties of the \emph{average} member galaxy of each system
(see \S\ref{composite} for a detailed explanation on the meaning of our use of ``average"). A quick inspection of the composite 
spectra of the members in the three cluster systems (A, B, \& D) and the five groups (C, F, G, H, \& I) reveal significant differences in the level of ongoing star-formation
(based on the strength of the [\ion{O}{2}] $\lambda 3727$\AA\ nebular emission feature), the luminosity-weighted fraction of older stellar
populations [based on the strength of the \ion{Ca}{2} and $G$-band $\lambda$4305\AA\ features and $D_n(4000)$, a quantitative measure of
the magnitude of the continuum break at 4000\AA], and the luminosity-weighted fraction of relatively young stars (based on the strength
of the H$\delta$ $\lambda$4101\AA\ and higher order Balmer absorption lines just blueward of CaII).

The average galaxy in cluster A, a system dominated by RSGs (see \S\ref{colormag}), is, not surprisingly, comprised
primarily of an older stellar population [large $D_n(4000)$] with moderate
signatures of recent star formation activity. What is perhaps surprising, however, is the [\ion{O}{2}] emission feature
is stronger in the average galaxy in cluster A than in cluster B, a system with a much lower
fraction of RSGs. This is likely due to non-star-forming processes and will be discussed in more
detail later. The spectrum of the average galaxy in cluster D is significantly different than
its counterpart in either of the higher mass clusters. The typical stellar population 
in D is several Gyr younger [$D_n(4000)_{\rm{D}}=1.25\pm0.018$ versus $D_n(4000)_{\rm{A}}=1.50\pm0.037$ and
$D_n(4000)_{\rm{B}}=1.47\pm0.024$]. The average galaxy in D also shows a higher level of current star formation than 
that of either of the more massive clusters. The spectrum of the average group galaxy similarly varies
from structure to structure, ranging from young systems that are dominated by A stars and ongoing star-formation (group F) to
systems comprised of extremely old stellar populations (group I). In Table \ref{tab:EWnD40001} we list the composite spectral
properties of the member galaxies of the eight Cl1604 groups and clusters.

\begin{figure}
\plotone{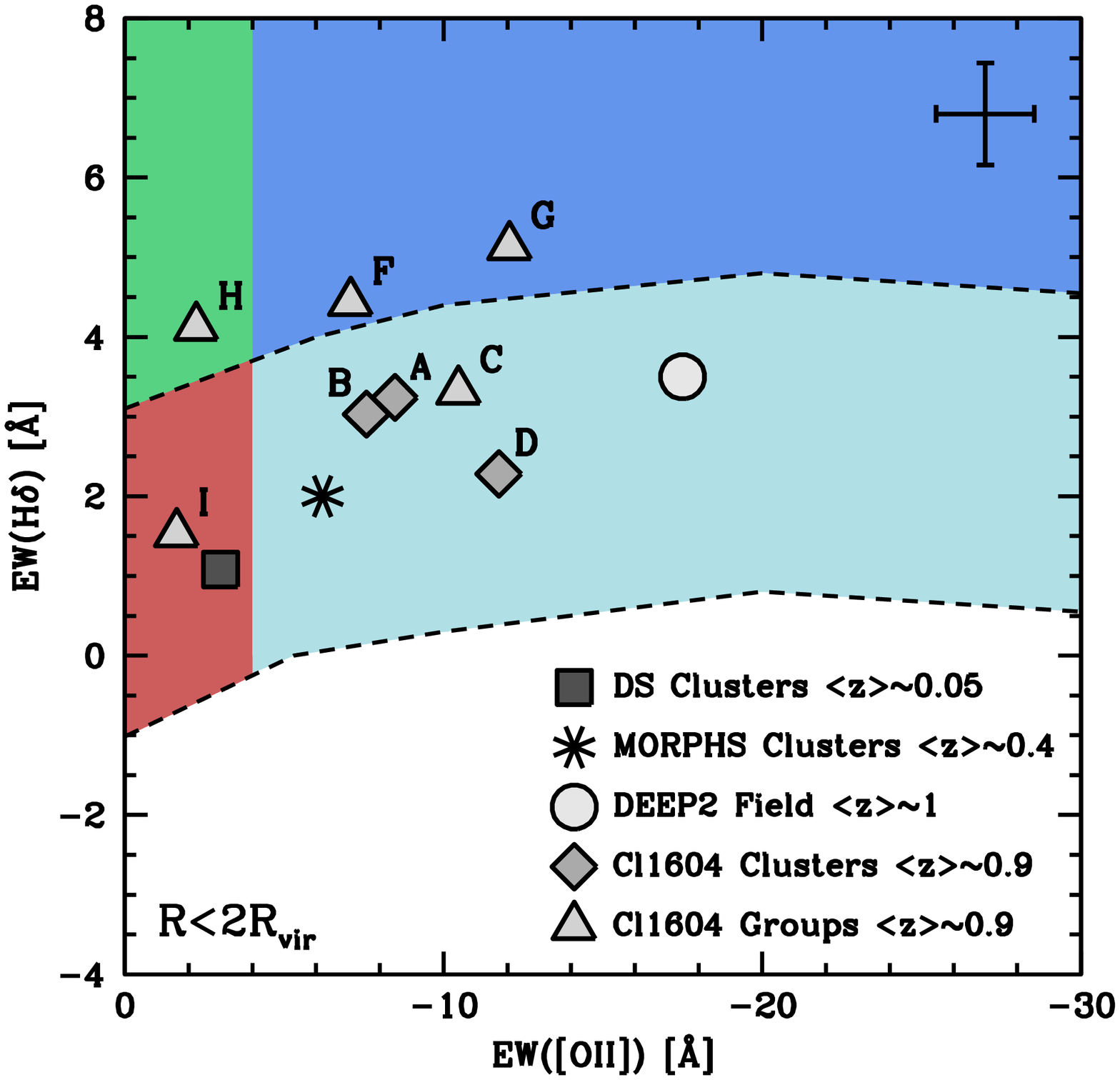}
\caption{Measurements of the equivalent width of the [\ion{O}{2}] and H$\delta$ spectral features from composite spectra of the
member galaxies of the eight groups and clusters which comprise the Cl1604 supercluster. Dashed lines indicate the
area in this phase space which is found to contain 95\% of \emph{normal} star-forming galaxies observed at $z\sim0.1$
(Goto et al.\ 2003; Oemler et al. 2009). The red, light blue, green, and dark blue shaded regions
correspond to quiescent, normal star-forming, post-starburst, and starbursting galaxies, respectively. The average error on
each measurement (which includes incompleteness errors, see \S\ref{bootstrap} and Appendix C) is shown in the upper right corner. Also plotted
are the average EW([\ion{O}{2}]) and EW(H$\delta$) values of field galaxies at a similar redshift as the Cl1604 supercluster
as well as those of lower redshift cluster populations. Significant variations are observed in the average spectral
properties of the Cl1604 cluster and group member populations.}
\label{fig:dresslergeneral}
\end{figure}

To further quantify this variance we plot in Figure \ref{fig:dresslergeneral} the EW of the [\ion{O}{2}] and H$\delta$
spectral features as measured from the composite spectra.
With spectral features that provide us with information on both the level of instantaneous star formation
(in the form of [\ion{O}{2}]) and the level of recent ($\lsim1$ Gyr) star-formation activity (in the form of H$\delta$),
such a diagnostic diagram is useful both to separate star-forming galaxies from quiescent populations
and to determine the manner in which active (i.e., star-forming) galaxies are forming their stars. Also
plotted in Figure \ref{fig:dresslergeneral} are the average properties of $z\sim1$ field galaxies from the DEEP2
redshift survey\footnote{A. Dressler (private communication, 2008).} (Davis et al.\ 2003, 2007), as well as the average
properties of selected cluster galaxies at $z\sim0.4$
(Dressler et al.\ 2004) and $z\sim0.05$ (Dressler \& Shectman 1988). Shaded regions correspond to quiescent, post-starburst,
starburst, and ``normal" (i.e., continuous) star-forming galaxies (red, green, dark blue, and light blue, respectively).

\begin{deluxetable*}{cccc}
\tablecaption{Composite Equivalent Width and $D_n(4000)$ Values of the Galaxy Populations of the Cl1604 Groups and Clusters \label{tab:EWnD40001}}
\tablehead{\colhead{} & \colhead{EW([\ion{O}{2}])\tablenotemark{a}} & \colhead{EW(H$\delta$)\tablenotemark{a}} & \colhead{} \\
\colhead{Name} & \colhead{(\AA)} & \colhead{(\AA)} & \colhead{$D_n(4000)$\tablenotemark{a}}}
\startdata
Cluster A & $-$8.47$\pm$0.16$\pm$1.16  & 3.24$\pm$0.14$\pm$0.79 & 1.501$\pm$0.005$\pm$0.037 \\[4pt]
Cluster B & $-$7.58$\pm$0.15$\pm$1.09  & 3.03$\pm$0.15$\pm$0.50 & 1.472$\pm$0.006$\pm$0.024 \\[4pt]
Group C   & $-$10.47$\pm$0.24$\pm$1.95  & 3.32$\pm$0.23$\pm$0.96 & 1.403$\pm$0.008$\pm$0.038 \\[4pt]
Cluster D & $-$11.74$\pm$0.12$\pm$1.02 & 2.28$\pm$0.13$\pm$0.51 & 1.249$\pm$0.003$\pm$0.018 \\[4pt]
Group F   & $-$7.08$\pm$0.16$\pm$1.72  & 4.43$\pm$0.18$\pm$0.79 & 1.171$\pm$0.004$\pm$0.025 \\[4pt]
Group G   & $-$12.07$\pm$0.23$\pm$1.60  & 5.13$\pm$0.22$\pm$0.52 & 1.339$\pm$0.007$\pm$0.044 \\[4pt]
Group H   & $-$2.24$\pm$0.33$\pm$2.09  & 4.12$\pm$0.30$\pm$0.87 & 1.650$\pm$0.011$\pm$0.038 \\[4pt]
Group I   & $-$1.62$\pm$0.29$\pm$1.60  & 1.53$\pm$0.23$\pm$0.49 & 1.889$\pm$0.013$\pm$0.058 \\[4pt]
Groups\tablenotemark{b} & $-$7.92$\pm$0.21$\pm$1.90  & 4.15$\pm$0.20$\pm$0.63 & 1.381$\pm$0.005$\pm$0.041 \\[4pt]
\tablenotetext{a}{Random and incompleteness errors are reported for EW([\ion{O}{2}]), EW(H$\delta$), and $D_n(4000)$ separately. The
second error given in each column is the uncertainty due to completeness effects (see \S\ref{bootstrap} and Appendix C).}
\tablenotetext{b}{Measurements made a composite spectrum comprised of all Cl1604 group galaxies}
\enddata
\end{deluxetable*}

Prior to investigating the results of Figure \ref{fig:dresslergeneral} for the Cl1604 systems, as well as 
for the DEEP2 field and lower redshift cluster populations, it is necessary to discuss the physical interpretation of EW([\ion{O}{2}]).
While [\ion{O}{2}] is traditionally associated with nebular star formation activity, other process relating to
AGNs or low-ionization nuclear emission-line regions (LINERs) 
generate significant [\ion{O}{2}] emission (Yan et al.\ 2006; L10; Kocevski et al.\ 2011b; Hayashi et al.\ 2011).
This is particularly an issue for [\ion{O}{2}]-emitting RSGs that have no other
indicators of current star-formation activity, as in a large fraction ($\sim90$\%) of such galaxies [\ion{O}{2}] emission originates
from a LINER/AGN. We will discuss the level of contamination in the composite [\ion{O}{2}] emission
from this population later in the section. Interpreting the EW([\ion{O}{2}]) for dust-reddened systems is also
complicated by certain dust geometries, which can non-trivially decrease the measured values of EW([\ion{O}{2}]).
Dust-reddened starbursts can appear in both the blue cloud and on the red sequence, with
differential reddening playing an increasingly significant role the redder such galaxies become. As there is a
large 24$\mu$m-bright starburst population observed in the Cl1604 supercluster (K11), we take care to account
for this population throughout this paper.

For systems primarily comprised of blue-cloud or quiescent\footnote{Quiescent here refers to both star-formation processes and
LINER or other AGN processes} RSGs, 
the relationship between EW([\ion{O}{2}]) and the global SFR of a galaxy requires knowledge of that galaxy's rest-frame UV brightness.
Since our spectral measurements come from composite spectra rather than a single galaxy,
translating the composite EW([\ion{O}{2}]) to an average SFR for each group and cluster galaxy population involves the
rest-frame UV brightness of the average member galaxy in each system. The median absolute $B$-band magnitude, $M_B$,
of the constituent galaxies of the eight Cl1604 groups and clusters varies between 
$M_B=-20.07$ and $M_B=-20.64$. This is a difference of only a factor of $\sim1.5$ in luminosity for the most extreme cases.
These values of $M_B$ are roughly consistent with the median $M_B$ of the DEEP2 field galaxy sample
(see Cooper et al.\ 2007) and that of the cluster galaxy samples at $z\sim0.4$ and $z\sim0.05$ 
(assuming $B-V=0.5$; see Dressler et al.\ 1999, 2004). Thus, we ignore this point
for the remainder of this section and will speak of EW([\ion{O}{2}]) in such systems as being directly proportional to the
global SFR.

With these caveats in mind we examine the properties of the average member galaxies of the eight
Cl1604 groups and clusters. From our measurements of the members of clusters A, B, and D we
find that the \emph{average cluster galaxy at $z\sim0.9$ is a normal star-forming galaxy}, in stark
contrast with the average cluster galaxies at $z\sim0.05$, which appears to be devoid of any star-formation
activity. Furthermore, cluster galaxies at $z\sim0.9$ appear to be forming stars at, on average, roughly half
the rate as those galaxies in the field at similar redshifts, but roughly twice the rate as those
in $z\sim0.4$ clusters.

Measuring only the EW([\ion{O}{2}]) feature from a composite spectrum comprised of only RSGs in both cluster B and the group systems,
we find a negligible contribution to the [\ion{O}{2}] EW from this population (see \S\ref{discussion}). [\ion{O}{2}]-emitting 
LINER/AGN are apparently not prevalent in the RSGs in cluster B or the group systems. The red-sequence population
in both clusters A and D, however, exhibit significant levels of [\ion{O}{2}] emission. In cluster A
this is likely due to contamination from LINERs or AGN (see \S\ref{discussion}). Thus,
for cluster A we interpret the SFR derived from the composite EW([\ion{O}{2}]) measurement as an upper limit. 
In cluster D, much of the [\ion{O}{2}] emission is likely due to residual star formation in galaxies that have recently transitioned to
the red sequence (see \S\ref{discussion}). Furthermore, cluster D, the least massive cluster of the Cl1604 complex,
has a large fraction of dust-reddened starburst galaxies, a population that is less prevalent in the two massive
cluster systems. Thus, for cluster D the
EW([\ion{O}{2}]) value as measured from the composite spectrum is considered a lower limit. Even without these
considerations we observe a trend of decreasing SFR of the average cluster member with increasing halo mass.

If we instead make a correction for the [\ion{O}{2}] emission originating from non-star-forming processes in cluster A,
the EW([\ion{O}{2}]) of the average galaxy in cluster A drops to $\langle EW($[\ion{O}{2}]$)\rangle_{\rm{A}}=-5.4$\AA. This correction is
made by artificially setting the EW([\ion{O}{2}]) of RSGs in this system to $\langle EW($[\ion{O}{2}]$)\rangle_{\rm{RS,A}}=-2.0$\AA,
a value typically associated with no star formation. This corrected value of EW([\ion{O}{2}]) places
the average cluster A member in line with member galaxies of lower-redshift ($z\sim0.4$) clusters. In cluster D,
the SFR derived from the composite EW([\ion{O}{2}]) value is an underestimate due to the large number of
$24\mu m$-bright galaxies observed in the system. To correct for this, we extinction correct the spectra of
the $\sim$25\% of cluster D members that are observed in $24\mu m$ (assuming an $E(B-V)=0.5$ and a Calzetti et al.\
2000 reddening law). The EW([\ion{O}{2}]) from this ``corrected" composite spectrum is $\langle EW($[\ion{O}{2}]$)\rangle_{\rm{D}}=-20.5$\AA, 
consistent with the EW([\ion{O}{2}]) observed for average field galaxies at $z\sim1$. Since H$\delta$ is observed in these 
spectra primarily in absorption, the resulting ``corrected" is statistically identical to the uncorrected case. While 
there is significant uncertainty in this process, it is clear that \emph{the average cluster galaxy at $z\sim0.9$ in the Cl1604 supercluster
is (i) undergoing normal star formation, (ii) has an SFR that lies somewhere between the average SFR of galaxies in lower redshift clusters and that of
the average field galaxy at $z\sim1$, and (iii) the level at which the cluster galaxy is forming stars is related to 
the host halo mass and the dynamics of the cluster system in which it is embedded.}

In contrast, only one of the group systems (group C) has an average member galaxy that is undergoing continuous star
formation. The average level of star formation in this group is roughly consistent with the
average SFR in the Cl1604 cluster galaxies. This is perhaps not surprising, as the color$-$magnitude properties of
group C are the most ``cluster-like" of any of the group systems; this group contains both bright RSGs,
a significant population of bright blue galaxies, and a red-sequence fraction that is nearly identical to cluster B.
The other group systems exhibit large differences in the spectral properties of their member galaxies. The average
member galaxies in groups H and I have an EW([\ion{O}{2}]) consistent with no ongoing star formation. In group H, the
average member galaxy is classified as a post-starburst (i.e., K+A; Dressler et al.\ 1999; Poggianti et al.\ 1999),
suggesting significant recent ($\lsim1$ Gyr) star formation has occurred. In the two remaining
group systems, groups F and G, which have the highest observed fraction of 24$\mu$m bright dusty starburst galaxies (see
K11), the average member is a starburst galaxy. If we instead consider the composite group properties by
combining all group galaxies in a single population, the average measured EW values, $\langle EW($[\ion{O}{2}]$)\rangle_{\rm{groups}}=-7.92$\AA\ and
$\langle EW(H\delta)\rangle_{\rm{groups}}=4.15$\AA, imply that \emph{the average Cl1604 group galaxy is undergoing a starburst}. This
conclusion is somewhat surprising given the large number of bright (and, as we will show later, massive and early-type) RSGs 
observed in the group systems. All these results suggest that significant processing of galaxies is occurring in group
environments before such systems are formed into clusters, consistent with the conclusions of several other studies (e.g., Zabludoff
\& Mulchaey 1998; Jeltema et al.\ 2007; Kautsch et al.\ 2008; Tran et al.\ 2009; Bai et al.\ 2010; Balogh et al.\ 2009, 2011).

\subsection{Red-sequence Luminosity Function}
\label{lumfunc}

Much work has been done on observing the properties of the red-sequence luminosity function (LF) in high redshift clusters ($z\sim0.8$-$1.6$).
These studies confirm the existence of bright (or massive) RSGs in overdense environments at $z>1.2$
(e.g., Stanford et al. 2005, 2006; Tanaka et al.\ 2007; Papovich et al.\ 2010; Stott et al.\ 2010; Tran et al.\ 2010) and show a significant deficit in the population
of the low-luminosity RSGs at such redshifts (Tanaka et al.\ 2005, 2007; De Lucia et al.\ 2007; Koyama et al.\ 2007; Stott et al.\ 2007; Lerchster et al.\ 2011; but see
Andreon 2006, 2008 for a different view). While the latter point is well established by such studies, the lack of dense spectroscopic sampling forces these works
to rely primarily on photometric redshifts or statistical field subtraction techniques (as in, e.g., Pimbblet et al.\ 2002), which leaves significant uncertainty in
the magnitude of this deficit for individual cluster systems. With the wealth of spectroscopic data on the Cl1604 supercluster we present here
for the first time the LF of a deep, magnitude limited survey of high-redshift cluster RSGs using \emph{solely}
spectroscopically confirmed members (though see Zucca et al.\ 2009 for a similar survey of overdense regions in the COSMOS field).

\begin{figure*}
\plotonespecial{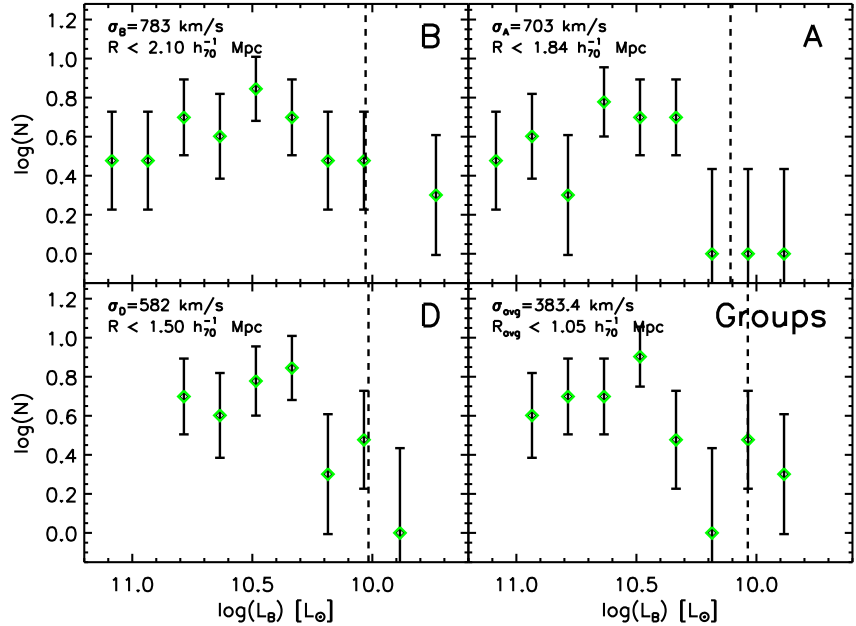}
\caption{Red-sequence $B$-band LF of the group and cluster members of the Cl1604 supercluster. The
name of each system along with the velocity dispersion and projected radial cutoff used to determine membership are
given in each panel.
Errors are derived through a combination of bootstrap techniques and Poisson statistics. The dashed line denotes our rough
spectroscopic completeness limit for RSGs in each system. A significant decrease in the number of red cluster and
group galaxies is observed at low luminosities in all systems with the possible exception of cluster A.}
\label{fig:RSlumfunction}
\end{figure*}

In Figure \ref{fig:RSlumfunction} we plot the rest-frame $B$-band red-sequence LF of the confirmed members of the three clusters
and five groups that comprise the Cl1604 supercluster. Both here and for the bulk of our remaining analysis
we combine all the group galaxies into a single ``Groups" population. This facilitates 
comparisons to the cluster populations and to create a sample of group galaxies that is similar in number to
members in each of the Cl1604 clusters. Transformations to the rest-frame B band are made using our ACS photometry and
the relationship derived by Homeier et al.\ (2006b) specifically for cluster galaxies in the Cl1604 system:

\begin{align}
M_{B} = -0.16(F606W-& F814W)  + 0.75  \nonumber \\
 &  + F814W - 5\log(\frac{d_L}{10\rm{pc}}) 
\label{eqn:Bband}
\end{align}

\noindent where $d_L$ is the luminosity distance to each source as determined by its spectroscopic redshift and our choice of cosmology.
The absolute rest-frame $B$-band luminosity of each galaxy was translated to $L_B$ using the $B$-band luminosity of the
sun\footnote{http://www.ucolick.org/$\sim$cnaw/sun.html}. No correction was made for internal dust extinction, as the extinction values 
derived from the SED process are only precise enough to use in a statistical manner. The average fitted extinction of the RSGs
presented here is $E(B-V)=0.2$, which translates to a difference of $\sim20$\% in luminosity. While this is a non-trivial absolute 
uncertainty, the RSGs of each system span an order of magnitude in luminosity, and, thus, this uncertainty is much smaller than the bin size
used for this analysis. Furthermore, we find no significant difference between the average $E(B-V)$ values of the brighter 
[$\log(L_{B}) > 10.6$] RSGs in the sample than that of the fainter [$\log(L_{B}) < 10.6$] RSGs, which is crucial to our analysis. We therefore
ignore the effects of extinction for the remainder of the section.

What is immediately noticeable in Figure \ref{fig:RSlumfunction} is the level of development in the group red sequence. The bright end of the red-sequence
LF in the group systems appears nearly identical that of the two most massive
Cl1604 clusters (clusters A and B). The only exception is the few extremely bright [$\log(L_{B}) > 11$] RSGs present
in clusters A and B that are lacking in the group systems. Conversely, in our lowest mass cluster (cluster D) we observe no RSGs with
$\log(L_{B}) > 10.8 $. While this cluster is still quite young (as determined by the average stellar ages of its massive RSGs, see 
\S\ref{discussion}), it appears that galaxies at the bright end of the red sequence were not ``embedded" into the system at an early time in its formation history.
While it is not necessarily the case that such galaxies were embedded into the potentials of the two more massive clusters, the presence of bright
(and, as we will show later, massive) RSGs in clusters A and B allows for this possibility. Furthermore, the presence of such galaxies in
the group systems (except for the very brightest end, a distinction which will become important later) argues strongly in favor of a scenario where the bulk of
the bright end of the red sequence is formed primordially through ``early quenching" (Poggianti et al.\ 2006; Kriek et al.\ 2006; Faber et al.\ 2007; Cooper et al.\ 2007). In this scenario star-forming galaxies
are transformed into massive quiescent ellipticals at early ($z\gsim2.5$) times. However, since we observe no such galaxies in cluster D, it is puzzling
to consider how such a system might form if early quenching processes are universal for bright (massive) red-sequence cluster galaxies. Since there are
no bright RSGs in cluster D, a different scenario is required to explain how such galaxies will\footnote{While it is possible that 
cluster D represents a special case of a cluster where bright/massive RSGs do not form by, e.g., $z\sim0$, we assume that its galaxy
population will eventually resemble that of a ``typical" $z\sim0$ cluster.} arise in cluster D. The formation of a system like cluster D requires very specific progenitors,
as every group system (with the exception of group F) has at least one RSG that is brighter than the brightest RSG
in cluster D. We will return to the issue of what processes are likely responsible for building up the bright (massive) end of the red sequence in cluster D
and the other Cl1604 structures in \S\ref{discussion}.

At the faint end of the red-sequence LF a noticeable decrement occurs in the number counts of RSGs at
luminosities of $\log(L_{B}) \lsim 10.35 $ ($M_B>-20.55$). The red-sequence completeness limit (indicated by the dashed line in each panel of
Figure \ref{fig:RSlumfunction}) is determined from the blueward envelope of the red-sequence in each system and the magnitude where we have obtained high
quality spectroscopic redshifts for 90\% of RSGs in any particular structure. 
This completeness limit is roughly $\log(L_{B})\sim10.0$ ($M_B\sim-19.7$) for all systems or $\sim$0.3$L^{\ast}_B$
(where $M^*_B$ is adopted from the red galaxy sample in Willmer et al.\ 2006). For all structures
except the most isolated and
relaxed system (cluster A; see \S\ref{radial}), the deficit in the number counts of RSGs occurs at significantly brighter
luminosities than our completeness limit. This suggests that the paucity of faint RSGs in these systems is real and not a result of
our spectral sampling. In cluster A we observe a flattening out of the red-sequence number counts persisting nearly to the completeness limit
in this system and only marginal evidence for a decrease in the number count at luminosities consistent with our completeness limit. These
results are identical to the photometric analysis of cluster A by Crawford et al.\ (2009), in which no decrease in faint RSGs 
was observed to their completeness limit. However, in an evolved system such as Coma, the number counts of RSGs are seen to
increase nearly monotonically with decreasing luminosity (Terlevich et al.\ 2001; De Lucia et al.\ 2007). This suggests that, although the deficit of
low-luminosity RSGs is not as pronounced in cluster A as in the less-evolved Cl1604 systems, cluster A still has a significant
decrement in the low-luminosity end of the red-sequence LF despite being the most evolved system in the supercluster. These results are
consistent with the observations of De Lucia et al.\ (2007) and Koyama et al.\ (2007), who found that the deficit of
low-luminosity red-sequence cluster galaxies is strongly tied to the evolutionary state of the cluster. Clusters which are more evolved
or observed at lower redshift (and, therefore, generally more evolved than those at high redshift) were found in both studies to contain
a larger fraction of faint RSGs than their younger counterparts.

The noticeable lack of faint RSGs within the bounds of the Cl1604 structures initially seems somewhat difficult to reconcile
with the observation of a large number of bright, RSGs in the two most massive clusters and in a majority of the
group systems. We previously argued that early quenching processes were strongly favored by the presence of such bright (and massive; see 
\S\ref{colormass}) RSGs. The process that transformed these bright RSGs early in their formation histories cannot,
however, generally be responsible for the transformation of their low-luminosity counterparts since the majority of this population is not formed by
$z\sim0.9$. The fraction of low-luminosity red galaxies is also quite low in the field at these redshifts, both in the Cl1604 superfield population and
in larger field surveys (e.g., Cooper et al.\ 2007). This further suggests that early quenching processes are not responsible for the formation of such 
galaxies. As there are few low-luminosity RSGs in the field for clusters to ``passively" accrete at $z\sim0.9$, it is
likely that late-time transformation of low-luminosity (or low-mass) blue cluster and group galaxies is responsible for comprising the low-luminosity
end of the red-sequence LF at low redshifts. As we will show later, a large number of faint (and low-mass) blue galaxies are
observed outside the core ($R>0.5R_{\rm{vir}}$) of all the Cl1604 structures (see \S\ref{radial}), suggesting that such galaxies have yet to be quenched
by the cluster or group environment.

\subsection{Color$-$Stellar-mass Properties}
\label{colormass}

\begin{figure*}
\plottwospecial{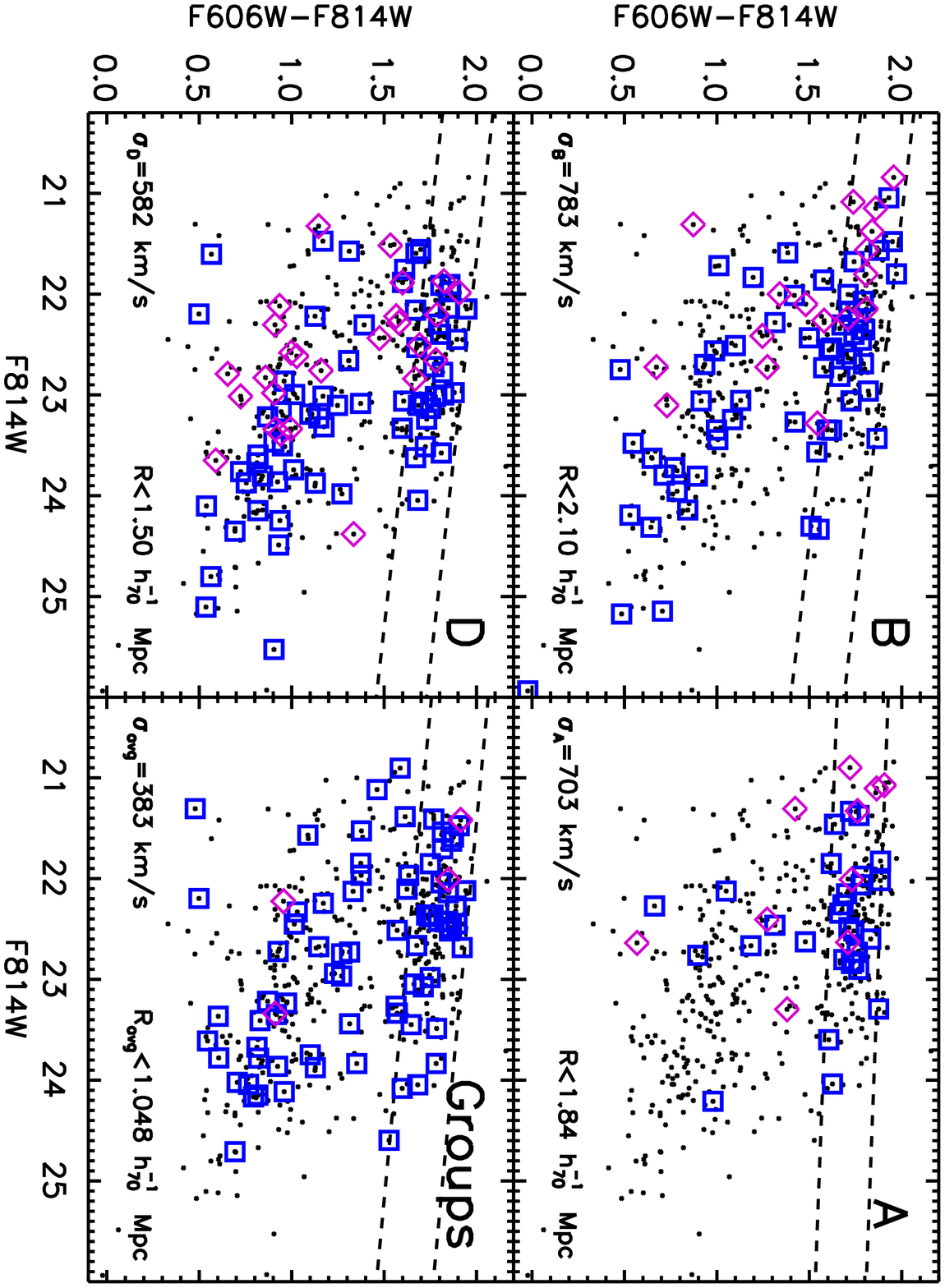}{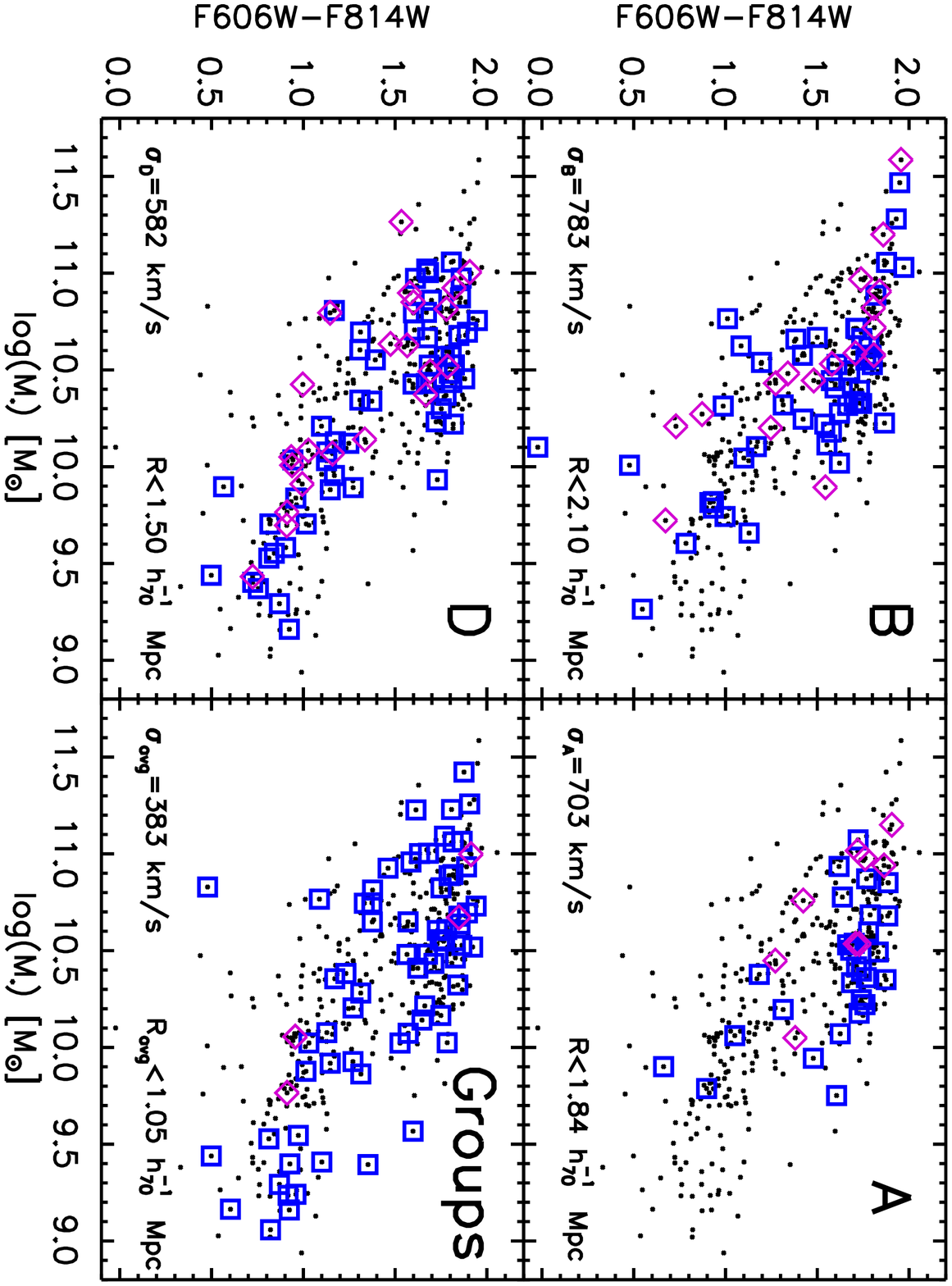}
\caption{\emph{Left:} As in Figure \ref{fig:individualCMDs}, 
ACS color$-$magnitude diagram of the 460 spectroscopically confirmed members of the Cl1604
supercluster detected in both ACS bands. All group member galaxies have been combined into a single ``Groups" sample. The meanings of the symbols are
identical to those of Figure \ref{fig:individualCMDs}. 
\emph{Right}: ACS color$-$stellar-mass diagram of the $\sim400$ member galaxies of the Cl1604
supercluster with well-defined stellar mass which are detected in both ACS bands. The $\sim$60 galaxies present in the left panel but absent in the right
panel are almost exclusively at faint $F814W>23.5$ magnitudes and roughly half ($\sim25$) lie within $R<2R_{\rm{vir}}$ of a Cl1604 group or cluster.
Notice that nearly all of the bright and massive red-sequence
galaxies in the supercluster are contained within either the cluster or group environment. While both bright and massive blue-cloud galaxies are
observed in the lower-mass systems (cluster D and the groups), the bright blue-cloud galaxies present in cluster B are much less massive.}
\label{fig:CmDnCMD}
\end{figure*}

In Figure \ref{fig:CmDnCMD} we plot the CMDs and color$-$stellar-mass diagrams (CSMDs) for the member galaxies of the
three clusters and five groups of the Cl1604 supercluster. One of the most striking observations from both the CMDs and the CSMDs
is both the level of development of the group red sequence and the number of massive RSGs present in the 
five Cl1604 groups. The mass range of the confirmed red-sequence members of the groups is nearly identical
to that of the most massive cluster (B), the only exception being the highest mass galaxies
[$\log(\mathcal{M}_{\ast})>11.5$] in cluster B which are absent in the groups. The presence of these 
massive red galaxies in the groups suggests that either $(i)$ there is a large population of massive dust-reddened
starbursts in the groups populating the red sequence, or $(ii)$ significant
pre-processing is occurring in the group environments in the supercluster. This question will be addressed when we
discuss their morphologies in \S\ref{massmorph}.

A dramatic shift is observed in specific subsets of galaxies in the cluster and group populations when the CMDs and
CSMDs are compared. In the two most massive clusters (A \& B), what was already a reasonably tight
color$-$magnitude relation has become an even tighter relationship between color and mass (though the average error in 
stellar mass is roughly seven times that of the average $F814W$ error). This phenomenon is particularly noticeable
in cluster B, where the color$-$stellar-mass (CSM) relation is observed with virtually no scatter for over order of magnitude
beginning at $\log(\mathcal{M}_{\ast})\sim10.5$ and extending to higher stellar mass. The low scatter of the RSGs
observed in the CMDs and CSMDs of the two most massive Cl1604 clusters is typical of 
systems that have formed their RSGs at much earlier epochs (see Mei et al.\ 2009 and references therein).
Another dramatic shift when comparing the CMDs and CSMDs occurs at the bright end
of the red sequence in clusters A \& B. While the bright ends of their red sequences look nearly identical,
there exist significant disparities between the masses of these galaxies. In particular, the most massive
[$\log(\mathcal{M}_{\ast})>11 $] RSGs that are observed in both cluster B and the group systems are
largely absent in cluster A.

In the lower mass systems (cluster D and the groups)
significant scatter is observed in the CSM relation for RSGs at nearly all masses. This scatter
is the result of a large population of massive [$\log(\mathcal{M}_{\ast})>10.7$] blue-cloud galaxies 
that have colors just blueward of the red sequence. This is a
population that is virtually absent in the two most massive clusters. A comparison between bright blue-cloud galaxies
($F814W>22.2$) in the low-mass systems and the high-mass systems (clusters A \& B) reveals a factor of two disparity in
their average stellar masses, with bright blue-cloud galaxies in the low-mass systems being, on average, twice as massive. If we
assume that rest-frame $B$-band luminosity is roughly proportional to the SFR in blue-cloud galaxies
(James et al.\ 2008), the bright blue-cloud galaxies in the two most massive clusters have optical specific star formation
rates (SSFR) that are, on average, a factor of two higher than
the analogous galaxy population in the low-mass systems. This will be investigated further in the next section.

In both the X-ray bright clusters (clusters A \& B), as well as the low mass cluster and group systems, we do not observe
the extremely massive RSGs [$\log(\mathcal{M}_{\ast})\sim12 $] that exist in local clusters (e.g.,
Stott et al. 2010). The most massive galaxy observed in the supercluster (a member of cluster B) is required to double its mass
by $z\sim0$ to reproduce the mass of a typical BCG at low redshift. In
clusters A \& D this disparity is more pronounced. The most massive galaxies observed in these two systems are roughly
a factor of 10 lower in stellar mass than typical low-redshift BCGs. Through these comparisons we are inherently assuming
that the galaxy population of the Cl1604 clusters and groups are typical progenitors of the galaxy populations of modern
clusters. However, the presence of a very massive [$\log(\mathcal{M}_{\ast})\gsim12 $] BCG is a common occurrence in
\emph{average}, X-ray bright clusters at $z\sim0$ (Stott et al.\ 2010), suggesting that such galaxies are a consequence
of a wide variety of formation histories. Thus, it is likely that most massive red galaxies observed at $z\sim1$ in the clusters
and groups of the Cl1604 supercluster will experience significant buildup over the next $\sim7$ Gyr. We will return to
this point in \S\ref{discussion}.

\subsection{Radial Distributions}
\label{radial}

\begin{figure*}
\begin{center}
\centerline{\includegraphics[width=0.95\columnwidth]{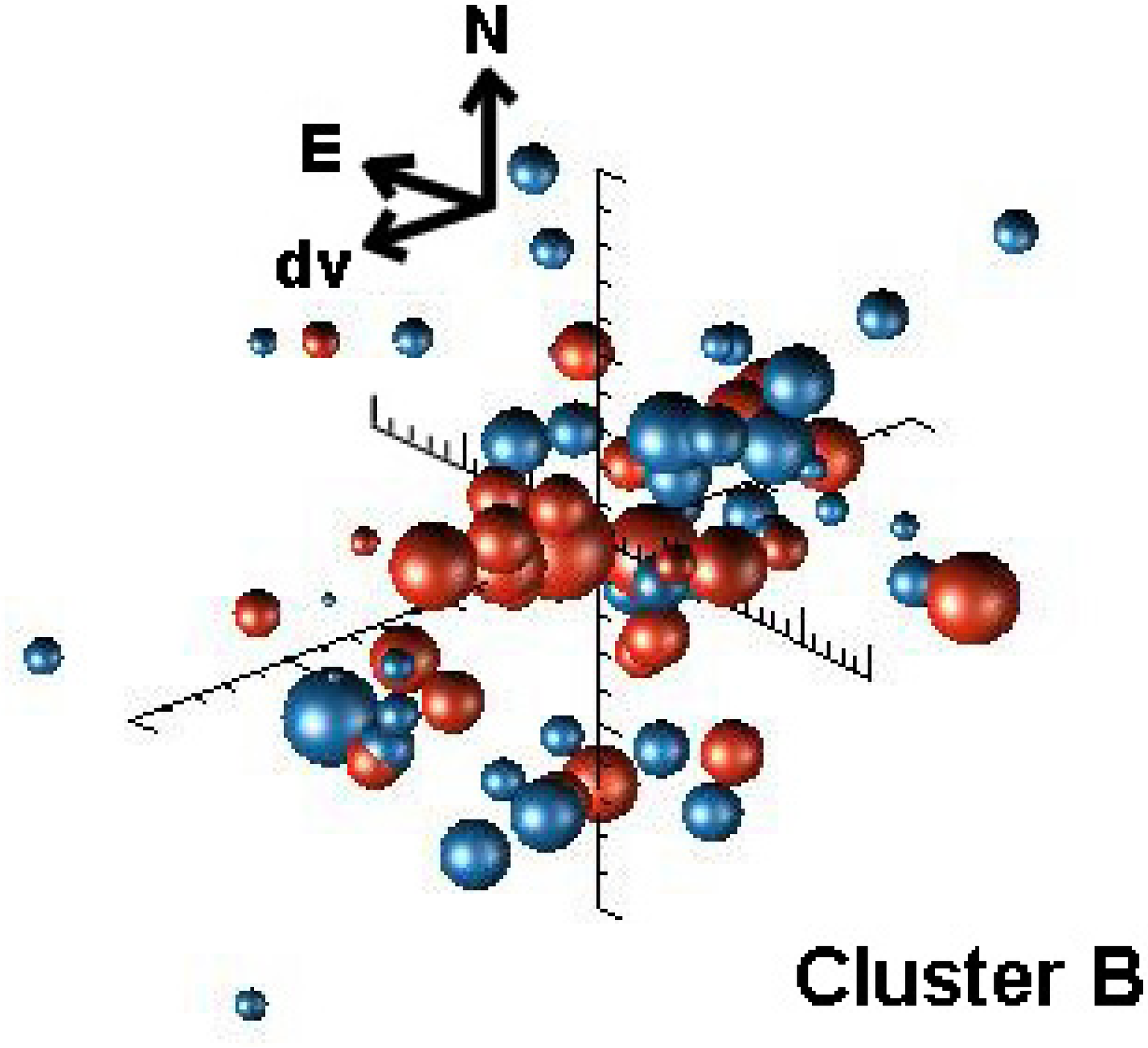}\includegraphics[width=1.0\columnwidth]{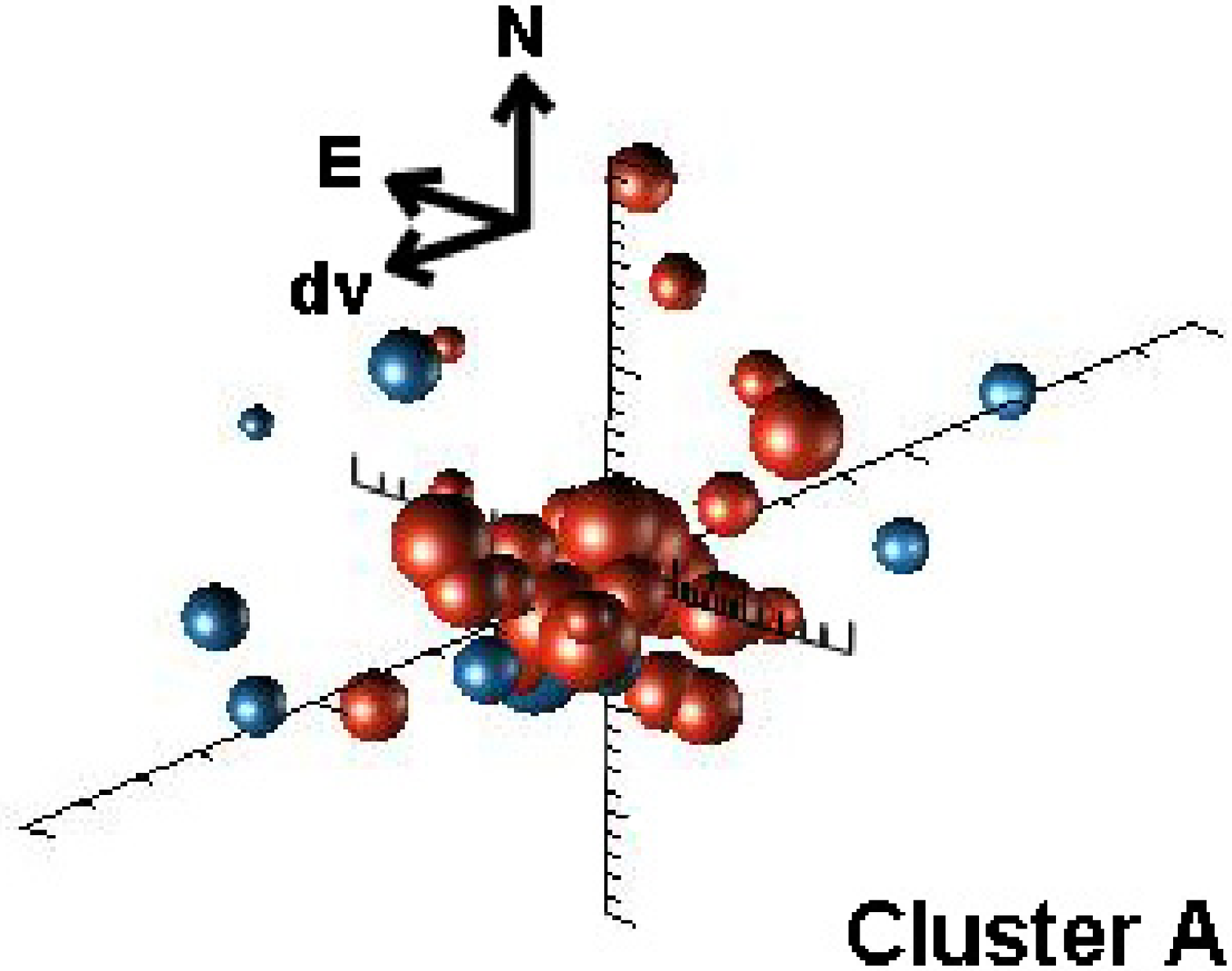}}
\centerline{\includegraphics[width=1.0\columnwidth]{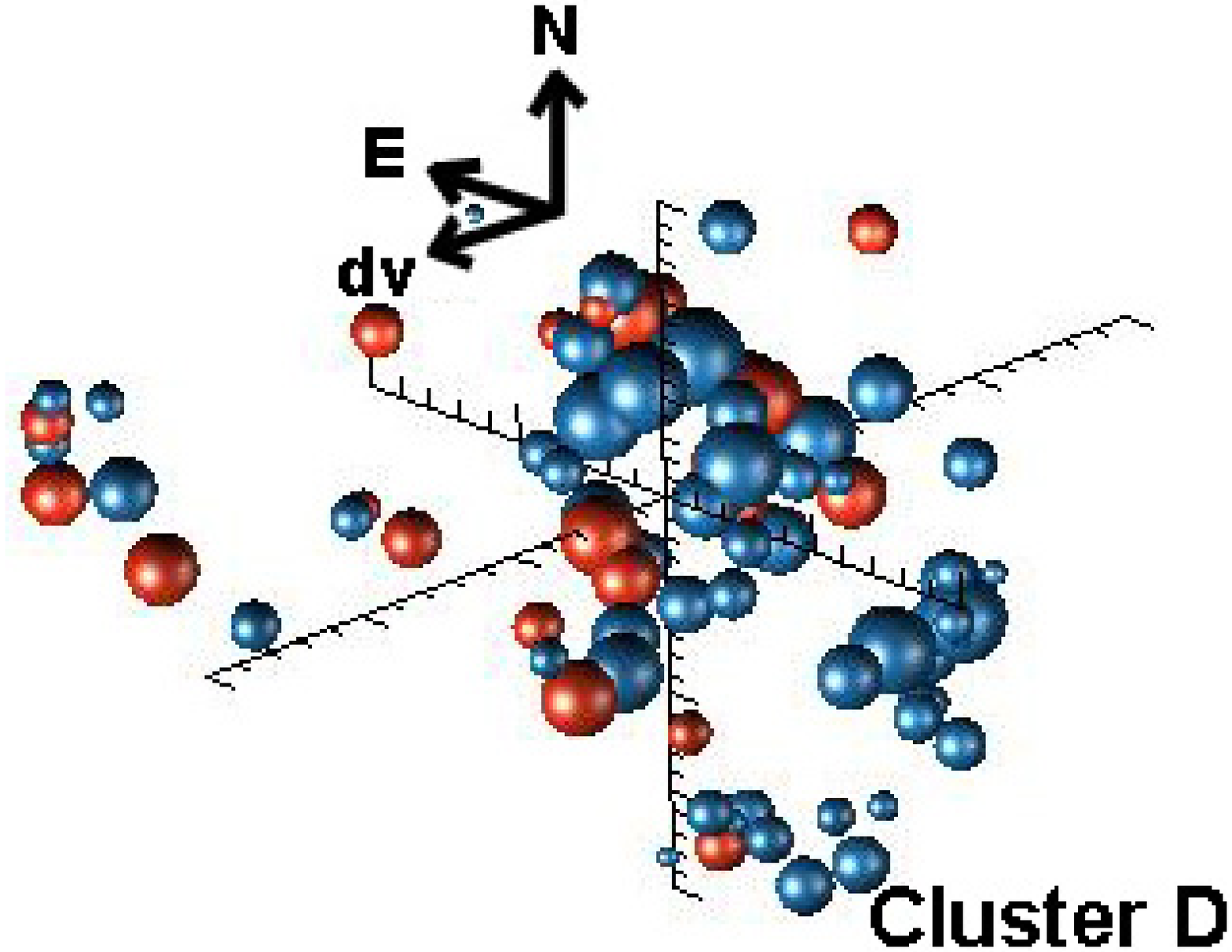}\includegraphics[width=0.85\columnwidth]{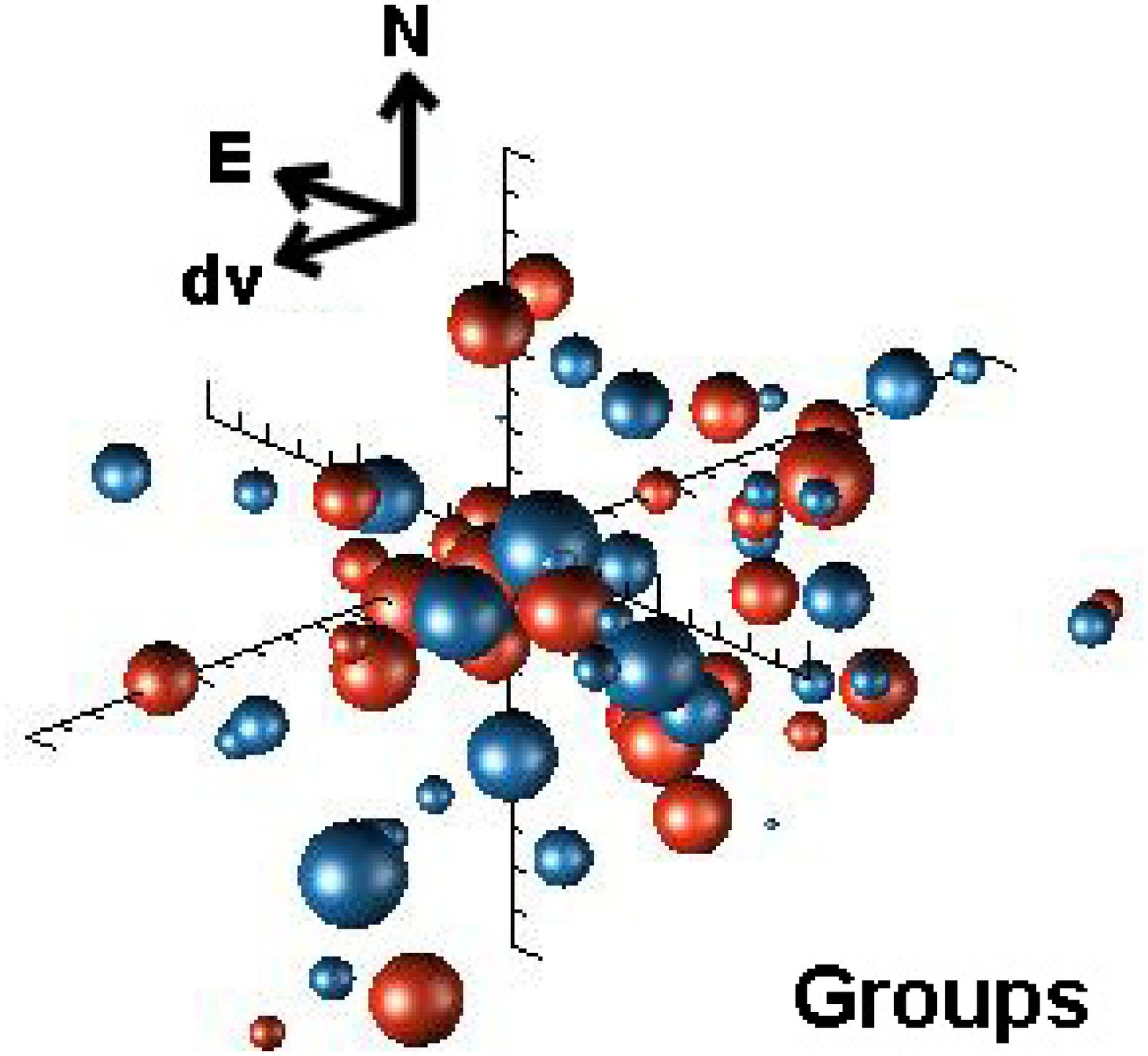}}
\end{center}
\caption{Plot of the spatial and velocity distribution of the member galaxies of each of the constituent Cl1604 clusters and groups. The large tickmarks
on the spatial axes of each panel denote $R_{\rm{vir}}$ and $2R_{vir}$. The velocity axis shows the differential velocity of each galaxy with respect to the
group or cluster mean. Large tickmarks on the velocity axis of each panel denote $\sigma$, 2$\sigma$, and 3$\sigma$, where $\sigma$ is the measured
velocity dispersion of that system. Plotting in this way allows us to combine all group galaxies into a single panel. The size of each sphere
is scaled (linearly) with the B-band luminosity of each galaxy and color coded such that red spheres correspond to RSGs and
blue spheres to blue-cloud galaxies. Differences in the dynamical states of the red and blue galaxies populations of each system (or
collection of systems in the case of the groups sample) can be clearly seen.}
\label{fig:3dsphere}
\end{figure*}

In this section we examine the radial distributions of galaxies in the Cl1604 clusters and groups. In Figure \ref{fig:3dsphere}
we present a ``three dimensional" plot of the member galaxies of each system. The two spatial dimensions are plotted (normalized by the virial radius of
each system) and the third dimension is represented by the differential velocity of each galaxy with respect to the mean velocity of its parent cluster
or group (normalized by the velocity dispersion of each system). Galaxies are separated into blue-cloud and RSGs following the definitions in
\S\ref{RSfitting}. The size of each sphere is scaled linearly by the rest-frame B-band luminosity (see \S\ref{lumfunc}) of each galaxy.
We will formally quantify the (projected) radial distributions of certain subsets of cluster and group galaxies in the various systems later in the
section. Prior to that, however, Figure \ref{fig:3dsphere} provides us with a useful diagnostic to quickly assess the overall
populations and dynamics of each system (or composite of systems in the case of the groups).

We begin this discussion with cluster A, the second most massive cluster in the Cl1604 complex, and the cluster that lies most securely on the optical-X-ray
cluster scaling relations (Kocevski et al.\ 2009a; N.\ Rumbaugh et al.\ 2012, in preparation). Earlier we asserted that cluster A was the most relaxed of the Cl1604 clusters. 
From Figure \ref{fig:3dsphere} we can see that this is quite obviously the case; nearly all of the galaxies in the system are red and a large fraction
of these lie at small (projected) radii and low differential velocities with respect to the cluster center. Nearly all of the blue galaxies 
(faint and luminous) are observed at either large projected radii or large velocity offsets. Considering cluster B and then cluster D, we see a clear trend in
both the galaxy populations and the level of relaxation. Cluster D contains a galaxy population that is both the bluest (on average) and 
the least centrally concentrated of any of the Cl1604 clusters. In cluster B we see that, 
as in cluster A, a bulk of the faint blue-cloud galaxies lie at large clustocentric
distances or velocity offsets (or both). In cluster D and the group systems this does not seem to be the case; faint blue galaxies are distributed relatively
evenly, indistinguishable from the spatial and kinematic distributions of the general galaxy population. The spatial distribution of all constituent
galaxies of cluster D is consistent with the interpretation of
a large filamentary structure intersecting the cluster core (G08; K11). The Cl1604 group systems seem, on average, to be in an intermediate stage relative
to clusters B \& D in both their dynamical evolution and the evolution of their constituent galaxies. As in cluster A and to a lesser extent in clusters B \& D,
a large fraction of luminous RSGs in the group systems appear at low projected radii and small differential velocity. 

\begin{figure*}
\plottwospecial{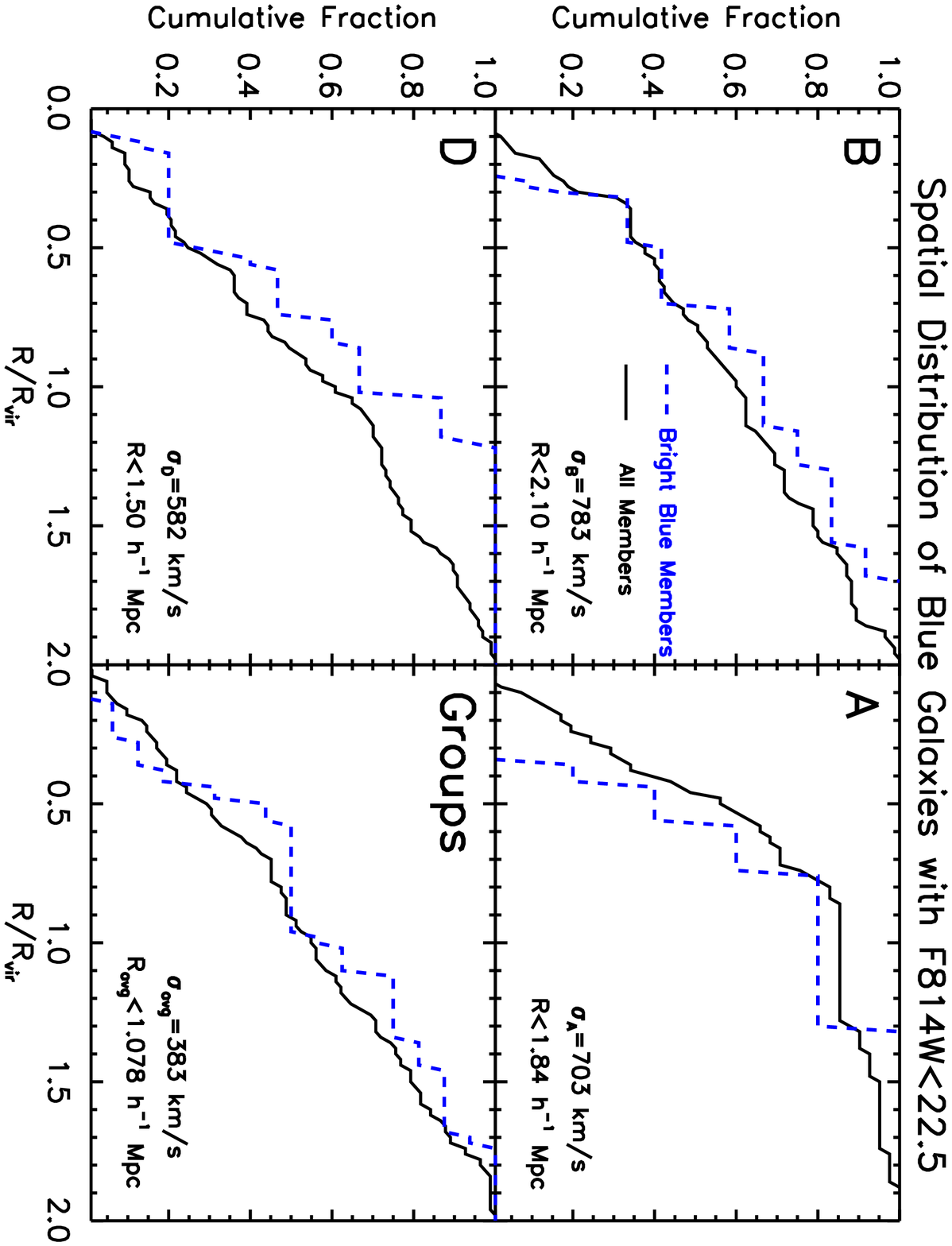}{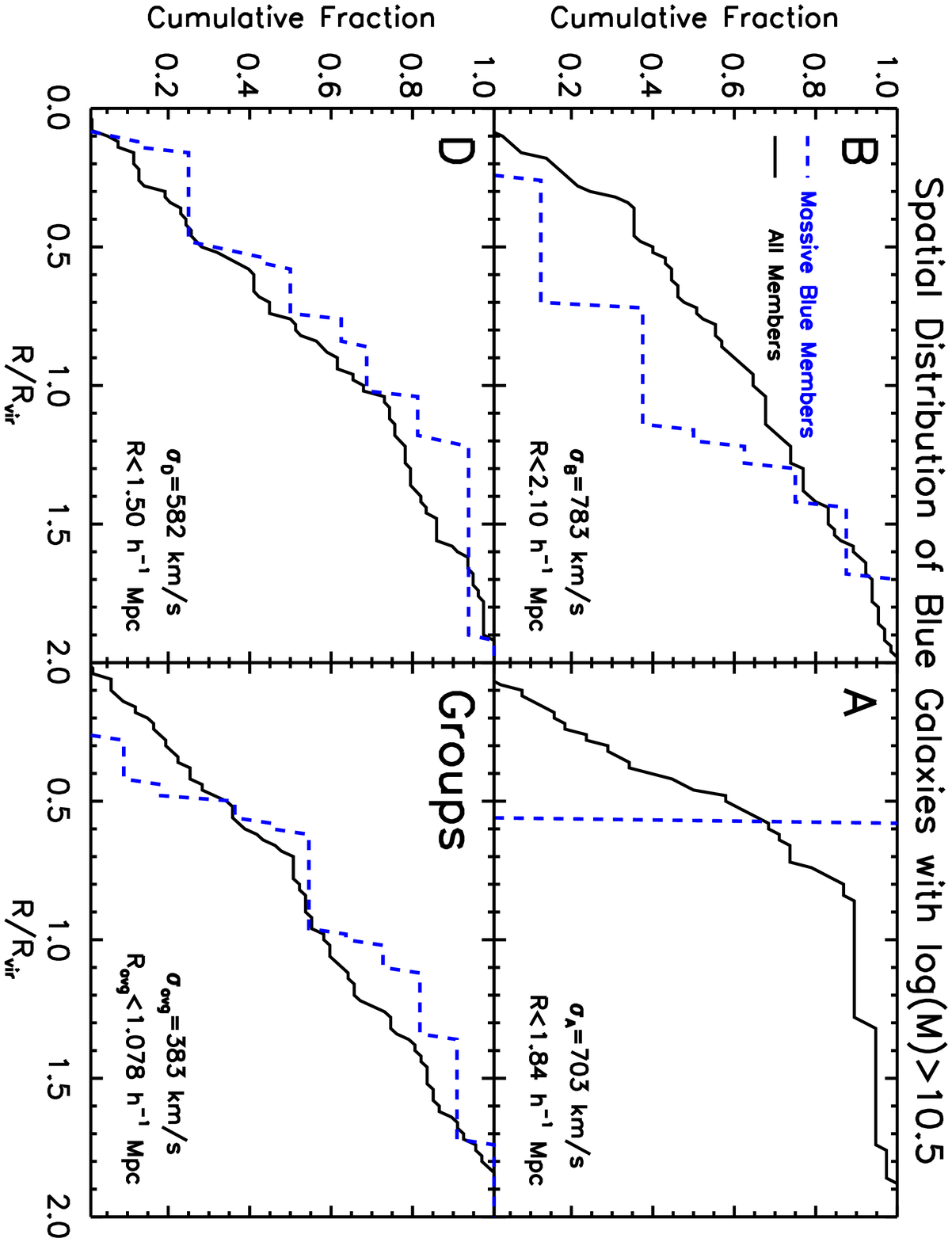}
\caption{\emph{Left:} Cumulative distribution of the projected radial distributions of bright (i.e., luminous) blue-cloud galaxies in the three Cl1604 clusters
and combined groups sample. Also plotted are the distributions of all member galaxies detected in the ACS data (solid
black line). Projected distances from the group/cluster centers are normalized by $R_{\rm{vir}}$. The name of each system (or collection
of systems) is given in each panel along with the velocity dispersion and radius used to determine membership. \emph{Right:}
Same as left panel, but now considering massive blue-cloud galaxies. The number of galaxies considered in this plot and the left
plot is generally not the same. The solid black line now shows the cumulative radial distribution of all ACS detected members with well-defined
stellar masses.} 
\label{fig:BCradial}
\end{figure*}

We now discuss the projected radial distributions
of both bright and massive red and blue galaxies in each system. As we will show later (\S\ref{massmorph}), a large population of transition galaxies is
observed in the Cl1604 groups and clusters. By analyzing and comparing the (projected) spatial distribution of different types of galaxies in each of
the clusters and groups we can begin to discuss the nature of such transformations. In Figure \ref{fig:BCradial} we plot the projected radial distributions
of both luminous and massive blue-cloud members of the Cl1604 clusters and groups. As before, member galaxies of the five groups are combined
to create a single composite population by normalizing the projected distance of each member galaxy by the virial radius of its parent group. Examining the
radial distribution of \emph{bright}\footnote{Since all of the Cl1604 systems are essentially at the same redshift, bright here, and throughout the paper, is equivalent to luminous.} 
($F814W<22.5$) blue-cloud galaxies in each system, we see that their distribution is generally consistent
with that of the overall galaxy population. The only possible exception to this trend is cluster D, where
bright blue members show preference towards lower clustocentric radii than the overall galaxy population, with no bright blue galaxies observed at radii greater than
$R\sim1.2R_{\rm{vir}}$.

In the right panel of Figure \ref{fig:BCradial} we plot the radial distributions of \emph{massive} (as opposed to simply luminous) blue-cloud galaxies for the same systems. Immediately it is
apparent that the number of massive blue-cloud galaxies in clusters A \& B is considerably different than the bright blue population. Less
than half of galaxies that comprised the bright blue galaxy population in clusters A \& B are at high enough stellar mass to be considered in the
right panel of Figure \ref{fig:BCradial}. This is best exemplified by the cumulative distribution of massive blue galaxies in cluster A, which
is simply a single vertical line marking the location of the one massive blue cloud galaxy in the system. In contrast, the numbers of bright blue
galaxies in cluster D and the groups are nearly identical to the number of massive blue galaxies in these systems. The radial distributions
of the massive blue galaxies in clusters A \& B also differ significantly from that of their bright counterparts. In cluster A there is
only a single massive blue galaxy, meaning that nearly all of the bright blue-cloud galaxies in cluster A are at lower mass. In cluster B,
massive blue galaxies tend to avoid the cluster core ($R<0.5R_{\rm{vir}}$) and a majority of these galaxies are observed at large projected radii
$R>R_{\rm{vir}}$. Conversely, a large fraction ($\sim40$\%) of 
bright blue galaxies in clusters A \& B are located within the cluster core and a
majority of these galaxies are situated within $R_{\rm{vir}}$ in both systems. These results imply that there exists \emph{a large population of bright,
low-mass blue galaxies in the cores of the two most massive clusters in the Cl1604 supercluster.} To supplement this analysis 
we have performed a Kolmogorov-Smirnov (KS) test to determine how similarly bright blue and massive blue galaxies are distributed 
(in projection) in cluster B\footnote{This same test is not performed in cluster A because only a single massive blue galaxy is present in this cluster.}. 
The KS test confirms that the two distributions differ at $>99.99$\% confidence level, further reinforcing the conclusions reached from our visual inspection.
In the lower mass systems (cluster D and the
groups), the distribution of massive and bright member galaxies is nearly identical, with a majority of both bright and massive blue cloud
located within $R<R_{\rm{vir}}$. Thus, the bright blue galaxies in the cores of the massive Cl1604
clusters have optical SSFRs that are considerably higher than both their counterparts at larger (projected) clustocentric radius and the analogous population
in the lower mass cluster and group systems. The cores of these massive clusters appear to be active in regulating star formation of low-mass blue
galaxies.

This result is in apparent contradiction to the findings of K11, where 24$\mu$m bright starbursting cluster galaxies were preferentially found
at larger projected radii from the cluster center. However, a large fraction of the low-mass bright blue galaxies considered here are not 24$\mu$m bright,
suggesting that whatever process is regulating star formation in blue galaxies in the cluster core is different than the process which is responsible for the dusty
starburst population observed in the clusters. In K11, we suggested that merging or other galaxy-galaxy tidal interaction processes were largely responsible
for the formation of dusty starbursts in the groups and cluster systems. Since these high SSFR blue cloud galaxies are observed largely in the inner regions
of the most massive clusters, clusters which contain a hot ICM, it is likely that some cluster specific process is responsible for regulating
star formation in this population.

\begin{figure*}
\plottwospecial{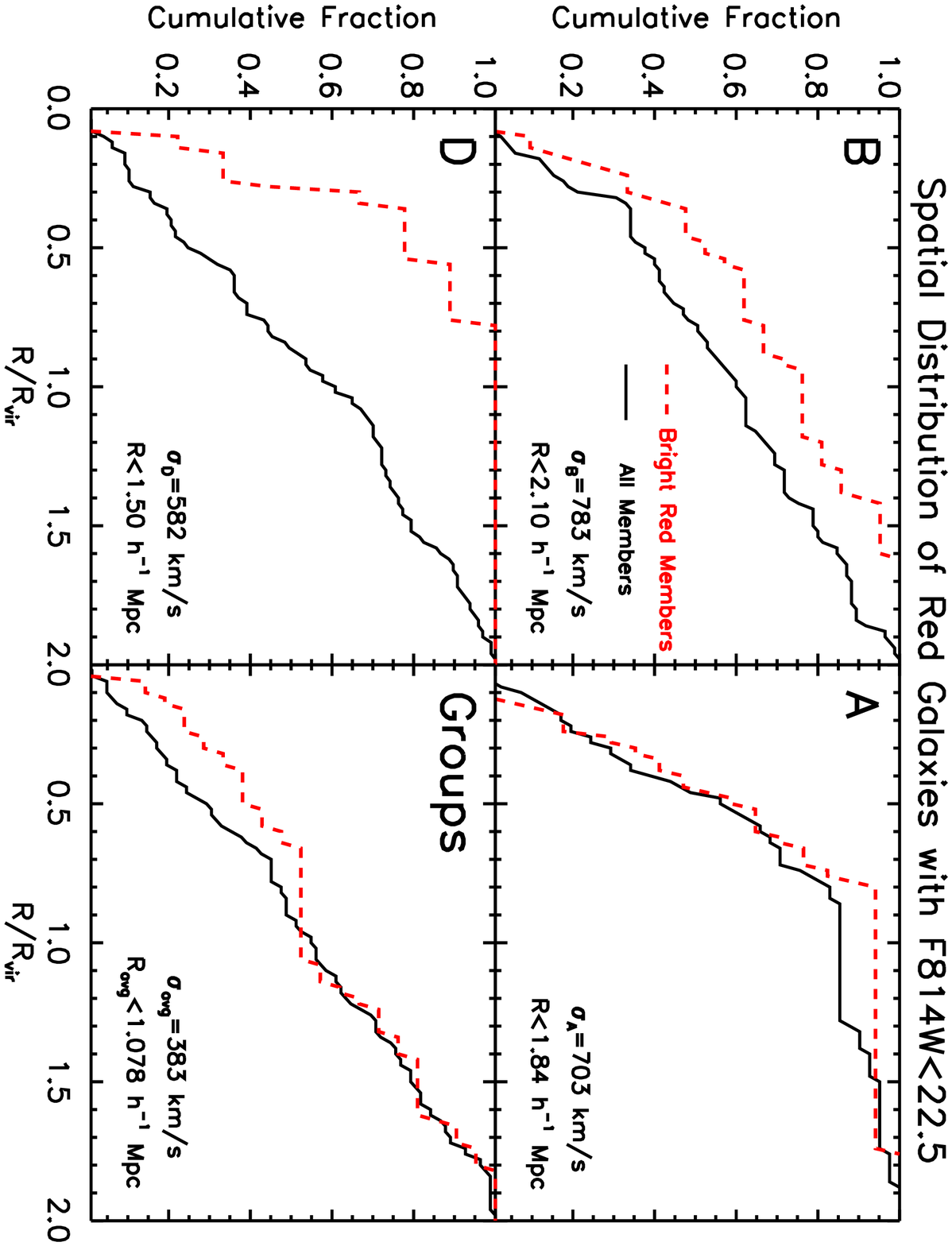}{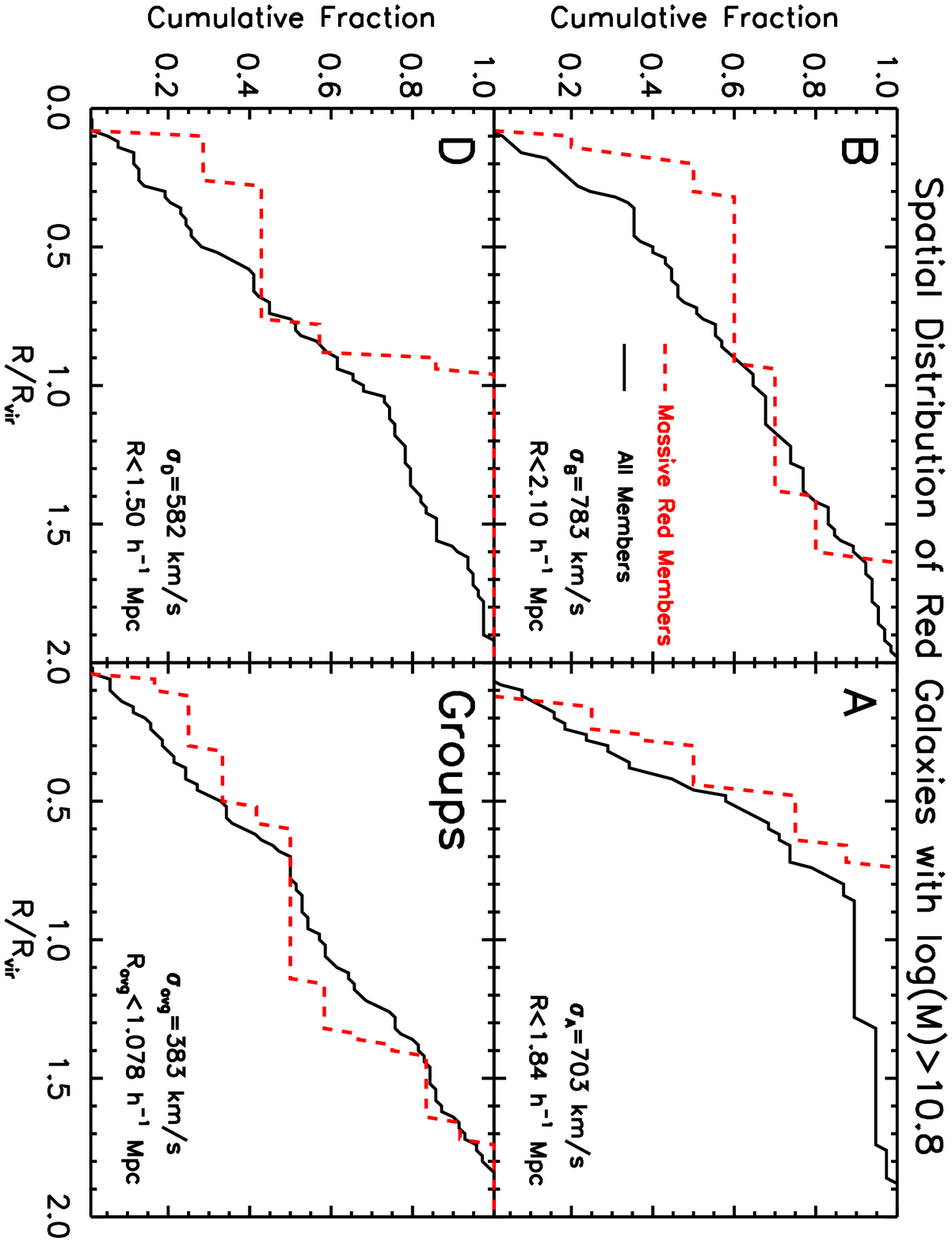}
\caption{\emph{Left:} Cumulative projected radial distribution of the bright (i.e., luminous) red-sequence galaxies of the member galaxies of the
Cl604 clusters and group systems. The solid black lines are identical to those in Figure \ref{fig:BCradial}. 
A majority of bright red galaxies lie in the cores ($R<0.5R_{\rm{vir}}$) of the three clusters, but are considerably more spread out in the groups. 
\emph{Right:} Identical to the right panels of Figure \ref{fig:BCradial}, except with massive red-sequence galaxies. The meanings of the solid black 
lines are identical to that of the solid black lines plotted in the right panels of Figure \ref{fig:BCradial}.} 
\label{fig:RSradial}
\end{figure*}

In Figure \ref{fig:RSradial} we repeat this analysis for both bright and massive RSGs. 
The radial distributions of both the massive and bright RSGs confirm the general picture from 
Figure \ref{fig:3dsphere}; a majority of the bright and massive RSGs in all systems are observed at low (projected) radii (i.e., $R<R_{\rm{vir}}$).
In the group systems we observe a slow, continuous increase in the number of bright and massive RSGs
out to $2R_{\rm{vir}}$. This is in contrast to the rapid increase in both bright and massive RSGs observed in cluster A to $0.8R_{vir}$, past
which there are essentially no such galaxies, highlighting the difference in the dynamical states of the two populations. With the
exception of cluster D, the radial distributions of the bright red galaxies are nearly identical to those of the massive red galaxies, suggesting the red-sequence
populations in the two most massive clusters and the low mass group systems have similar $\mathcal{M}/L$ ratios. In cluster D, however, there is a
significant difference in the radial distributions of bright red members relative to their massive analogues (confirmed by a KS test at $\gg99.99$\% confidence level). 
While both populations are observed solely within
$R_{\rm{vir}}$, their radial distributions within the core of the cluster ($R<0.5R_{vir}$) are dramatically different. Nearly all
($\sim90$\%) of the bright red population is observed within a projected radius of $0.6R_{\rm{vir}}$, while only $\sim40$\% of the massive red population is contained
within this radius. The bright galaxies in cluster D that are not observed in the radial distribution of massive red galaxies must be less massive than the mass limit of the plot [$\log(M_{\ast})=10.8$], and, by definition,
have lower $\mathcal{M}/L$ ratios than their counterparts in any of the other systems. Thus, \emph{there are a large number of bright, low-mass RSGs
present in cluster D that are either still star forming or have formed stars in the past $\sim1$ Gyr and have only recently transitioned to the cluster red sequence.} We will discuss this population further in
\S\ref{discussion}.

\subsection{Mass-Morphology Properties}
\label{massmorph}

Thus far we have discussed the color, magnitude, mass, spectral, and radial properties of the member galaxies of the eight groups
and clusters in Cl1604. While several conclusions have been reached from investigating
these properties, open questions still remain regarding the nature of RSGs in the groups and cluster
systems as well as the processes responsible for placing them on the red sequence. It is with these questions in mind that we
explore in this section the morphological properties of member galaxies of the Cl1604 supercluster. 

\begin{figure*}
\plottwospecial{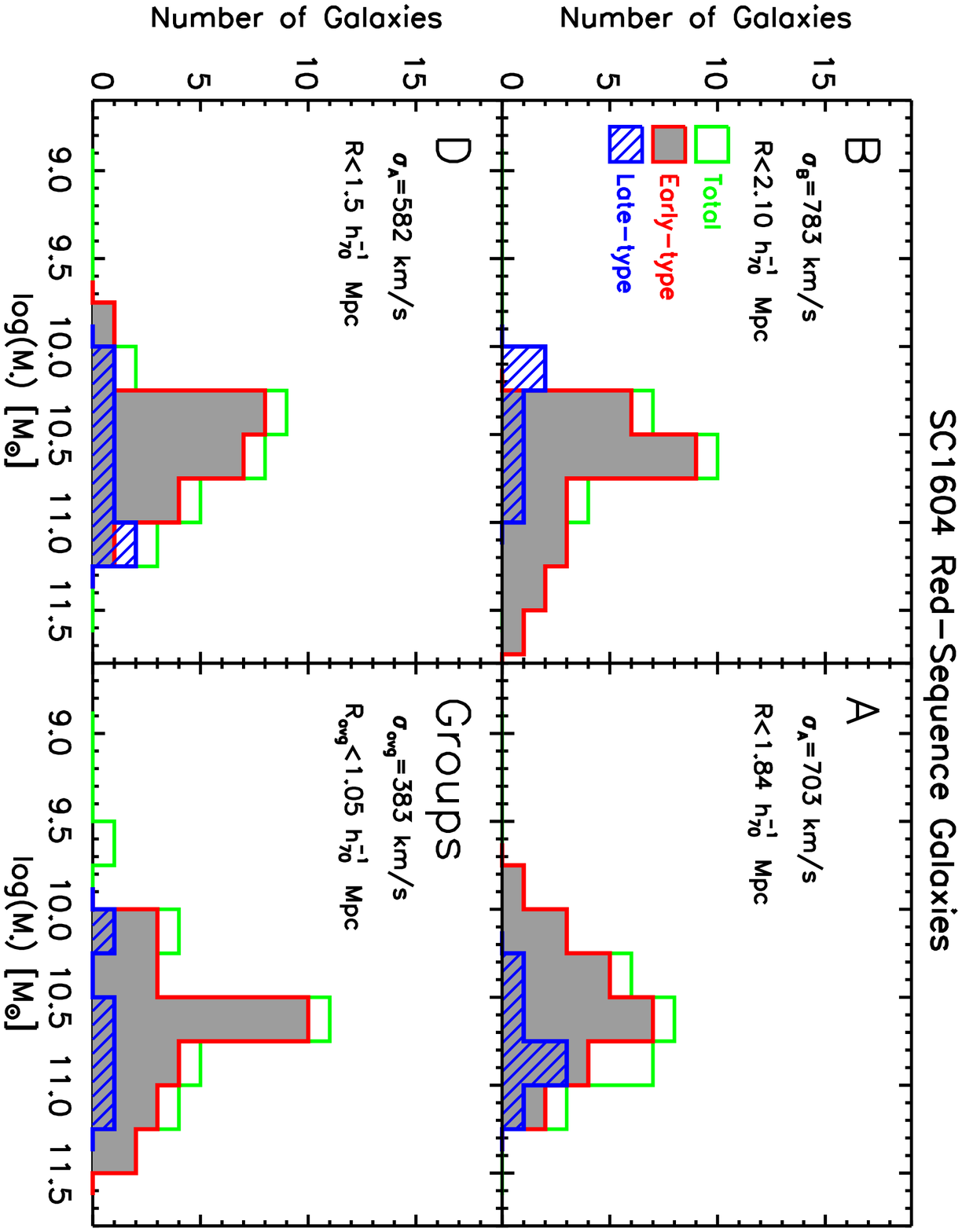}{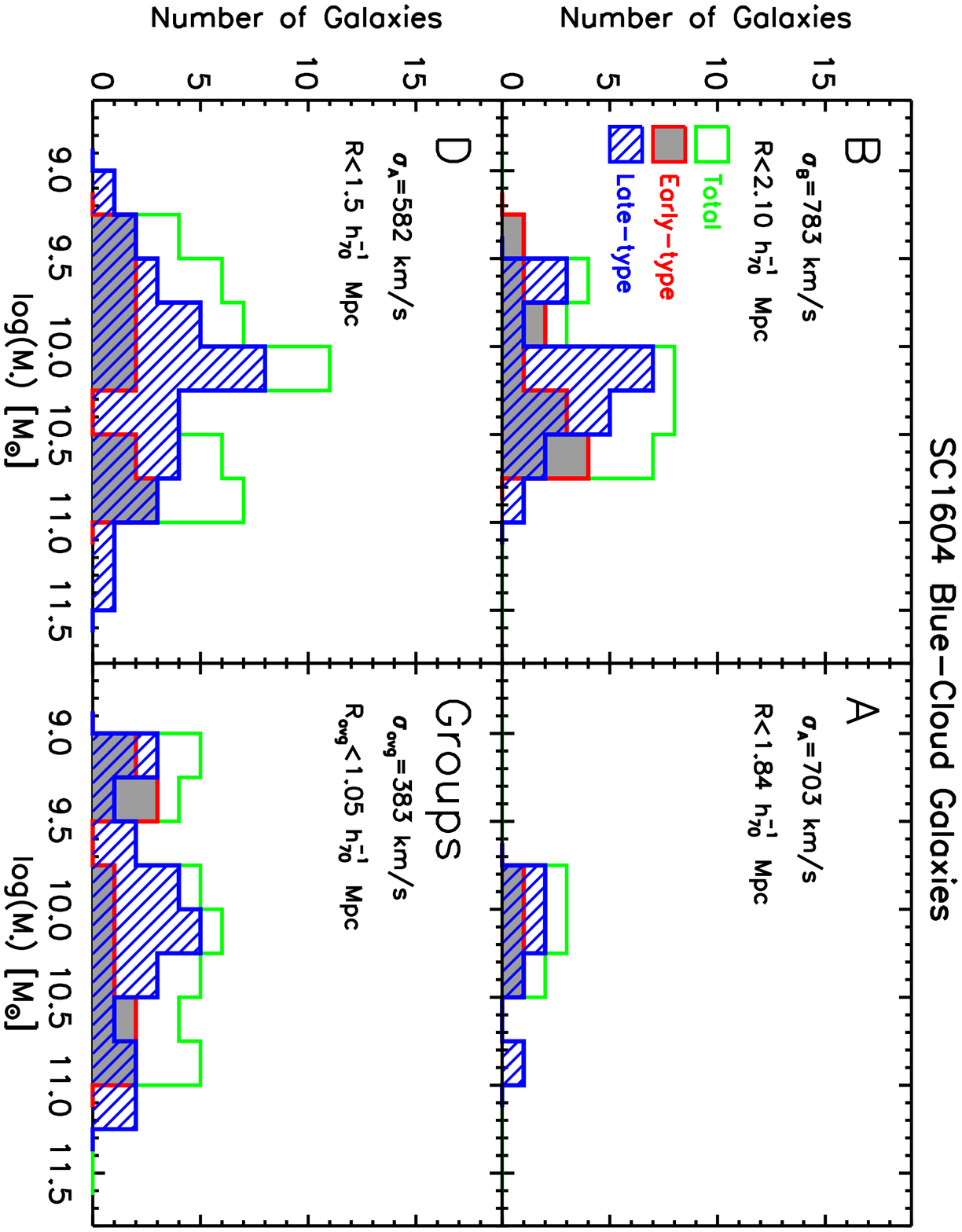}
\caption{\emph{Left:} Histogram of morphological type vs. stellar mass for red-sequence member galaxies of the Cl1604 clusters and groups. The name of
each system as well as the associated velocity dispersion and projected radial cutoff used for membership is shown in each panel (the groups subset is
a combined sample of the members of the five Cl1604 groups). Early-type galaxies include both ellipticals and S0 galaxies. Galaxies for which we were not
able to determine a morphological type are included only in the ``Total" histogram. \emph{Right:} Similar to left panels except now plotted for
blue-cloud member galaxies. While red-sequence galaxies are primarily early-type systems and blue-cloud galaxies are primarily late-type systems,
there are important exceptions. This is especially noticeable in the mass range of $\log(\mathcal{M}_{\ast})\sim10.25$-$10.75$, where several transitional
populations exist (see \S\ref{massmorph}). Late-type galaxies observed on the red sequence are composed of quiescent disks and dusty $24\mu m$ bright
star-forming galaxies. The latter are particularly prevalent at high masses [$\log(\mathcal{M}_{\ast})>10.75$].}
\label{fig:massmorphology}
\end{figure*}

In Figure \ref{fig:massmorphology} we present mass histograms of late-type and early-type member galaxies 
for both the blue cloud and red sequence of each system (or composite systems in the case of group members). Recall that in our classification
scheme, S0 galaxies are considered ETGs. Examining the red sequence mass histogram first (left panels), it is immediately clear
that a majority of galaxies on the cluster and group red sequences are elliptical or S0 galaxies. In the previous section we discussed the possibility
of that the group red sequence contained either a large number of dust-reddened starbursts or a large number of massive quiescent galaxies.
The high fraction of ETGs present in the red sequence of the groups strongly favors the latter interpretation. Furthermore,
in the study of K11 we found that only a small fraction ($\sim7$\%) of ETGs in the groups are detected in $24\mu m$ at
the starburst or luminous infrared galaxy (LIRG) level. These galaxies are also not likely to be forming stars at sub-starburst (i.e., ``normal")
levels, as the average [\ion{O}{2}] EW of the massive group red sequence population is consistent with no current star formation (see \S\ref{discussion}).
Thus, it appears that \emph{the presence of massive, quiescent, ETGs is a common phenomenon in the Cl1604 groups}, further reinforcing
our previous claim that significant pre-processing is occurring in such environments.

In cluster environments a similar trend is observed; a majority of the RSGs have early-type morphologies. However, both in the
group and cluster environment a small number of late-type galaxies are observed on the red sequence. A large fraction (76\%) of these
galaxies, as well as nearly all of the massive [$\log(\mathcal{M}_{\ast})>10.75$] red-sequence late-type galaxies, are the 24$\mu$m-bright dusty starburst galaxies
studied in K11. The remaining population, which is primarily observed at lower stellar masses [$\log(\mathcal{M}_{\ast})\sim10.25$-$10.5$], is comprised of 
late-types that are not observed at 24$\mu$m and have weak [\ion{O}{2}] emission (SFR([\ion{O}{2}])$<1\mathcal{M}_{\odot}$ yr$^{-1}$, where the [\ion{O}{2}] luminosity is measured using methods nearly identical 
to that of L10 and translated to an SFR using the relationship of Kennicutt et al.\ 2009). Because such galaxies are on the red sequence and not detected at 24$\mu$m, and because there
exists a strong correlation between SFR and the colors of dusty starbursts (in that systems with a higher SFR are redder, see K11), it is unlikely that 
these galaxies are lower luminosity analogs to the 24$\mu$m-bright dusty starburst population, but are rather truly quiescent. 
Though rare in the Cl1604 supercluster, this red ``passive disk" population is of particular interest, as it is thought to be one of the main progenitors
of S0 galaxies that are found in large numbers in low-redshift clusters, and thus potentially representing an intermediate stage in the transformation of a star-forming
late-type galaxy to a quiescent S0 (Moran et al.\ 2006, 2007; Bundy et al.\ 2010). We will return to consider this population later in the section.

In the right panel of Figure \ref{fig:massmorphology} we plot the mass histograms of early- and late-type systems for blue-cloud member
galaxies. While the fraction and mass distribution of these galaxies changes
dramatically from system to system, a large majority of such galaxies are late-type galaxies.
Again, we observe a significant difference at the massive end of the blue-cloud galaxy mass function between the two most massive clusters (A and B)
and the lower mass systems (cluster D and the groups); the large population of massive [$\log(\mathcal{M}_{\ast})>10.75$] blue-cloud galaxies observed
in cluster D and the groups are largely absent from clusters A and B. At all masses we also observe a non-trivial fraction of blue
ETGs. The bulk of this population lies at a mass range identical to that of the passive disks discussed earlier and also
at the mass range where the number counts of red-sequence ETGs begins to decrease dramatically. The stellar mass range of $\log(\mathcal{M}_{\ast})\sim10.25$-$10.75$
seems to be an important threshold in group and cluster environments; a significant number of galaxies of this mass appear to be
transforming from blue late-types into red ETGs at $z\sim1$. This mass range is also roughly the stellar mass
where a majority of the star formation takes place in blue-cloud galaxies and where a large fraction of the stellar mass is added to the RSG mass function 
from $z\sim0.9$ to $z=0$ (Bell et al.\ 2004).

\begin{figure*}
\plottwospecial{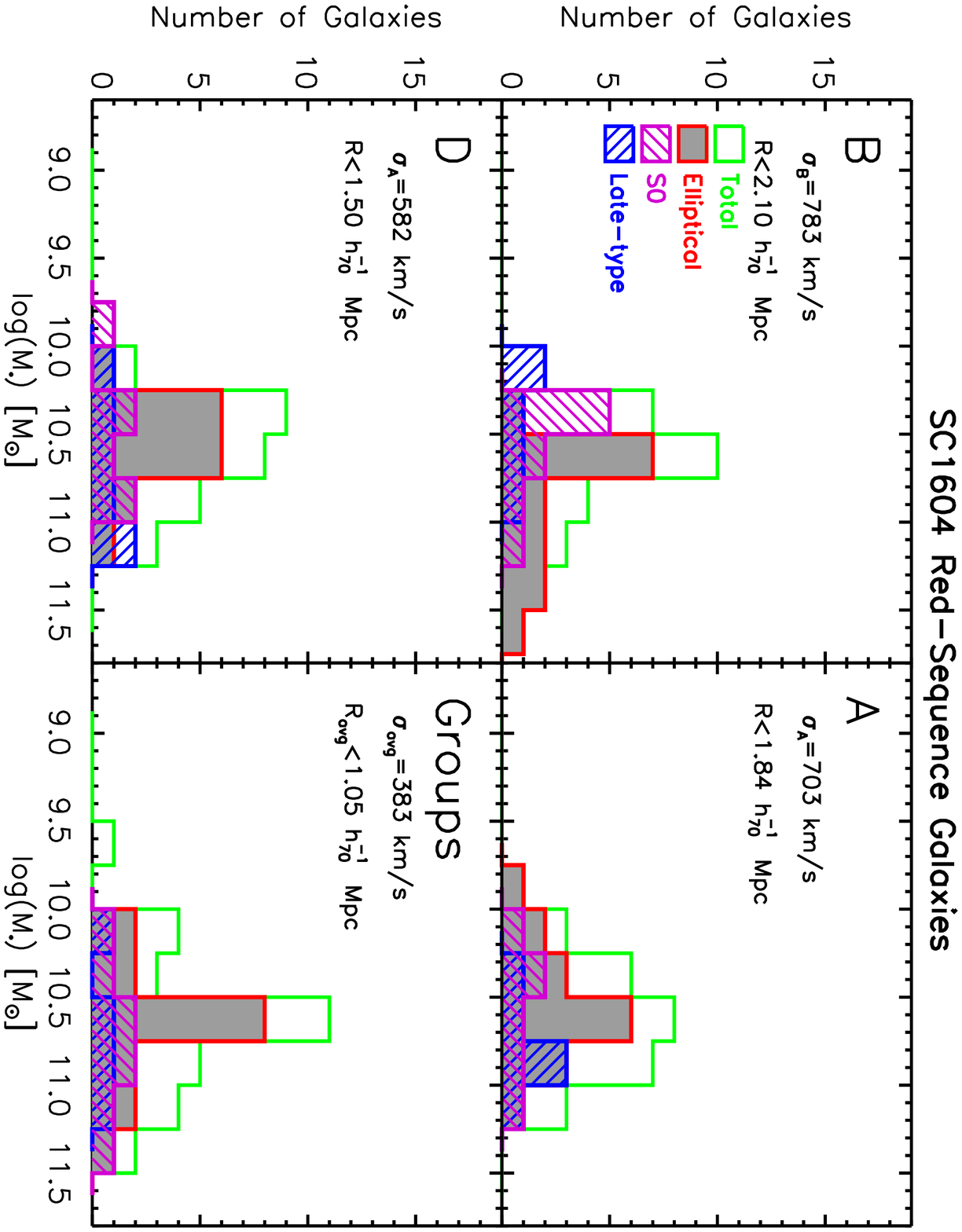}{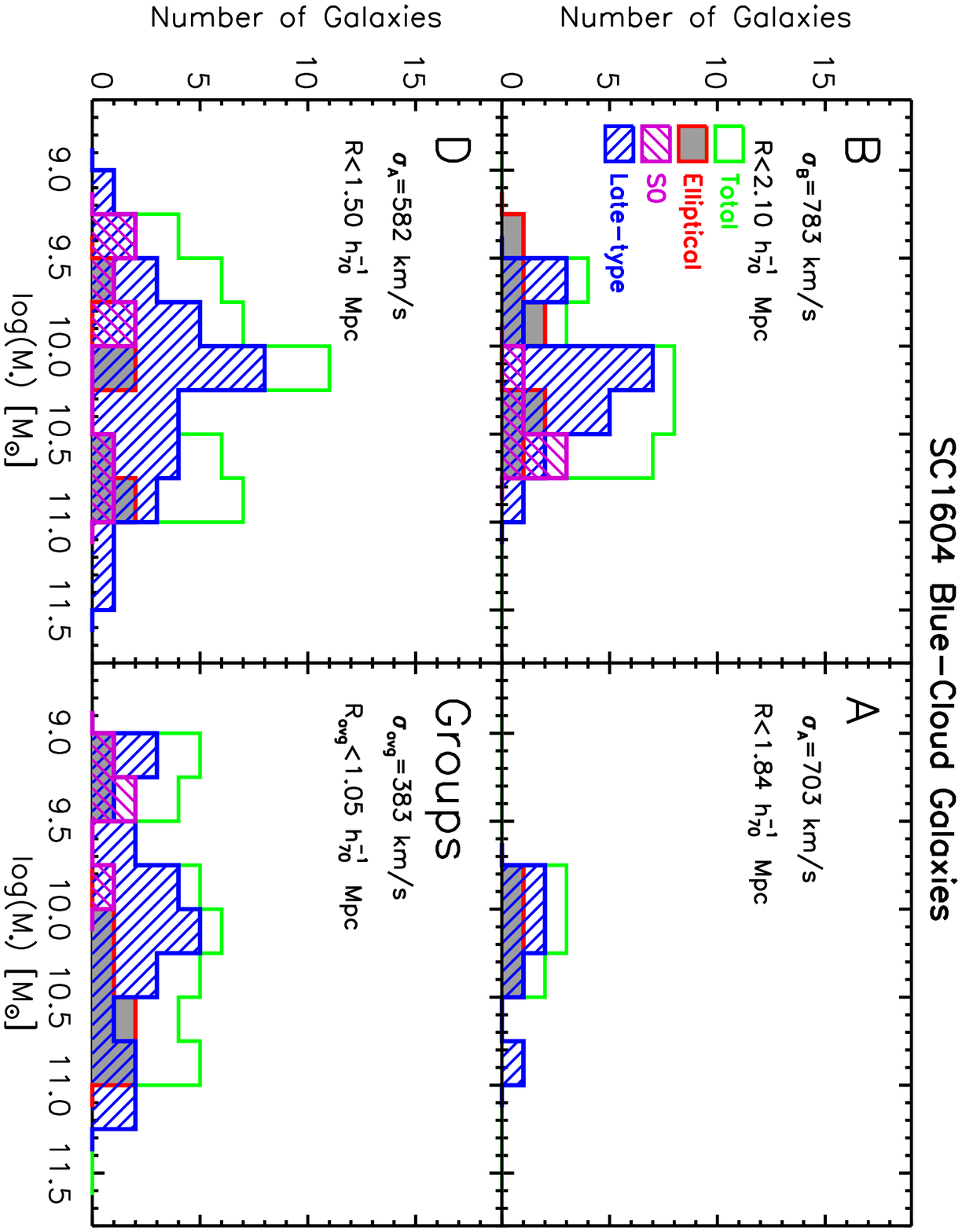}
\caption{\emph{Left:} Identical to the left plots of Figure \ref{fig:massmorphology}, 
except that we now separate the early-type classification into
ellipticals and S0 galaxies. \emph{Right:} Same as left panels, except plotted for blue-cloud galaxies only. Note the relatively large number
of S0 galaxies present in cluster B in both the red sequence and blue cloud populations in the mass range $\log(\mathcal{M}_{\ast})\sim10.25$-$10.75$. The mass and color
distribution of the three morphological types suggest that in cluster B member galaxies typically have their star formation truncated prior to
morphological transformation. In the other clusters and the group systems the two phenomena appear to be roughly coeval (see \S\ref{massmorph}).}
\label{fig:massmorphologywS0}
\end{figure*}

If we instead consider the two components of the ETG population, ellipticals and S0 galaxies, separately\footnote{In some cases, discriminating
between elliptical and S0 galaxies is extremely difficult, both in visual inspections and with statistical measurements.
The number of ETGs in our sample is large enough to average out the ambiguities in the classification
process (see \S\ref{morphology}), however, we urge the reader to keep in mind the difficulty of separating these populations when conclusions based on the differences
in the two populations are presented.}, the evidence becomes somewhat clearer. As
in Figure \ref{fig:massmorphology}, Figure \ref{fig:massmorphologywS0} shows mass histograms for red-sequence and blue-cloud galaxies. This time,
however, we differentiate between elliptical and S0 galaxies. Considering again the stellar mass range of $\log(\mathcal{M}_{\ast})\sim10.25$-$10.75$, we
observe an obvious trend in cluster B. Both the massive end of the blue cloud and the low-mass end of the red sequence
(in both cases the same mass of $\sim\log(\mathcal{M}_{\ast})\sim10.5$) are dominated by passive disks and S0 galaxies. Red ellipticals are
found at higher stellar masses, while blue late-type galaxies are primarily at lower masses. These blue S0 galaxies are
likely the progenitors of the more massive red sequence S0 galaxies in this system and will not re-assemble into star-forming late-types 
(as in Kannappan et al.\ 2009). This is a reasonable assumption given the plethora of quenching processes a galaxy is subject to in cluster
environments. Since fading of the disk alone cannot create elliptical galaxies from disk galaxy populations (Faber et al.\ 2007),
\emph{in cluster B it appears that the processes responsible for quenching star formation in a cluster galaxy largely occur prior to those
responsible for morphologically transforming a passive disk or S0 into an elliptical}, something that is also observed in the field and in intermediate
density environments at $z\sim1$ (see Bolzonella et al.\ 2010 and references therein). If merging events are the primary
mechanism responsible for morphologically transforming disk galaxies to ellipticals, as is generally thought the case (see, e.g., Hopkins et al.\
2010), such mergers must either \emph{(i)} occur once the S0 population is quenched and situated on the red sequence, or \emph{(ii)} be directly
or indirectly responsible for both the quenching and the morphological transformation of this population, with
the morphological transformation occurring over a longer timescale than the quenching of star formation. While it has been shown that
in field environments at $z\sim1$ merging activity is largely not responsible for quenching processes associated with populations transitioning onto the
red sequence (Mendez et al.\ 2011), it is unclear whether such results extend to high-redshift group and cluster environments. It has also
been suggested both in observational studies and in simulations that a two-stage process such as the possibility raised in \emph{i)}
may be favored to explain the transformation of star-forming late-types into quiescent ellipticals (see discussions in S{\'a}nchez-Bl{\'a}zquez et al.\
2009 and Bundy et al.\ 2010). Though such questions are outside the scope of the current work, these topics will be the focus of a future study.

In the remaining systems the picture is somewhat different; ellipticals and S0 galaxies are observed at a large range of masses in both the
red sequence and the blue
cloud (with the exception of cluster A). Though the number of galaxies in each sub-population is somewhat small (i.e., 10-15 galaxies), the fractions of red,
passive disk/S0 galaxies and blue elliptical galaxies with $\log(\mathcal{M}_{\ast})\sim10.25$-$10.75$ in the two lower mass clusters and the group
systems is nearly identical. If we assume that the progenitors of these galaxies are late-types, a reasonable assumption given that
such galaxies comprise the bulk of field galaxies at $z\sim1$ (see, e.g., Scarlata et al.\ 2007), the relative number of blue ellipticals and red passive spiral/S0
galaxies in these systems suggests that \emph{in lower mass clusters and groups the quenching of star formation occurs prior to morphological
transformation in only roughly $50$\% of cases}. 

It is not clear what exactly makes the morphological and color transformations of the constituent galaxies of cluster B different from those of 
the other clusters and groups. Cluster B is the only system in the supercluster which exhibits both a bright ICM and a galaxy population 
that is dynamically unrelaxed. It is possible that the confluence of these two physical conditions results in transformations that do not
occur when only a dominant ICM (as in cluster A) or a dynamically unrelaxed galaxy population (as in cluster D) is present. However, 
a full characterization of the processes responsible in each case requires detailed analysis of the spectral
properties of the cluster and group systems as a function of, e.g., clustocentric radius, as well as detailed analysis of the morphological properties of
these transition populations. This will also be explored in a future study. What is clear, however, is that there appears to be a large population of transition
galaxies in Cl1604 group and cluster environments at $\log(\mathcal{M}_{\ast})\sim10.25$-$10.75$. We will continue to discuss this population over the course of the following section.

\section{The Buildup of the Red Sequence at $z\sim1$}
\label{discussion}

Throughout the previous sections we have approached the analysis of the Cl1604 member galaxies from a
variety of different observational standpoints. The sheer amount of data presented in this paper on the supercluster
member galaxies makes the task of drawing a cohesive picture of galaxy evolution within the supercluster environment
somewhat daunting. As such, before going further in our discussion, we begin this section by highlighting those results from
previous sections that are relevant to obtaining such a picture.

\begin{itemize}
\renewcommand{\labelitemi}{$\circ$}

\item In \S\ref{colormag} we discussed the color and magnitude properties of galaxies comprising the three clusters and five group
systems in the Cl1604 complex, finding that the cluster and group environment is instrumental in creating bright red galaxies,
but that significant variation exists in the fraction of red galaxies that comprise each cluster and group system.

\item In \S\ref{spectral} we found again large variations in the spectral properties of average member galaxies of the Cl1604 groups,
ranging from populations dominated by quiescent galaxies to galaxies largely comprised of starbursts. The average galaxy populations of
the three cluster systems were, however, relatively homogeneous in their star-forming properties. The average cluster galaxy in each
system was classified as a normal star-forming galaxy with an SFR at an level intermediate to an average field galaxy at $z\sim1$ and
an average low-redshift cluster member.

\item In \S\ref{lumfunc} a significant deficit of low-luminosity RSGs was observed in all systems. In
cluster A, the most relaxed cluster in the Cl1604 supercluster, this deficit is seen to be significantly less than the deficit
observed in the other Cl1604 systems. This result suggests that low-luminosity galaxies have largely not transitioned to the
cluster or group red sequence at $z\sim1$, but that the processing of such galaxies is beginning to occur in the environments of
more advanced systems at these redshifts.

\item In \S\ref{colormass} we determined that massive $\mathcal{M}_{\ast}\gsim10^{11} \mathcal{M}_{\odot}$ RSGs are observed
in all clusters and the composite group population. The most massive RSGs are housed in the most
massive Cl1604 cluster (cluster B) and, surprisingly, the group systems. The most massive galaxies observed in all systems
are, however, less massive than typical $z\sim0$ BCGs by a factor of two or greater, suggesting that significant evolution will occur in such
galaxies. In this section we also showed that a large population of high-mass, blue-cloud galaxies are observed in both cluster D
and the groups systems. This population is largely absent in the two most massive clusters (clusters A \& B). We also showed
that the bright blue galaxies observed in clusters A \& B had higher SSFRs than the analogous population observed in cluster D and
the group systems.

\item In \S\ref{radial} we discussed the radial distributions of the galaxies comprising the Cl1604 cluster and composite
group populations. These high SSFR blue-cloud galaxies observed in the two most massive clusters are found primarily
at low projected clustocentric radius. In the lower mass systems (cluster D and the groups) we find the majority of both bright
and massive blue cloud member galaxies are also observed at low clustocentric radius. The radial distribution of RSGs 
in these systems revealed a large population of unusually bright red galaxies in the core of cluster D.

\item Finally, in \S\ref{massmorph} we considered the morphology of red-sequence and blue-cloud galaxies that comprise
the Cl1604 cluster and group systems. Several populations of ``transition" galaxies were found in all systems at intermediate
stellar masses [$\log(\mathcal{M}_{\ast})\sim10.25-10.75$], though a majority of this population was observed in cluster B. These
transition populations included passive spiral galaxies, blue and red S0s, and massive blue ellipticals.

\end{itemize}

We are now in a position to bring all our observational evidence to bear on the question of galaxy evolution in high-redshift
clusters and groups. While it is important to note that the Cl1604 supercluster represents only one set of clusters and groups
at high-redshift, we stress here that we are observing all of these systems at virtually the same epoch. In such a way, we are
able to cleanly separate out redshift dependent galaxy evolution from galaxy evolution driven largely by environment.

\begin{deluxetable*}{ccccc}
\tablecaption{Composite Equivalent Width and $D_n(4000)$ Values of the Red Galaxy Populations of the Cl1604 Groups and Clusters \label{tab:EWnD40002}}
\tablehead{\colhead{} & \colhead{Stellar Mass Range\tablenotemark{b}} & \colhead{EW([\ion{O}{2}])\tablenotemark{c}} & \colhead{EW(H$\delta$)\tablenotemark{c}} & \colhead{} \\
\colhead{Subset\tablenotemark{a}} & $(\log(\mathcal{M}_{\odot})$ & \colhead{(\AA)} & \colhead{(\AA)} & \colhead{$D_n(4000)$\tablenotemark{c}}}
\startdata
Cluster A &  $>9.8$ & $-$5.75$\pm$0.29  & 2.22$\pm$0.25 & 1.593 $\pm$0.010 \\[4pt]
Cluster B &  $>9.8$ & $-$2.52$\pm$0.25  & 2.92$\pm$0.20 & 1.646 $\pm$0.010 \\[4pt]
Cluster B low-$\mathcal{M}$ & $<10.5$ & 1.03$\pm$0.54 & 3.14$\pm$0.38 & 1.445$\pm$0.023  \\[4pt]
Cluster D &  $>9.8$ & $-$7.05$\pm$0.46 & 2.25$\pm$0.28 & 1.365$\pm$0.016 \\[4pt]
Groups\tablenotemark{d} & $>9.8$ & $-$1.67$\pm$0.23  & 1.25$\pm$0.20 & 1.703$\pm$0.009 \\[4pt]
\tablenotetext{a}{Subsets include only those galaxies brighter than $F814W<23.5$ that are within $R<1.2R_{\rm{vir}}$ of the cluster/group center.}
\tablenotetext{b}{Range of red-sequence galaxies included in each subsample. Only galaxies with well-defined (see \S\ref{SEDfitting}) stellar masses are included.}
\tablenotetext{c}{Only random errors are reported for EW([\ion{O}{2}]), EW(H$\delta$), and $D_n(4000)$ as incompleteness errors are negligible.}
\tablenotetext{d}{Measurements made a composite spectrum comprised of all Cl1604 group galaxies}
\enddata
\end{deluxetable*}

\begin{figure*}
\plottwosuperspecial{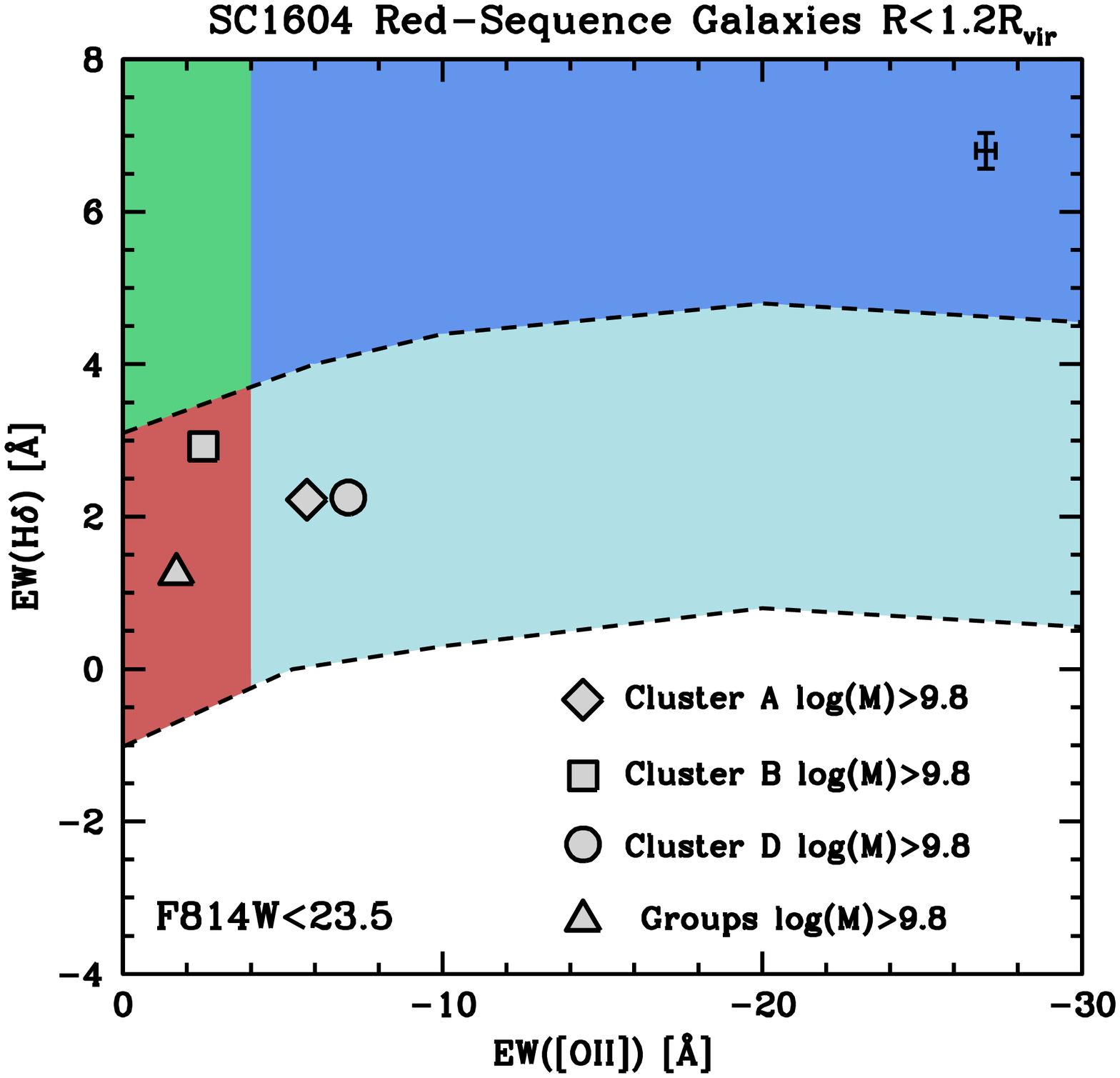}{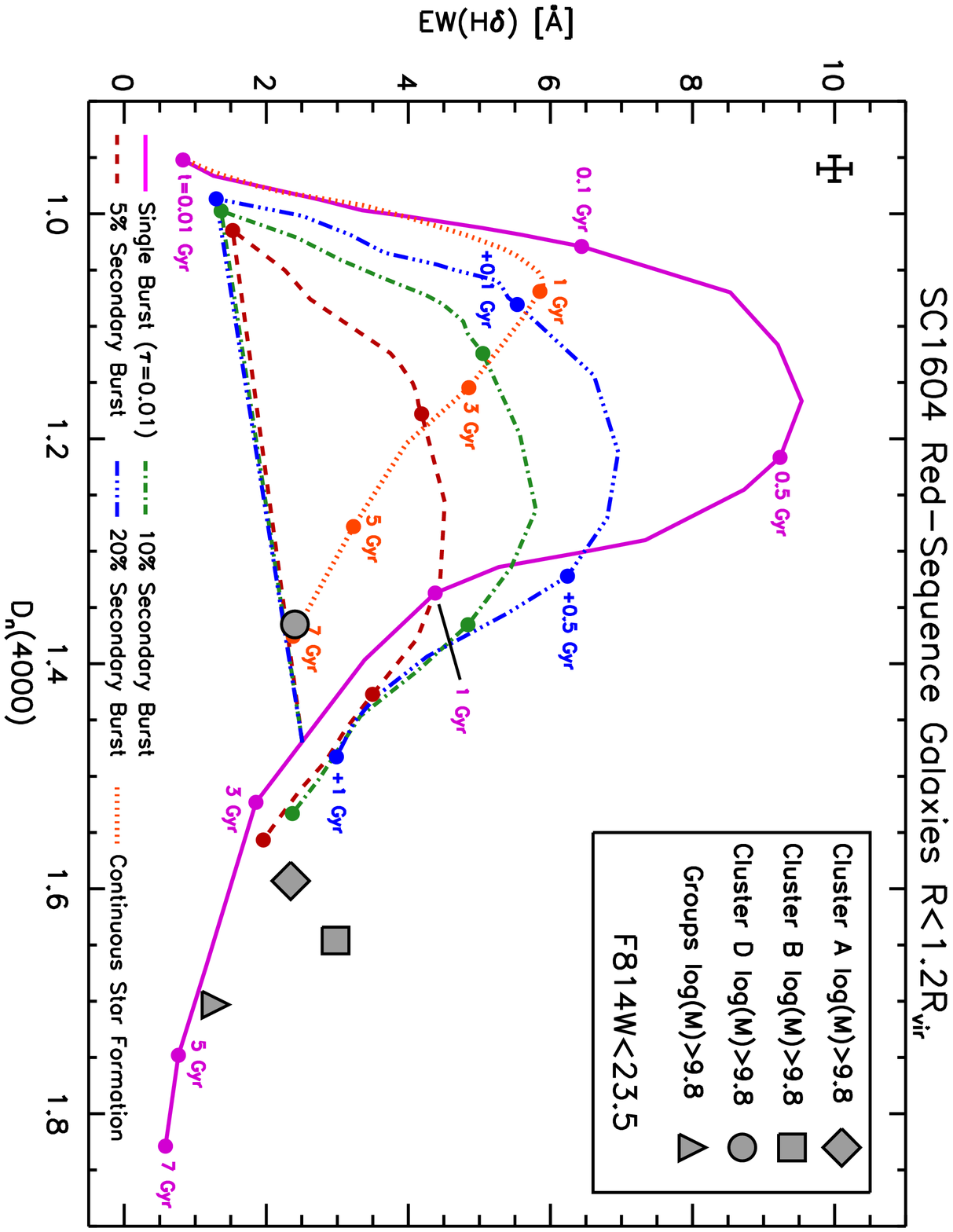}
\caption{\emph{Left:} Plot of the equivalent width of the [\ion{O}{2}] and H$\delta$ spectral features measured from the composite
spectra of red-sequence members galaxies of the Cl1604 clusters and groups. Only those galaxies that are brighter than $F814W<23.5$,
have well-defined stellar masses, and are within a projected radial distance of $R<1.2R_{\rm{vir}}$ are included in each composite spectrum. The
meanings of the dashed lines and shaded regions are identical to Figure \ref{fig:dresslergeneral}. 
The average measurement error
is shown in the upper right corner. These errors do not include spectroscopic completeness effects, as we are relatively complete
for RSGs to these magnitudes. The elevated [\ion{O}{2}] levels in the average RSG in cluster A is likely
due to LINER/Seyfert activity, while in cluster D this is likely due to residual or recent star-formation (see \S\ref{massmorph}). 
\emph{Right:} Plot of measurements of the $D_n(4000)$ and H$\delta$ spectral features from the same composite spectra as in the 
left panel. Overplotted are measurements of $D_n(4000)$ and H$\delta$ from several different Bruzual (2007) models at a variety 
of different times. The synthetic spectra are generated using solar-metallicity models with $A_{\nu}=1.01$. Age tickmarks
on the secondary burst models are labeled on the blue (20\% Secondary Burst) line and have identical meanings to those on the 
green dot-dashed and red dashed lines. The
average errors on the measurements of the Cl1604 composite spectra are given in the upper left corner. While
the stellar populations of the average RSG in clusters A \& B and the group systems (filled diamond, square, and triangle, respectively) 
are consistent with being formed at early times ($z_f\sim3$), the average RSG stellar population in cluster D (filled triangle) 
appears at significantly lower $D_n(4000)$.}
\label{fig:redEWnD4000}
\end{figure*}

We begin by focusing on the properties of the galaxies in each system that have already transitioned 
to the red sequence at $z\sim1$. In Figure \ref{fig:redEWnD4000} we plot
average spectral properties of certain subsets of Cl1604 RSGs against 
measurements made from solar-metallicity synthetic spectra (Bruzual 2007) with a variety of different SFHs. 
Each synthetic spectra is ``extincted" using a Calzetti et al.\ (2000) reddening law and an $E(B-V)= 0.25$,
approximately equal to the average $E(B-V)$ of both red-sequence and blue-cloud members estimated using our SED fitting.
Plotted are EW([\ion{O}{2}]), EW(H$\delta$), and $D_n(4000)$ measured from composite spectra of RSGs in each cluster and the combined group sample 
that are within $R<1.2R_{\rm{vir}}$ and brighter than $F814W=23.5$. The projected radial limit is chosen in order to minimize contamination from 
galaxies at large projected radii that may have had no interaction with the central regions of the group/cluster and that are dynamically removed 
from the central regions by $\gsim1$ Gyr. Additionally, the majority of all galaxies in both the red-sequence galaxies considered here and the 
blue-cloud galaxies considered later are within this radius in all systems. The magnitude limit ensures that any effects from completeness are 
minimal when making our comparisons.

The spectral properties of the observed RSGs in clusters A \& B (filled diamond and square, respectively) are broadly consistent with the
conclusions reached in other sections: \emph{the average RSG in the two most massive clusters is consistent with forming
through a single burst at $z_f=2.5-3$.} The significantly elevated EW(H$\delta$) of the average RSG in cluster B
suggest that at least some of these galaxies have had recent ($\lsim1$ Gyr) star formation activity. In 
\S\ref{massmorph} we suggested that this population was largely comprised of low-mass galaxies on the red sequence.
We confirm that suggestion here; a composite spectrum comprised of only low-mass [$\log(\mathcal{M}_{\ast})<10.5$] RSGs
at $R<1.2R_{\rm{vir}}$ in cluster B reveals a relatively young stellar population with recent (i.e., $\lsim1.5$ Gyr) star formation [$D_n(4000)=1.445$;
EW(H$\delta$)=3.14\AA]. \emph{In the most massive cluster in the Cl1604 supercluster low-mass galaxies have only recently arrived on
the red-sequence.} In cluster A this does not seem to be the case. The composite spectrum of a similar
population of RSGs in cluster A reveals even the lowest-mass RSGs in this system have, on average, moderately old
(i.e., $\sim3-3.5$ Gyr) stellar populations [$D_n(4000)=1.558$].

In the group systems the picture is largely similar. Red-sequence galaxies in group environments (filled triangle) have the oldest average
stellar population of any population in the supercluster, consistent
with a formation epoch of $z_f\gsim3$. However, in contrast with cluster B, we observe no significant
excess of EW(H$\delta$) for the average group RSG relative to the single burst model plotted in the
right panel of Figure \ref{fig:redEWnD4000}. This suggests that, unlike cluster B (and to a lesser extent cluster A), there
are essentially no new arrivals to the group red sequence in the last $\sim1$ Gyr. While significant pre-processing has obviously occurred
in these systems prior to $z\sim1$, as evidenced by the large number of bright and massive galaxies observed in the groups,
it appears that RSGs in group environments are largely in place at early times.

In cluster D the average RSG (filled circle) has a much smaller $D_n(4000)$ than the analogous population of any of
the other systems. In \S\ref{radial} we discussed a large population of unusually bright RSGs in the core
of cluster D. The observed EW([\ion{O}{2}]), EW(H$\delta$), and $D_n(4000)$ values of cluster D RSGs imply
that this population has low residual levels of star formation and is comprised of relatively young stellar populations. This
strongly suggests \emph{that the red-sequence galaxies in the center of cluster D, which includes nearly all of the most massive
galaxies in the system, have only recently transitioned onto the red sequence.} Values of EW([\ion{O}{2}]), EW(H$\delta$),
and $D_n(4000)$ for each of the red-sequence samples discussed here are given in Table \ref{tab:EWnD40002}.

\begin{deluxetable*}{ccccc}
\tablecaption{Composite Equivalent Width and $D_n(4000)$ Values of the Blue Galaxy Populations of the Cl1604 Groups and Clusters \label{tab:EWnD40003}}
\tablehead{\colhead{} & \colhead{Stellar Mass Range\tablenotemark{a}} & \colhead{EW([\ion{O}{2}])\tablenotemark{b}} & \colhead{EW(H$\delta$)\tablenotemark{b}} & \colhead{} \\
\colhead{Subset\tablenotemark{c}} & $(\log(\mathcal{M}_{\odot})$ & \colhead{(\AA)} & \colhead{(\AA)} & \colhead{$D_n(4000)$\tablenotemark{b}}}
\startdata
Cluster A All Masses&  $>9.8$ & $-$13.90$\pm$0.91  & 6.15$\pm$0.74 & 1.250 $\pm$0.023 \\[4pt]
Cluster B All Masses&  $>9.8$ & $-$16.29$\pm$0.47  & 3.03$\pm$0.30 & 1.342 $\pm$0.010 \\[4pt]
Cluster D high-$\mathcal{M}$ & $>10.5$ & $-$8.04$\pm$0.47 & 3.75$\pm$0.31 & 1.342$\pm$0.011  \\[4pt]
Cluster D low-$\mathcal{M}$&  $<10.5$ & $-$24.03$\pm$1.03 & 4.52$\pm$0.81 & 1.051$\pm$0.019 \\[4pt]
Groups\tablenotemark{d} high-$\mathcal{M}$ & $>10.5$ & $-$8.78$\pm$0.18  & 6.15$\pm$0.20 & 1.112$\pm$0.005 \\[4pt]
Groups\tablenotemark{d} low-$\mathcal{M}$ & $<10.5$ & $-$13.27$\pm$0.58  & 5.14$\pm$0.81 & 1.208$\pm$0.017 \\[4pt]
\tablenotetext{a}{Range of blue-cloud galaxies included in each subsample. Only galaxies with well-defined (see text) stellar masses are included.}
\tablenotetext{b}{Only random errors are reported for EW([\ion{O}{2}]), EW(H$\delta$), and $D_n(4000)$ (see text).}
\tablenotetext{c}{Subsets include only those galaxies brighter than $F814W<22.5$ that are within $R<1.2R_{\rm{vir}}$ of the cluster/group center.}
\tablenotetext{d}{Measurements made a composite spectrum comprised of all Cl1604 group galaxies}
\enddata
\end{deluxetable*}

\begin{figure*}
\plottwosuperspecial{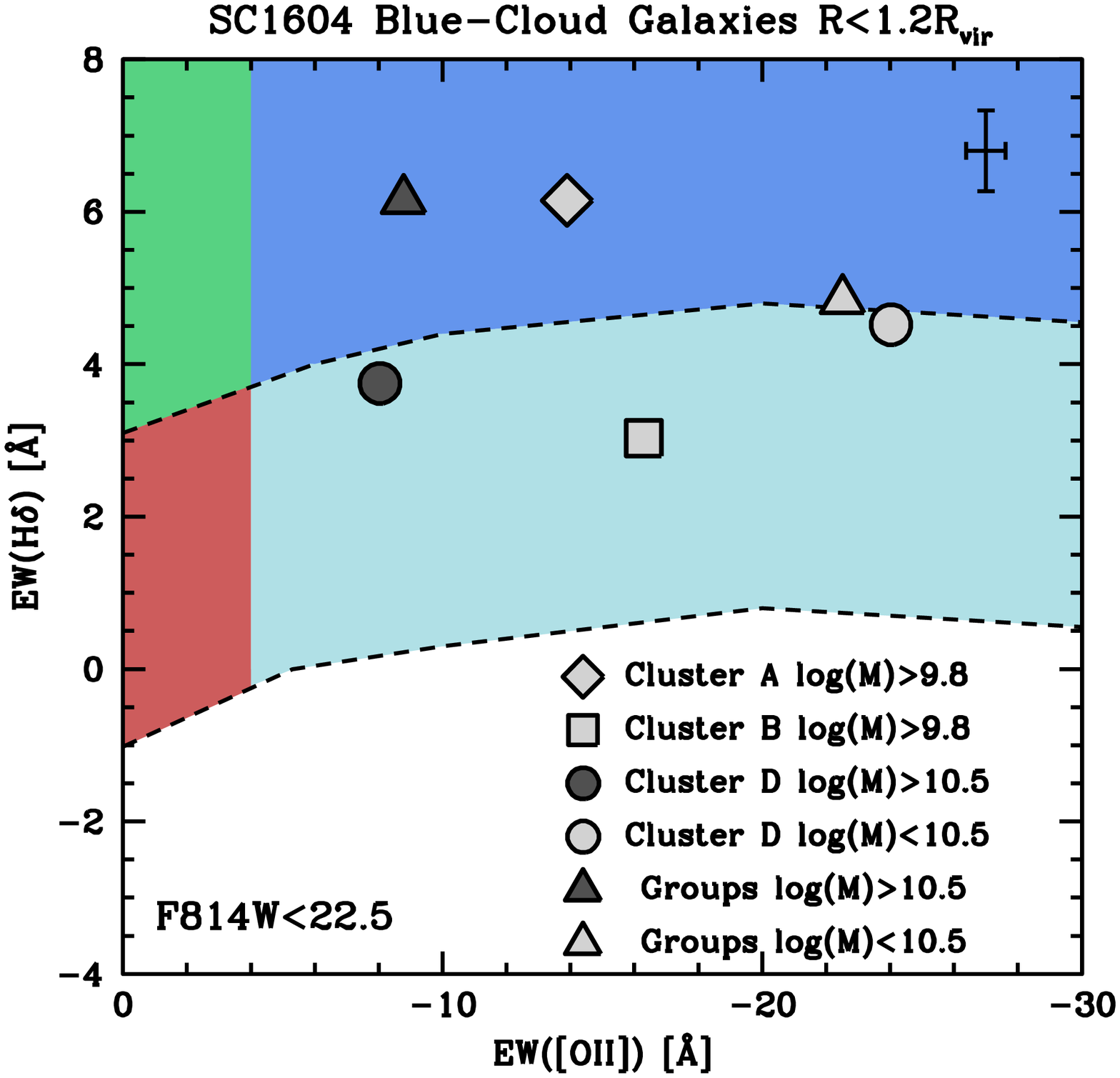}{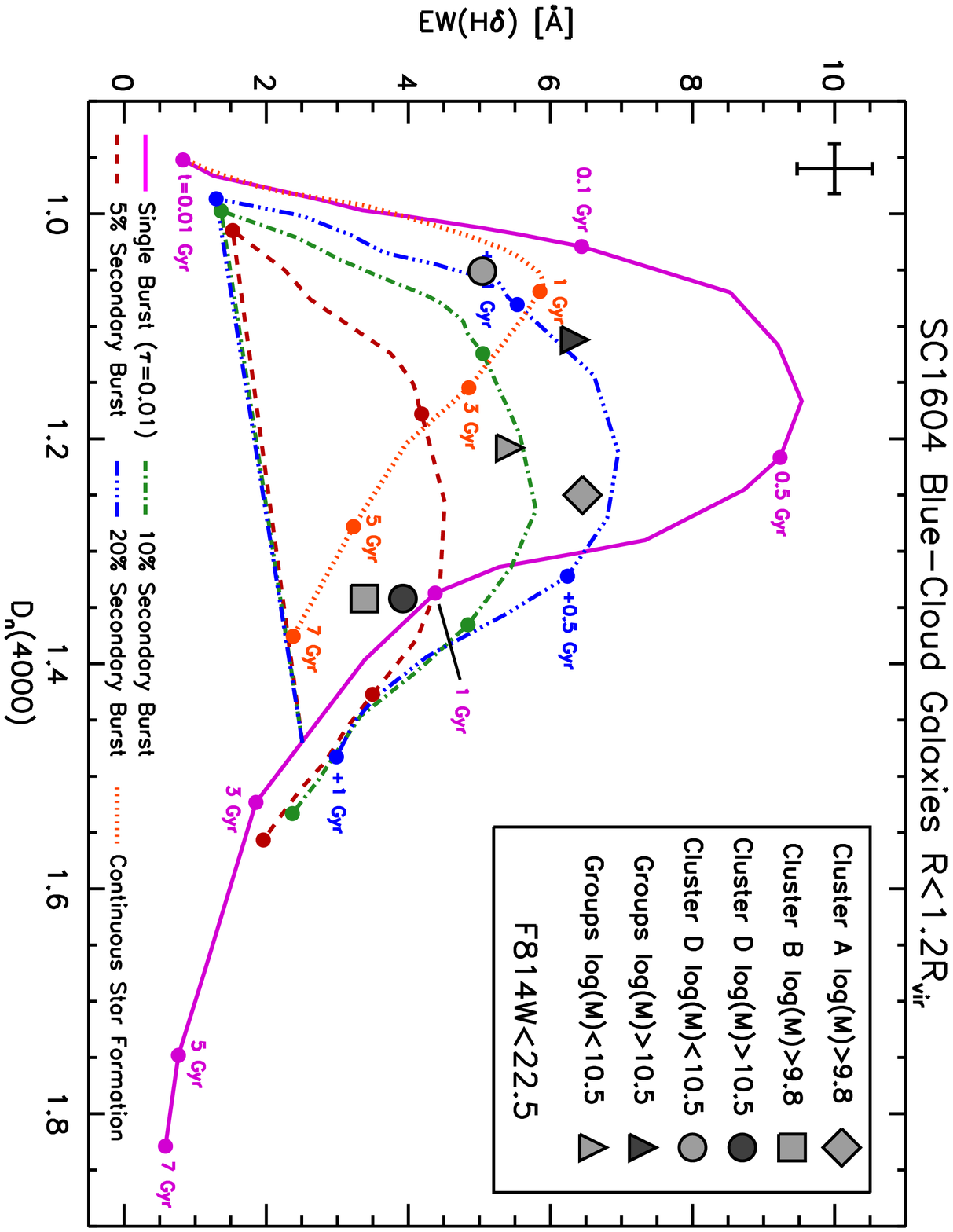}
\caption{\emph{Left:} EW([\ion{O}{2}]) and EW(H$\delta$) measurements made from composite spectra of various subsets of blue-cloud member
galaxies of the eight Cl1604 clusters and groups. The meanings of the dashed lines and shaded regions are identical to Figure
\ref{fig:dresslergeneral}. Average measurement errors are given in the top right corner. These errors do not include spectroscopic
completeness effects, as we are relatively complete for blue-cloud galaxies brighter than $F814W<22.5$. \emph{Right:} Measurements
of $D_n(4000)$ and H$\delta$ from the same composite spectra as in the left panel. Meanings of the overplotted measurements from
stellar synthesis modeling are identical to those in Figure \ref{fig:redEWnD4000}.
Significant differences in the spectral properties of blue-cloud member galaxies are apparent. Bright, high-mass blue-cloud galaxies
in cluster D (dark filled circle), as well as all bright, blue-cloud galaxies in cluster B (filled square), appear to be transitioning on to the red sequence, while low-mass
blue-cloud galaxies in cluster D (light filled circle) require several Gyr before transitioning on to the cluster red sequence. Though the average blue-cloud galaxy in cluster
A (filled diamond) is starbursting and still relatively young, most of the galaxies observed in this system are situated on the cluster red sequence by
$z\sim0.9$. All blue-cloud galaxies in the group systems (filled triangles) appear to be relatively young and still several Gyr away from transitioning
onto the red sequence.}
\label{fig:blueEWnD4000}
\end{figure*}

If the picture we have given so far is correct, the progenitors of galaxies recently transitioning onto the cluster red sequence
should be observable in the cluster galaxy populations. The two main galaxy populations for which we have the most evidence of
these transitions are the moderate- to high-mass red galaxies in cluster D and the low-
to moderate-mass red galaxies in cluster B. In Figure \ref{fig:blueEWnD4000} we plot the spectral
properties of composite spectra generated from various subsets of blue galaxies within $R<1.2R_{\rm{vir}}$ and $F814W<22.5$ in the group and cluster
environments. These limits are again chosen to minimize contamination from field galaxies and spectroscopic completeness effects for cluster and group blue-cloud galaxies.
Table \ref{tab:EWnD40003} also lists the composite spectral properties for this and other blue-cloud populations considered in this section. 
Bright, massive [$\log(\mathcal{M}_{\ast})>10.5$] blue-cloud galaxies in the center ($R<1.2R_{\rm{vir}}$) of cluster D (dark filled circle) share almost identical properties
with the red-sequence population in cluster D. Of the blue-cloud populations observed in the supercluster, the bright, massive blue-cloud
galaxies in cluster D have on average both the second highest $D_n(4000)$ and the second lowest EW(H$\delta$) values. In fact, the 
$D_n(4000)$ of such galaxies in the center of cluster D is
consistent with being \emph{identical} to that of the average cluster D RSG. While the stellar mass and brightness limits differ
between the blue-cloud and red-sequence galaxies being considered, it is not necessary for this analysis that they be the same, as 
we are comparing this blue-cloud population to the average cluster D RSG. Thus, these results strongly suggest that
\emph{bright, massive blue-cloud galaxies at the center of cluster D are in the process of transitioning on to the red sequence at $z\sim1$}. 
There is some concern that a large population of
dusty starburst galaxies, as is observed in cluster D (see K11), could artificially inflate the $D_n(4000)$ value, causing the
average stellar population to appear artificially older. However, in K11 we determined that the majority of the 24$\mu$m-bright galaxies
observed in the center of the Cl1604 clusters are decaying in their starburst activity, consistent with these results. 

In contrast to the analogous population in cluster D, the average $D_n(4000)$ and EW values of the bright, massive blue-cloud galaxies in the group 
systems (dark filled triangle) reveal 
a starbursting population that is very young. Even with the instant cessation of star-formation activity, it would require
$\sim1$ Gyr to move these galaxies onto the group red sequence. This result is consistent with the results of K11, in which we
found that 24$\mu$m-bright starbursting galaxies were comprised of extremely young stellar populations. It appears that bright,
massive blue-cloud galaxies in the group systems are generally not transitioning to the red sequence at $z\sim1$.

The bright, massive blue-cloud population observed in cluster D and the groups is largely absent in clusters A \& B. Assuming these clusters were formed from
progenitors with similar galaxy populations to cluster D and the groups, the processing of massive blue-cloud galaxies that
is occurring in cluster D at $z\sim1$ must have occurred at earlier times in the two most massive Cl1604 clusters.
Since such galaxies do not exist in the two most massive Cl1604 clusters, we now consider instead the progenitors of the low- to moderate-mass galaxies that recently transitioned 
onto the red sequence in cluster B. In Figure \ref{fig:blueEWnD4000} we show the average spectral properties of the all bright blue-cloud
galaxies in cluster B that have well-defined stellar masses [$\log(\mathcal{M}_{\ast})>9.8$; light filled square]. The stellar mass limit here, and for 
blue-cloud galaxies in cluster A, differs from the limit imposed on cluster D galaxies due to the lack of massive blue galaxies in 
clusters A \& B discussed earlier. While we include the few blue-cloud galaxies in clusters A \& B in this analysis with $\log(\mathcal{M}_{\ast})>10.5$, 
the bulk of the population considered is comprised of the bright, high-SSFR galaxies discussed in \S\ref{colormass} and \S\ref{radial}, 
which share similar color and mass properties with the low- to moderate-mass RSGs in cluster B.
From the moderately high levels of [\ion{O}{2}] emission seen in the left panel of Figure \ref{fig:blueEWnD4000} we 
confirm the results of the \S\ref{colormass}; this population has, on average, higher optical SFRs and SSFRs than
the massive blue galaxies observed in cluster D or the group systems.

However, the right panel of Figure \ref{fig:blueEWnD4000}
presents a significantly different picture. The bright, moderate-mass blue cloud galaxies in cluster B have the both the highest
$D_n(4000)$ and the lowest EW(H$\delta$) values of any blue-cloud population considered in the supercluster. Both values are nearly
identical to those measured from the composite spectra of the lower-mass RSGs in the cluster, suggesting
that \emph{bright, low- to moderate-mass blue galaxies at the center of cluster B appear to be transitioning to the red sequence
at $z\sim1$}. In cluster D and the group systems, this is not the case for similar mass galaxies. The bright, moderate-mass 
[$\log(\mathcal{M}_{\ast})<10.5$]\footnote{While this is not the same mass limit as is used for cluster B, the average mass of 
the bright, moderate-mass blue-cloud galaxies in cluster D and the groups using this limit is nearly identical to the [$\log(\mathcal{M}_{\ast})>9.8$]
blue-cloud population considered in cluster B.} blue-cloud populations
in these systems (light filled circle and triangle, respectively) are comprised, on average, of young stellar populations with significant ongoing star-formation activity.
In cluster A there exist only four analogous galaxies (light filled diamond), two of which are the
decaying 24$\mu$m-bright starbursts discussed earlier in the section. Thus, the paucity of bright, blue intermediate-mass galaxies in the
center of cluster A suggests that much of the processing that is occurring in cluster B at $z\sim1$ occurred in cluster A at early
times. This is further evidenced by the relatively old stellar populations in the low-mass RSGs and
the smaller observed deficit of low-luminosity RSGs in cluster A as was discussed in \S\ref{lumfunc}.

We have now presented a picture in which the characteristic mass transitioning on to the group or cluster red sequence at $z\sim1$
is highly dependent on the environment in which that galaxy resides. In all systems (perhaps with the exception of cluster D) it
appears that early quenching is important in building up the red sequence at early times. However, late quenching of
cluster or group blue-cloud galaxies also appears to be prevalent in these environments at $z\sim1$. In the most relaxed and isolated
system in the supercluster (cluster A), both low-mass and high-mass galaxies have largely been processed and moved on to the cluster
red sequence prior to $z\sim1$. In the slightly more massive and slightly less relaxed cluster B, this processing appears to have occurred
for high-mass blue-cloud galaxies, but is only just beginning to occur at $z\sim1$ for moderate-mass [$\log(\mathcal{M}_{\ast})\lsim10.5$] blue galaxies. 
In the lower mass cluster D, a cluster shown in \S\ref{radial} to be extremely unrelaxed, we observe
the late quenching of massive [$\log(\mathcal{M}_{\ast})>10.5$] blue-cloud galaxies in the cluster center. The moderate-mass [$\log(\mathcal{M}_{\ast})\lsim10.5$] blue-cloud
galaxies in cluster D, however, seem generally unaffected by the quenching processes and are
likely still several Gyr from transitioning to the cluster red sequence. In the low-mass group systems the red sequence has been built up
significantly by $z\sim1$, but this buildup appears to have primarily occurred at early times ($z\gsim3$). Both the high-mass [$\log(\mathcal{M}_{\ast})>10.5$] and
moderate-mass [$\log(\mathcal{M}_{\ast})\lsim10.5$] blue-cloud populations in the group systems appear several Gyr from evolving into quiescent RSGs.

The process known as downsizing (Cowie et al.\ 1991, 1996), in which galaxies at fainter magnitudes (i.e., lower mass) are found to transition
on to the red sequence at later epochs than their brighter (i.e., higher mass) counterparts, is a process that is thought to be broadly
independent of environmental effects (see, e.g., Bundy et al.\ 2006). The picture we present here is in direct conflict with this claim. Since
all the systems in the Cl1604 supercluster are observed at virtually the same epoch, any effect that changes typical ``quenching mass"
($M_{Q}$, i.e., the mass of a typical blue-cloud galaxy that is transitioning on to the red sequence) in
the various Cl1604 systems must be largely due to environmental effects. From our data it appears, however, that the mass of the cluster
or group potential is not the key quantity in determining the typical $M_Q$ for a given system. Rather, it appears that the global dynamical
state of the group or cluster system is the primary determinant in setting the quenching mass in each system. Instead of
traditional (redshift-dependent) downsizing, in the Cl1604 supercluster \emph{we are observing dynamical downsizing, in which massive
blue-cloud galaxies are quenched earliest in the massive, relaxed clusters and the quenching of higher mass blue-cloud galaxies
occurs at progressively later times in lower mass and less relaxed clusters and groups.} This picture of dynamical downsizing
is consistent with the study of Baldry et al.\ (2006), in which the fraction of $\log(\mathcal{M}_{\ast})=10$-$11$ galaxies observed
on the red sequence at $z\sim0.1$ was found to be a strong function of environment.

Despite the wide variation in the formation history of RSGs in Cl1604,
galaxies with the characteristics of BCGs observed in relaxed clusters at $z\sim0$ appear to be missing from all of the systems.
These characteristics include extremely large stellar masses [$\log(\mathcal{M}_{\ast})\gsim12$; Stott et al.\ 2010], extremely
high visual luminosities [$\langle M_v\rangle = -23$; Ascaso et al.\ 2011], and large offsets in magnitude from the second brightest
cluster galaxy [$\langle\delta(m_{12}\rangle\sim1$; Smith et al.\ 2010; Ascaso et al.\ 2011]. As noted in \S\ref{colormass}, BCGs with
these properties are common in low-redshift clusters, suggesting a large variety of cluster progenitors at $z\sim1$
result in relaxed clusters dominated by BCGs. If we assume that the groups and clusters of the Cl1604 supercluster
are the progenitors of a ``typical" low-redshift cluster, significant evolution of the Cl1604 high-mass RSGs between $z\sim1$ and
$z\sim0$ is required\footnote{In fact, both the brightest and most massive galaxies in cluster D as well as the the brightest galaxy in cluster A are \emph{spirals}, strongly hinting
that significant evolution of the BCG will occur in these systems between $z\sim1$ and the present day.}. At $z\sim1$ the most massive RSGs in the Cl1604 systems range 
from $\log(\mathcal{M}_{\ast})\sim11$ to $\log(\mathcal{M}_{\ast})\sim11.6$. If the BCGs in modern day clusters are built from such galaxies, 
then these galaxies must more than double their mass from $z\sim1$ to $z\sim0$. This scenario is in
good agreement with the simulations of De Lucia \& Blaizot (2007) and the observations of Arag\'{o}n-Salamanca et al.\ (1998), in
which it was found that 50-80\% of the mass of BCGs is assembled between $z\sim1$ and $z\sim0$ (but see also Stott et al.\ 2010, 2011
for an opposing view).

This result is somewhat surprising given the early formation epoch of the average RSG in clusters A \& B and the group systems.
If these galaxies were in place at an early epoch, there seems to be no reason why the assembly process should be so vigorous between
$z\sim1$ and $z\sim0$ but languid at higher redshifts. However, our data does not preclude the possibility of early dry mergers,
in which two gas-depleted galaxies of roughly equal age merge into a single galaxy. Indeed, the old stellar populations
observed in low-redshift BCGs favors a scenario in which the most massive galaxies observed at $z\sim1$ accrete mass through dry mergers or
in mixed mergers with low mass ratios. Additionally, it has been shown that dry
mergers are known to become more frequent at later times (Lin et al.\ 2008) and that merging occurs more frequent and generally
involves both more massive galaxies and higher merger mass ratios in high density environments (see, e.g., McIntosh et al.\ 2008; de Ravel
et al.\ 2011). Thus, it seems reasonable to appeal to dry merging processes to buildup the massive end of the cluster red sequence
from $z\sim1$ to $z\sim0$ (though see Kaviraj et al.\ 2011 for an alternative view of ETG mass buildup involving mixed minor merging).
Such a scenario is also supported by the studies of Bell et al.\ (2006a,b), who found that a
majority of both massive and luminous galaxies have experienced a merger over the last $\sim7$ Gyr. Preliminary analysis of the most
massive galaxies in each of the Cl1604 cluster systems also supports this scenario. Each of the most massive galaxies in clusters A, B, \& D
has at least one other nearly equal mass companion within a small projected radius ($r\sim35-200$ kpc) and at a small differential velocity
($\delta_v\sim200-500$ km s$^{-1}$). This result is suggestive of such galaxies eventually merging to create galaxies with similar masses to
$z\sim0$ BCGs (as in, e.g., Tran et al.\ 2008; Jeltema et al.\ 2008). This result will be explored further in a future paper (Ascaso et al.\ 2012, in preparation).
For now, we simply state here that our data is
broadly consistent with a scenario in which the most massive RSGs gain mass through dry merging processes. Taken as
a whole, the picture presented in this section is consistent with the ``mixed" galaxy evolution scenario favored by Faber et al.\ (2007), in
which galaxies on the red sequence transition there through both early and late quenching processes and buildup mass
through dry mergers.

The relative importance of these early and late quenching processes seems, however, to vary greatly from system
to system in the Cl1604 supercluster. While the mass of the typical galaxy affected by late quenching processes 
correlates well with the global dynamical state of a system, the efficiency of early quenching does not seem to correlate well with either
the global dynamical state or the optically derived mass of the system. In all systems it appears that dry merging or minor mixed merging
likely plays a significant role in building up the mass of the BCG from $z\sim1$ to $z\sim0$, but it is unclear how important such
processes are in the mass evolution of low- to intermediate-mass RSGs. To fully characterize the relative importance
of quenching and merging processes at $z\sim1$, as well as to investigate the physical mechanisms responsible for these processes,
it is necessary to study a statistical sample of high-redshift
groups and clusters. In a future paper we will use the full ORELSE sample, consisting of over 40 high-redshift group and cluster systems,
to investigate these questions.

\section{Summary and Conclusions}

In this paper we studied the properties of the 525 spectroscopically confirmed members of the Cl1604 supercluster at
$z\sim0.9$, focusing in particular on the 305 member galaxies of the eight clusters and groups that comprise the
supercluster. With this spectroscopic sample, unprecedented for large scale structures at this redshift, we
explored the magnitude, color, stellar mass, spectral, morphological, and radial properties of cluster and group galaxies at
$z\sim0.9$. Through this exploration we gained a cohesive picture of galaxy evolution and the buildup of the
red sequence in the Cl1604 supercluster. Our main conclusions are:

\begin{itemize}
\renewcommand{\labelitemi}{$\bullet$}

\item Considering the color and magnitude properties of the Cl1604 members, we found that a large
fraction of the RSGs (and nearly all of the bright ones) are contained within the
group and cluster environments. The red-sequence fraction of \emph{both} the cluster and composite group populations
is 47\%, compared with only 23\% in the supercluster ``field". Many bright RSGs are observed in several
of the group systems, suggesting significant pre-processing is occurring in these environments at $z\sim0.9$.

\item Measuring the composite spectral properties of member galaxies of the Cl1604 clusters, we found the average
cluster galaxy at $z\sim0.9$ exhibited features indicative of a star-forming galaxy, forming stars at a level
roughly between field galaxies at $z\sim1$ and
cluster galaxies at $z\sim0.4$. While the average galaxy in the groups of the Cl1604 supercluster exhibited a large
variety of spectral properties, combining all group galaxies in to a single population, we found that the average group
galaxy at $z\sim0.9$ was undergoing a starburst.

\item An analysis of the red-sequence LF of the member galaxies of the three Cl1604 clusters
and a composite population of the member galaxies of the five Cl1604 groups (i.e., the ``groups" sample)
revealed differences in the number of bright galaxies in each of the systems. Bright ($L_{B}\sim10^{11}
L_{\odot}$) red galaxies were observed in the two most massive clusters (A \& B) as well as the groups.
However, such galaxies were noticeably absent from the least massive cluster (cluster D). We also observed a significant deficit
of low-luminosity ($L_{B}\lsim 2.5\times10^{10} L_{\odot}$) red-sequence galaxies in all systems except the
most relaxed cluster of the Cl1604 supercluster (cluster A), suggesting that low-luminosity galaxies have
largely not transitioned onto the red sequence by $z\sim0.9$.

\item While bright blue galaxies are observed in all systems, massive ($\log(\mathcal{M}_{\ast})\gsim10.5 \mathcal{M}_{\odot}$)
blue galaxies are observed almost exclusively in cluster D and the groups. The few massive blue galaxies
that belong to the galaxy population in cluster B tend to avoid the cluster core. The bright
blue galaxies observed in the two most massive Cl1604 clusters (A \& B) are primarily low-mass galaxies at
low clustocentric radii. This population has an average SSFR considerably higher
than the low-mass galaxies observed in cluster D and the groups, suggesting that the cores of these massive
clusters are inducing star formation in such galaxies.

\item A large fraction of brightest and most massive RSGs in the Cl1604 cluster systems are
observed at low clustocentric radii. In the group systems, massive and bright RSGs are observed
in a continuous distribution out to a projected distance $2R_{\rm{vir}}$ from the group centers. A large population
of bright, lower mass [$\log(\mathcal{M}_{\ast})\lsim10.8$] RSGs is observed at the center of cluster
D, suggestive of a population that has recently transitioned onto the cluster red sequence.

\item A large fraction of the RSGs observed in both the clusters and groups was found to be
morphologically early type. Transitional populations of red passive disks and blue ETGs are observed
in differing amounts in all systems, primarily in the stellar mass range $\log(\mathcal{M}_{\ast})\sim10.25$-$10.75$,
suggesting that this mass range is instrumental in the buildup of the red sequence at $z\sim0.9$.
In cluster B we found that galaxies transitioning to the red sequence were quenched of their star formation
largely prior to their morphological transformation. In the lower mass cluster and the group systems
the quenching of star formation was found to occur prior to morphological transformation for only
$\sim50$\% of transitional galaxies.

\item The average stellar populations of RSGs within $R<1.2R_{\rm{vir}}$ of the two most massive Cl1604 clusters are 
broadly consistent with formation through a single burst of star formation at $z_f=2.5-3$. Surprisingly, the average stellar
population of the group RSGs within $R<1.2R_{\rm{vir}}$ of the group centers are found to be consistent with formation 
through a single burst at even higher redshifts ($z_f\gsim3$). In contrast, the average red galaxy in the cluster D, the least massive Cl1604
cluster, is found to have only recently transitioned on to the cluster red sequence.

\item Massive cluster galaxies [$\log(\mathcal{M}_{\ast})\gsim12$] with properties similar to those observed
in BCGs at $z\sim0$ are absent from all Cl1604 clusters and groups at $z\sim0.9$.
We suggest that either dry merging or minor mixed merging may be important in building up
such galaxies from $z\sim0.9$ to $z\sim0$. This topic will be the subject of a future study.

\item Galaxies transitioning on to the red sequence were found to be at significantly different masses
in each of the cluster and group systems in the Cl1604 supercluster. Furthermore, this mass was found to correlate
well with the dynamical state of the system, in that the typical mass of such a galaxy decreased with 
increased virialization. We presented evidence for ``dynamical downsizing", a
process in which massive blue cloud galaxies are quenched earliest in the most dynamically relaxed systems
and at progressively later times in dynamically unrelaxed systems.

\end{itemize}

While this work represents only a case study of galaxy evolution
in dense environments at $z\sim0.9$, it is important to note that the supercluster structure contains three clusters
and five groups at essentially a single epoch. Furthermore, the Cl1604 clusters and groups 
are largely isolated from one another and are in very different stages of assembly. Though we stress that conclusions
drawn from a study of a single structure or even several groups and clusters at high redshift are limited in their
capacity to constrain the processes governing galaxy evolution, the
comprehensive dataset available for this system has allowed us to study the galaxy population in this particular collection
of groups and clusters in great detail. In future work we will extend this analysis to the remaining 19 ORELSE fields,
minimizing the effects of cosmic variance and allowing us to study galaxy evolution in a statistical sample of groups
and clusters at high redshift.

~~
~~

We thank Jeff Newman and Michael Cooper for guidance with the \emph{spec2d} reduction
pipeline and for the many useful suggestions and modifications necessary to
reduce our DEIMOS data. We also thank Adam Stanford, Bego\~{n}a Ascaso, and an anonymous referee 
for their careful reading of the manuscript and for several useful suggestions. B.C.L also 
thanks both Daisys for being supportive and patient throughout the entire process of this work, even
when it was not deserved. We also thank the Keck II support astronomers for their dedication, knowledge, and
ability to impart that knowledge to us at even the most unreasonable of hours.
Support for this research was provided by the National Science Foundation under grant 
AST-0907858. In addition, we acknowledge support from program number HST-GO-11003 which was provided by NASA through
a grant from the Space Telescope Science Institute, which is operated by the Association of Universities for Research
in Astronomy, Incorporated, under NASA contract NAS5-26555.  Portions of this work are also based in part on observations
made with the \emph{Spitzer Space Telescope}, which is operated by the Jet Propulsion Laboratory, California Institute of
Technology under a contract with NASA. Support for this work was provided by NASA through an award issued by JPL/Caltech.
The spectrographic data presented herein were obtained at the W.M. Keck Observatory, which is operated
as a scientific partnership among the California Institute of
Technology, the University of California, and the National Aeronautics
and Space Administration. The Observatory was made possible by the
generous financial support of the W.M. Keck Foundation. We wish to thank the indigenous
Hawaiian community for allowing us to be guests on their sacred mountain; we
are most fortunate to be able to conduct observations from this site.

\appendix

\section{\normalsize{\bf{Appendix A:} Multiband Photometry of the C\lowercase{l}1604 Supercluster}}

Initial photometry of the LFC $r\arcmin i\arcmin z\arcmin$ imaging of the Cl1604 field was performed with
SExtractor run in dual-image mode. This process involved the use of a deep combined $r\arcmin i\arcmin z\arcmin$
frame for object detection, while magnitudes were measured in the individual band images. Variable diameter
elliptical apertures were used, with major axis radius 2$r_{K}$, where $r_{K}$ is the
Kron radius (Kron 1980; Bertin \& Arnouts 1996), outputted as MAG\_AUTO in SExtractor.
These apertures are determined in the deep images and used to extract the photometry
in each of the individual band images. The virtue of this process is that a single, identical aperture
is used for the photometry in all three bands, which reduces biases introduced to color measurements
by integrating the light within the same physical scale for each galaxy (Lubin et al.\ 2000). A full discussion of
this process can be found in Gal et al.\ (2005).

Following these initial steps, the LFC data were calibrated to SDSS data release 5 (DR5;
Adelman-McCarthy et al.\ 2007) imaging, which spanned the entirety of our two LFC
pointings in the Cl1604 field. The LFC $r\arcmin i\arcmin z\arcmin$ magnitudes
were compared to SDSS ``\emph{modelmags}"\footnote{Detailed information can be found at
http://www.sdss.org/dr7/algorithms/photometry.html} for all objects that were
detected in both the LFC and SDSS imaging. The SDSS \emph{modelmags} have the advantage that,
while still using a single aperture to measure the photometry in each band, a S$\acute{e}$rsic model
(i.e., either an exponential disk or a de Vaucouleurs profile) is fit to each object in the
$r\arcmin$ band and applied to each individual band image to correct for the effects of
using a finite aperture. The model is truncated for each galaxy
measurement such that $>$99.3\% of the (model dependent) flux from each galaxy is recovered in
the $r\arcmin$ band, with a similar percentage recovered in the other bands. While the SDSS imaging does
not go nearly as deep as our own LFC imaging, this
calibration process allowed us to make bulk corrections to our LFC magnitudes, which results
in more self-consistent photometry and less aperture-induced bias when comparing our optical
magnitudes to those in the NIR. Further details on this calibration process
can be found in G08.

The photometry of the WIRC $K_s$ imaging was obtained by running SExtractor on the final images,
with object detection performed on a $5\times5$ pixel Gaussian smoothed image. All objects with more
than 8 contiguous pixels above $1.1\sigma$ were cataloged. The photometric zero point for each deep image
was found by comparison to the Two Micron All Sky Survey (2MASS; Skrutskie et al. 2006). The overlap
regions between pointings were checked, and
in the case of duplicate detections the one with the higher S/N was retained.  Based on the overlap
regions, typical astrometric errors are $\sim0.1''$ in each coordinate.  The 2MASS $K$-band magnitudes were
transformed to the AB system assuming $K_{s,\rm{AB}}=K_{\rm{2MASS}}+1.84$ (Ciliegi et al.\ 2005). Magnitudes
from the WIRC data were drawn from the SExtractor output MAG\_AUTO. It is important to note that, since
this reduction was done independently of the LFC reduction, the physical scale of the aperture used to
compute the WIRC magnitudes is not necessarily the same one used in the optical imaging. However,
a comparison of the $k$-corrected $K_{s}$ WIRC magnitudes and UKIRT $K$-band magnitudes (which are aperture
corrected, see below) for a subset of the spectroscopically confirmed members of the Cl1604 supercluster
exhibits only a small systematic offset ($\sim0.1$ mag) in the WIRC magnitudes for the very brightest
and faintest galaxies in the supercluster. For a majority of the supercluster members there is no systematic
offset between the two sets of magnitudes. We also make further attempts to correct zero-point offsets
between filters when performing the SED fitting (see \S\ref{SEDfitting}). Thus, it is likely that the small systematic
offsets between the $K_{s}$ band and magnitudes measured from our other imaging does not introduce significant
biases into our fitting process, especially in the case of supercluster members.

\emph{Spitzer} IRAC 3.6/4.5/5.8/8.0$\mu$m photometry was obtained with SExtractor run in
dual-image mode. For these data the 3.6$\mu$m image was used for object detection in
the region containing the supercluster, while the magnitudes were measured in the
individual band images. A stacked, four-band image was not used for
object detection in this case, as the point spread function (PSF) degrades significantly
in the bands longward of 3.6$\mu$m. A fixed aperture of 1.9$\arcsec$ was used to extract magnitudes
for all bands, roughly chosen to match the PSF in the 3.6$\mu$m image. For blended sources we used
the IRAF task \emph{daophot} to perform an iterative PSF fitting and subtraction in order to
remove contamination from neighboring objects. For all supercluster members a multiplicative
aperture correction of 1.36, 1.40, 1.65, and 1.84 was applied to the measured fluxes in the
3.6, 4.5, 5.8, and 8.0$\mu$m bands, respectively. These corrections are appropriate for a galaxy at $z\sim0.9$,
which appears as a PSF to \emph{Spitzer}, and a measurement aperture of 1.9$\arcsec$
\footnote{See http://swire.ipac.caltech.edu//swire/astronomers/public ations/SWIRE2\_doc\_083105.pdf, Table 9.}.

Photometry of the \emph{HST} ACS data was obtained by SExtractor run in dual-image mode. A
deep $F606W$+$F814W$ image was used for object detection, while magnitudes were obtained in the
single band images using an identical aperture for both $F606W$ and $F814W$ and drawn from the
MAG\_AUTO output of SExtractor (for more details see Kocevski et al.\ 2009b and references therein). Since the \emph{HST}
ACS data were not used in our SED fitting\footnote{The trouble of including \emph{HST} ACS data in our SED fitting process is
the significant wavelength overlap of the ACS filter set with the bands chosen for our ground-based optical imaging and
the higher resolution of the ACS imaging with respect our other imaging data of the supercluster. The latter
makes the process of matched aperture photometry extremely difficult. Significant
headway has been made in the astronomical community in this area, and it is likely that we will include ACS data in future
SED fitting.}, no attempt was made to match the apertures used here
to the ones used in the LFC/WIRC or \emph{Spitzer} imaging. For this study, the only important
feature of this photometry is that it be self-consistent between the two \emph{HST} ACS bands,
which we have ensured by choosing a common aperture to measure the magnitudes in each band.

The UKIRT $K$-band imaging, as with the \emph{HST} ACS data, was not used in our SED fitting.
As such, we only briefly describe the photometry. Photometry catalogs of the UKIRT
observations are provided by the Cambridge pipeline. An aperture corrected magnitude with
a fixed aperture of 2$\arcsec$ (``kAperMag3" in the Cambridge nomenclature) was chosen for
all Cl1604 member galaxies as it most closely resembled the methodology used for obtaining
magnitudes in the other imaging data. While we again did not make any attempt to match
the apertures of the UKIRT $K$-band magnitudes to those in other bands, the aperture correction
performed on these magnitude measurements should, in principle, allow for consistent
comparison between these data and those in other bands.

\section{\normalsize{\bf{Appendix B:} Red-Sequence Fitting of the C\lowercase{l}1604 Clusters and Groups}}

For each Cl1604 cluster, a $\chi^2$ minimization to a linear model of the form:

\begin{equation}
F606W-F814W = y0 + m\times F814W
\label{eqn:RS}
\end{equation}

\noindent was performed on the member galaxies within a certain range of colors and magnitudes (Gladders et al.\ 1998; Stott et al.\
2009). The color range was defined by an initial ``by eye" estimate of the color range of the red sequence of each
system, while the magnitude limits were defined as the limit at which photometric errors were reasonably small
($\sigma_{F814W}\leq0.05$). For most systems these resulted in a magnitude limit of $F814W<23.5$ and a
color range of $1.6 < F606W-F814W < 2.1$. Following the results of the fitting, the color slope of each system is
removed and the resulting ``corrected" color distribution of the galaxies in each system is fit to a single
Gaussian using iterative $3\sigma$ clipping. The formal red sequence is then defined as the color$-$magnitude envelope
bounded by a $\pm3\sigma$ departure from the relationship given in Equation \ref{eqn:RS}. The slope, intercept, and
width of the red sequence for each of the three Cl11604 clusters are given in Table
\ref{tab:RSparameters}. Fitting was performed on galaxies within a projected radius of $R_{\rm{vir}}$,
1.5$R_{\rm{vir}}$, and 2$R_{vir}$, but we find that there is little difference between the three fits. In
this paper we used the values derived using $R<R_{\rm{vir}}$, as there is minimal
contamination from bluer galaxies in most of the systems at these radii.

While we performed red sequence fitting on the individual populations of each of the three clusters
(i.e., A, B, \& D), there are too few members in any individual group system for us to perform the fitting on a
single group. Instead, the individual group populations were combined into one ``group" sample.
Red sequence fitting was performed on this composite population in each of the three radial bins
($R_{\rm{vir}}$, 1.5$R_{vir}$, and 2$R_{vir}$) using the methodology described above. A single ``group" red-sequence
fit is reported in Table \ref{tab:RSparameters}. This fit is used to discriminate red and blue
galaxies in each of the individual group systems.

Since the Cl1604 groups contain populations of galaxies at moderately different redshifts,
the \emph{HST} ACS observations probe slightly different
rest-frame colors for each group, which tends to artificially inflate the scatter in
the combined CMD. However, we compute a maximum difference of $\delta(F606W-F814W)\sim0.08$
for an elliptical template (Maraston 1998, 2005) spanning the redshift range of the Cl1604 groups. This value does not change
significantly when an equivalent model from Bruzual (2007) is used. While this effect is not trivial for precision measurements, here we simply 
adopted the best-fit relation in observed color$-$magnitude space for the combined group sample and used the $\pm2\sigma$
departure from this relation as the boundaries of the group red sequence. The choice of the latter is motivated
by the redshift effects discussed above, which artificially inflate the observed color scatter at roughly the $1\sigma$ level.
The $2\sigma$ width of the group red sequence corresponds to a slightly higher width than the $3\sigma$ width of
any of three Cl1604 clusters, which may be expected of systems still in the process of formation
(i.e., Homeier et al.\ 2006a; Mei et al.\ 2009).

\section{\normalsize{\bf{Appendix C:} Spectral Errors from Sampling and Incompleteness}}

To estimate the error and any possible biases spectroscopic selection effects and incompleteness might have on the quantities 
measured from composite spectra [i.e., EWs and $D_n(4000)$] we performed a bootstrap analysis on the composite spectra of all Cl1604 
groups and clusters. Bootstrap analysis was performed using a combination of the \emph{HST} ACS 
photometry and the DEIMOS/LRIS spectroscopic information in the
following manner. The observed ACS CMD for the Cl1604 supercluster was separated into bins spanning 0.5 mag in color and
1 mag in brightness. For the $i^{th}$ magnitude bin and the $j^{th}$ color bin, the redshift probability
distribution function, $P(z)$, is defined as:

\begin{equation}
P_{i,j}(z) = \frac{N_{\rm{mem},i,j}}{(N_{\rm{mem},i,j}+N_{\rm{non-mem},i,j})}
\label{eqn:Pz}
\end{equation}

\noindent where $N_{\rm{mem},i,j}$ is defined as any galaxy in that bin with a secure spectroscopic redshift
within the range of the supercluster, $0.84 < z < 0.96$, and $N_{\rm{non-mem},i,j}$ is defined as any object in that
bin with a secure redshift outside this range. This probability was calculated in
each bin over the color range $-0.5 \leq F606W-F814W \leq 3.0$ and a magnitude range of $18 \leq F814 \leq 24$.
The number of supercluster members missed by our spectroscopy is then:

\begin{equation}
N_{\rm{missed},i,j} = P_{i,j}(z)\times(N_{\rm{phot},i,j}+N_{\rm{bad},i,j})
\label{eqn:Nmissed}
\end{equation}

\noindent where $N_{\rm{phot},ij}$ is the number of photometrically detected objects within that color$-$magnitude
bin that were not targeted for spectroscopy and $N_{\rm{bad},i,j}$ is the number of low quality spectra in that bin
for which the redshift was uncertain. For almost all of the 42 color$-$magnitude bins defined in this manner
we have obtained spectral information on some non-zero fraction of the photometric objects in that
bin. For those color$-$magnitude bins that are highly populated with Cl1604 members
(i.e., $21 \leq F814W \leq 23$, $1 \leq F606W-F814W \leq 2.5$) the fraction of photometrically detected
objects for which we have spectral information is quite high, ranging from $\sim$30\% to greater than 80\%.

For each Cl1604 group and cluster, the effect of this incompleteness on
EW([\ion{O}{2}]), EW(H$\delta$), and $D_n(4000)$ values measured from our composite spectra is estimated in
the following way. For all color$-$magnitude bins in which $N_{\rm{missed},i,j}\neq0$, the spectra of $N_{\rm{missed},i,j}$
galaxies were randomly drawn from the observed supercluster members in that color$-$magnitude bin.
These randomly drawn galaxies are included in a new ``completeness-corrected" composite spectrum along
with the original members of that particular system such that the number of galaxies in the new composite
spectrum is:

\begin{equation}
N_{\rm{comp}} = \sum_{i=0}^n\sum_{j=0}^m N_{\rm{mem},i,j} + N_{\rm{missed},i,j}
\label{eqn:Ncomp}
\end{equation}

For each structure the completeness-corrected composite spectrum is generated 500 times and the EW([\ion{O}{2}]),
EW(H$\delta$), and $D_n(4000)$ are measured for each realization. While the number of randomly drawn supercluster members
used to make the completeness correction remains constant in each realization for a particular structure, the galaxies
used to make the completeness correction change for each realization. For each structure, a distribution of
EW([\ion{O}{2}]), EW(H$\delta$), and $D_n(4000)$ values is generated from the 500 realizations of the completeness corrected
composite spectrum. The ``incompleteness error" is calculated from the second moment of this distribution for each structure and
is assumed to be Gaussian. This error is added in quadrature to the random errors on the EWs and the $D_n(4000)$ measurements in 
cases where the full Cl1604 galaxy sample is used (i.e., not subsets of brighter galaxies, as in \S\ref{discussion}). 
In addition to performing this analysis for each of the eight groups and clusters of the supercluster, an identical analysis was
performed on a composite spectrum which included members of all five of the Cl1604 group systems (the ``Groups" sample).

The galaxies included when estimating the effects of completeness are drawn not only from the denser regions of the supercluster
(i.e., $R<2R_{\rm{vir}}$ from a group or cluster center) but from all environments. In such a way we attempt
to simulate the \emph{maximum possible variance} of the spectral quantities measured from the composite spectra due to
incompleteness. By using such a method we are inherently assuming that the spectral properties of observed supercluster members
in any given color$-$magnitude bin are similar to those supercluster galaxies for which we do not
have spectral information. If the spectroscopically undetected supercluster members have significantly different
spectra, this method loses its effectiveness. However, in any given color$-$magnitude bin we observe significant
variance of the spectral properties of confirmed member galaxies, which gives us confidence that this method is a reasonable
approximation of the true error due to our incomplete spectral sampling.

\end{document}